\documentclass[12pt,twoside]{article}
\usepackage{times,amssymb,cite,amscd}
\newcommand{\be}{\begin{eqnarray}}
\newcommand{\ee}{\end{eqnarray}}
\newcommand{\bez}{\begin{eqnarray*}}
\newcommand{\eez}{\end{eqnarray*}}
\newcommand{\pa}{\partial}
\newcommand{\la}{\lambda}
\newcommand{\A}{{\cal A}}
\newcommand{\res}{\mathrm{res}}
\newcommand{\hP}{\mathbf{\hat{P}}}
\newcommand{\tA}{\tilde{\cal A}}
\newcommand{\hA}{\mathbf{\hat{A}}}
\newcommand{\cA}{\mathbf{\check{A}}}
\newcommand{\hB}{\mathbf{\hat{B}}}
\newcommand{\cB}{\mathbf{\check{B}}}
\newcommand{\htimes}{\,{\hat{\times}}\,}
\newcommand{\ctimes}{\,{\check{\times}}\,}
\newcommand{\cP}{\mathbf{\check{P}}}
\newcommand{\bT}{\mathbf{T}}
\newcommand{\cQ}{\mathbf{\check{Q}}}
\newcommand{\vtr}{\vartriangle}

\renewcommand{\theequation} {\arabic{section}.\arabic{equation}}

\title{\bf An algebraic scheme associated with the noncommutative KP hierarchy \\
           and some of its extensions
\thanks{\copyright 2004 by A. Dimakis and F. M\"uller-Hoissen} }

\author{Aristophanes Dimakis \\
 Department of Financial and Management Engineering, \\
 University of the Aegean, 31 Fostini Str., GR-82100 Chios, Greece \\
 dimakis@aegean.gr
          \and
 Folkert M\"uller-Hoissen \\ Max-Planck-Institute for Dynamics and Self-Organization \\
 Bunsenstrasse 10, D-37073 G\"ottingen, Germany \\
 fmuelle@gwdg.de }

\date{}

\begin{document}

\newtheorem{theorem}{Theorem}[section]
\newtheorem{lemma}{Lemma}[section]
\newtheorem{proposition}{Proposition}[section]
\maketitle

\begin{abstract}
A well-known ansatz (`trace method') for soliton solutions turns the equations of
the (noncommutative) KP hierarchy, and those of certain extensions, into
families of algebraic sum identities. We develop an algebraic formalism, in particular
involving a (mixable) shuffle product, to explore their structure. More precisely,
we show that the equations of the noncommutative KP hierarchy and its extension (xncKP)
in the case of a Moyal-deformed product, as derived in previous work, correspond
to identities in this algebra. Furthermore, the Moyal product is replaced by
a more general associative product.
This leads to a new even more general extension of the noncommutative KP hierarchy.
Relations with Rota-Baxter algebras are established.
\end{abstract}


\small
\tableofcontents
\normalsize

\section{Introduction}
\setcounter{equation}{0}
Let $\mathbb{K}$ be a field of characteristic zero and
$(\mathcal{R}_0, \ast)$ the $\mathbb{K}$-algebra of differential polynomials
in (matrices of) functions $\{ u_{n+1} \, | \, n \in \mathbb{N} \}$ of
variables $t_n$, $n \in \mathbb{N}$, with an associative (and noncommutative)
product $\ast$ for which the operators of partial differentiation with respect to
$t_n$, $n \in \mathbb{N}$, are derivations.\footnote{Here and in the following
$\mathbb{N}$ denotes the natural numbers \emph{not} including zero.}
A formal pseudo-differential operator ($\Psi$DO) in the following means a formal series
in the operator\footnote{An expression like $\pa X$ has to be understood as a
product of operators, whereas $\pa_x X$ will be used for the partial derivative
of $X$ with respect to $x$, also denoted as $X_x$.}
$\pa$ of partial differentiation with respect to $x := t_1$ and its formal inverse
$\pa^{-1}$ with coefficients in $(\mathcal{R}_0, \ast)$. With elements $f \in \mathcal{R}_0$,
$\pa^{-1}$ satisfies the relation
\be
   \pa^{-1} f = f \, \pa^{-1} - f_x \, \pa^{-2} + f_{xx} \, \pa^{-3} - \ldots \; .
\ee
We will use $( \; )_{\geq 0}$ and $( \; )_{<0}$, respectively, to denote the projection
to that part of a $\Psi$DO which only contains non-negative, respectively negative,
powers of $\pa$.
Let $\mathcal{R}$ be the ring of $\Psi$DOs generated by
\be
    L = \pa + \sum_{n \geq 1} u_{n+1} \pa^{-n}    \label{L}
\ee
using the product $\ast$, the projections, and the defining relation for $\pa^{-1}$ as
the inverse of $\pa = L_{\geq 0}$. This is also a $\mathbb{K}$-algebra.
In the Sato framework, the noncommutative KP hierarchy (ncKP) is defined
by\footnote{Here $L^n$ stands for the $n$-fold product $L \ast \ldots \ast L$,
and $[ \, , \, ]$ is the commutator in the ring $(\mathcal{R} , \ast)$.}
\be
      L_{t_n}
  :=  \pa_{t_n} L
   = [(L^n)_{\geq 0}, L]
   = - [(L^n)_{<0}, L]    \qquad \quad n = 1,2, \ldots     \label{KP-Lax}
\ee
(see \cite{EGR97,Kupe00,Dick03,Hama03b,DMH04hier,DMH04ncKP,DMH04extMoyal}, for example).
Introducing a potential $\phi$ via
\be
        u_2 = \phi_x
\ee
one finds the following expressions for the commuting flows of the ncKP
hierarchy,\footnote{To be precise, here we need to supply $\mathcal{R}$ with the
operation of $x$-integration. The residue of a $\Psi$DO is the coefficient of its
$\pa^{-1}$ term.}
\be
    \phi_{t_n} = \res(L^n)    \qquad \quad n = 1,2, \ldots     \label{phi_tn-res}
\ee
\vskip.1cm

Let us now recall a method\footnote{For a different method in the noncommutative setting, see
\cite{EGR97}, for example.}
\cite{Okhu+Wada83} to obtain soliton solutions of the (potential) \emph{ncKP equation}
\be
   ( 4 \, \phi_{t_3} - \phi_{xxx} - 6 \, \phi_x \ast \phi_x )_x
    = 6 \, [\phi_y , \phi_x] + 3 \, \phi_{yy}   \label{ncKPpot}
\ee
where $y := t_2$. This is the first non-trivial member of the ncKP hierarchy.
Inserting the formal series
\be
  \phi = \sum_{N=1}^\infty \epsilon^N \phi^{(N)}   \label{eps-expansion}
\ee
in a parameter $\epsilon$, transforms it into the system of equations
\be
   4 \, \phi^{(N)}_{t_3 x} - \phi^{(N)}_{xxxx} - 3 \, \phi^{(N)}_{yy}
 = 6 \, \sum_{k=1}^{N-1} \Big( ( \phi^{(k)} \ast \phi^{(N-k)} )_x
   + [ \phi^{(k)}_y , \phi^{(N-k)}_x ] \Big)   \label{ncKPpotN}
\ee
which is solved by
\be
    \phi^{(k)}
  = \sum_{i_1,\ldots,i_k=1}^M \frac{\phi_{i_1} \ast \phi_{i_2} \ast \ldots \ast \phi_{i_k}}
    {(q_{i_1}-p_{i_2})(q_{i_2}-p_{i_3}) \cdots (q_{i_{k-1}}-p_{i_k})}
    \qquad \quad  k=1,\ldots,N    \label{phiN}
\ee
with
\be
  \phi_k = c_k \, e^{\xi(t,p_k)} \ast e^{-\xi(t,q_k)}     \label{phi_k_xi}
\ee
where $M \in \mathbb{N}$, $\xi(t,p_k) = \sum_{r \geq 1} t_r \, p_k^r$ (see also
\cite{Kupe00,Pani01,DMH04ncKP}).\footnote{$M$ is the soliton number. For $M=1$ we
can use the geometric series formula (in the domain of convergence of the series)
to obtain $\phi = \sum_{N=1}^\infty (\epsilon \phi_1/(q_1-p_1))^N
= (1-\epsilon \phi_1/(q_1-p_1))^{-1} -1$ from which one recovers a well-known expression
for the 1-soliton solution of the KP equation.}
Here $c_k, p_k, q_k$ are constants such that $c_k$ and the denominators in (\ref{phiN})
are different from zero.
Inserting (\ref{phiN}) with (\ref{phi_k_xi}) in (\ref{ncKPpotN}) first leads to a
sum which runs over all lists $(i_1, \ldots, i_N)$ where $i_k \in \{ 1, \ldots, M \}$.
But it actually results in \emph{separate} sum identities (of the same kind), involving
the constants $p_k, q_k$.
It is therefore sufficient to consider only the terms corresponding to one definite
representative list, say those proportional to $\phi_1 \ast \ldots \ast \phi_N$ (where some
of the $\phi_k$ may be equal).\footnote{For $N>M$ some of the $\phi_k$ necessarily have to
be equal, but for $N \leq M$ there are lists for which all factors $\phi_k$ are potentially
independent. We should consider such lists as `representative'. We thus assume that the
soliton number $M$ can be chosen arbitrarily large. The number $M$ then does not
enter the subsequent considerations any more.}
For example, the corresponding contribution of the expression
\be
    \phi^{(N)}_{t_r}
 = \sum_{i_1, \ldots, i_N =1}^M \sum_{k=1}^N (p_{i_k}^r - q_{i_k}^r) \,
   \frac{\phi_{i_1} \ast \ldots \ast \phi_{i_N}}{(q_{i_1}-p_{i_2}) \cdots (q_{i_{N-1}}-p_{i_N})}
\ee
is $T_r \, \phi_1 \ast \ldots \ast \phi_N/ \prod_{k=1}^{N-1} (q_k - p_{k+1})$ where
\be
       T_r := \sum_{k=1}^N (p_k^r - q_k^r) \; .
\ee
The $N$th order part (\ref{ncKPpotN}) of the ncKP equation is then mapped to the
following algebraic equation\footnote{We should stress again that each summand in
(\ref{phiN}) leads separately to this identity. The noncommutative terms on the right hand
side of (\ref{ncKP-T-id}) still remain present if we let the product $\ast$ become
commutative, i.e., in the case of the `commutative' KP hierarchy. In that case,
they disappear, however, via the summation in (\ref{phiN}). }
\be
    4 \, T_1 T_3 - T_1{}^4 - 3 \, T_2{}^2 = 6 \, T_1 (T_1 \times T_1)
    - 6 \, (T_1 \times T_2 - T_2 \times T_1)    \label{ncKP-T-id}
\ee
where
\be
    T_r \times T_s := \sum_{1 \leq i \leq j<k \leq N} (p_i^r-q_i^r) q_j (p_k^s-q_k^s)
    - \sum_{1\leq i<j\leq k\leq N} (p_i^r-q_i^r) p_j (p_k^s-q_k^s) \; . \label{TxT}
\ee
Equation (\ref{ncKPpotN}) is solved by (\ref{phiN}) if (\ref{ncKP-T-id}) is an identity,
which indeed turns out to be the case by closer inspection. Note that this identity not
only holds for arbitrary values of the $p_k, q_k$,
but also for arbitrary $N \in \mathbb{N}$.\footnote{Such identities typically decompose
into several identities since terms like $\sum_{1 \leq i \leq j<k \leq N} p_i p_j p_k$ and
$\sum_{1 \leq j<k \leq N} p_j p_k^2$, for example, are obviously linearly independent.
Accordingly, one can introduce a notion of \emph{length}, so that expressions decompose
into linearly independent parts of fixed length. In the step towards the algebra developed
in section~\ref{section:basic} we abstracted the above sums to products
$P \prec P \prec P$, respectively $P \prec P \bullet P$, from which they are recovered
via a representation $\Sigma_N$ (see section~\ref{section:ps}). The grading given
by $\prec$ takes care of the length. }
Inspection of the identity (\ref{ncKP-T-id}) suggests a way to obtain
such identities directly from ncKP equations.
The basic rules are\footnote{In the usual formulation of the ncKP hierarchy,
$\phi$ without derivatives acting on it does not appear, so we need not say to
what a bare $\phi$ should correspond.}
\be
     \phi_{t_{m_1} \ldots t_{m_k}} \mapsto T_{m_1} \cdots T_{m_k}
     \qquad \quad
     \phi_{t_r} \ast \phi_{t_s} \mapsto T_r \times T_s  \; .
\ee
Now (\ref{ncKP-T-id}) immediately follows from (\ref{ncKPpot}).
\vskip.1cm

Taking (\ref{eps-expansion}) with (\ref{phiN}) as an \emph{ansatz} to obtain solutions
of a partial differential equation involving the product $\ast$ and partial derivatives
of a field $\phi$ with respect to the variables $t_n$, turns it into an algebraic equation.
If this is an identity for all $N$, the respective equation has
KP-type soliton solutions.\footnote{Of course, one may think of modifications
of (\ref{phiN}) in the search for other equations admitting a soliton structure.}
Does the ncKP hierarchy exhaust the possibilities of such equations?
\vskip.1cm

In particular, we will be interested in the case where the product $\ast$ depends
on parameters. An example is given by the (Groenewold-) Moyal product
\cite{Groe46,Moya49,BFFLS78,Duet+Fred01}
\be
   f \ast g := \mathbf{m} \circ e^{P/2} (f \otimes g)
  \qquad\quad
   P := \sum_{m,n=1}^\infty \theta_{mn} \, \pa_{t_m} \otimes \pa_{t_n}
   \label{Moyal}
\ee
where $\mathbf{m}(f \otimes g) = f \, g$ for functions $f,g$, and
parameters $\theta_{nm} = - \theta_{mn}$.
Then there is another basic rule, namely
\be
 \pa_{\theta_{rs}} \mapsto \Theta_{rs} := {1 \over 2} \sum_{1 \leq j< k \leq N}
   [(p_j^r-q_j^r) (p_k^s-q_k^s) - (p_j^s-q_j^s) (p_k^r-q_k^r)]
 - {1 \over 2} \sum_{k=1}^N (p_k^r q_k^s - p_k^s q_k^r) \, . \quad
\ee
According to our correspondence rules, we have, for example,
\be
    \phi_{\theta_{rs} t_k } \ast \phi_{t_l} \ast \phi_{t_m}
    \quad \mapsto \quad
    (T_k \Theta_{rs}) \times T_l \times T_m
\ee
with a (rather obvious) generalization of (\ref{TxT}) which defines an associative product
(of sums of powers of $p_1, \ldots, p_N, q_1, \ldots, q_N$).
The first equation of the extension (in the sense of
\cite{DMH04hier,DMH04ncKP,DMH04extMoyal}) of the ncKP hierarchy (with Moyal $\ast$-product),
called \emph{xncKP hierarchy}, is
\be
    \phi_{\theta_{1,2}} = {1 \over 6} (\phi_{t_3} - \phi_{xxx}) - \phi_x \ast \phi_x \; .
       \label{xncKP-1eq}
\ee
This is mapped to
\be
    \Theta_{1,2} = {1 \over 6} (T_3 - T_1{}^3) - T_1 \times T_1  \label{xncKP-1id}
\ee
which indeed also turns out to be an identity.
\vskip.1cm

Hence, taking (\ref{eps-expansion}) with (\ref{phiN}) as an ansatz to obtain solutions of
a (in this case non-local) partial differential equation involving the Moyal product and
partial derivatives of a field $\phi$ with respect to the variables $t_r$ and $\theta_{mn}$,
converts it into an algebraic equation. If this is an identity for all $N$, the respective
equation has KP-type soliton solutions. The equations of the xncKP hierarchy provide us with
corresponding examples.
\vskip.1cm

The mapping of (x)ncKP equations to algebraic identities described above can
actually be reversed. From (\ref{ncKP-T-id}), respectively (\ref{xncKP-1id}),
we easily reconstruct the partial differential equations (\ref{ncKPpot}),
respectively (\ref{xncKP-1eq}). It should be clear that, in order to do this,
the sum calculus is not essential, but rather a certain algebraic abstraction.
This motivates to develop an algebraic scheme which allows us to prove
and to find identities of the kind we met above. The way in which we expressed
the identities (\ref{ncKP-T-id}) and (\ref{xncKP-1id}) already suggests some
main ingredients of such a scheme. A deeper analysis led us to the
algebra which we introduce in section~\ref{section:basic}.
A correspondence between identities holding in the abstract algebra and the equations
of the ncKP hierarchy and certain extensions is indeed established in this work.
In this context one should keep in mind that characteristic properties of the KP hierarchy
are indeed purely algebraic. In particular, this concerns the basic property of
commutativity of the flows. Writing (\ref{KP-Lax}) in the form
\be
     \pa_{t_n} L = \delta_n L    \qquad \quad
     \delta_n L := [ (L^n)_{\geq 0} , L ]
\ee
and extending $\delta_n$ to $\mathcal{R}$ according to the derivation rule (together
with $\delta_m X_{\geq 0} := (\delta_m X)_{\geq 0}$ for $X \in \mathcal{R}$), the
commutativity of the flows becomes equivalent to
\be
    [ \delta_m , \delta_n ] \, L = 0
\ee
which is a purely algebraic identity in the ring $\mathcal{R}$ (and in particular
makes no reference to the variables $t_n$, $n>1$).
Associated with the extension of the Moyal-deformed KP hierarchy are `generalized
derivations' which also commute as a consequence of algebraic identities.
We will meet even more generalized derivations in section~\ref{section:PsiDOs}. They
also define extensions of the KP hierarchy with a deformed product
(see section~\ref{section:XncKP}).
\vskip.1cm

The treatment of the xncKP hierarchy in \cite{DMH04hier,DMH04ncKP,DMH04extMoyal} heavily
relies on the fact that the underlying algebra $\mathcal{R}$ of $\Psi$DOs admits the
decomposition $\mathcal{R} = \mathcal{R}_{\geq 0} \oplus \mathcal{R}_{<0}$ into subalgebras,
whereas in the treatment of the ncKP hierarchy it is sufficient to have a corresponding
decomposition of Lie algebras (as common in integrable systems theory).\footnote{Also
$\mathcal{R} = \mathcal{R}_{\geq 1} \oplus \mathcal{R}_{<1}$ is a decomposition of the
algebra of $\Psi$DOs into subalgebras. This underlies the (extended) modified KP hierarchy
(see \cite{Kupe00,DMH04extMoyal}, for example).}
Such an algebra decomposition is equivalent to the existence of an idempotent
Rota-Baxter operator $R$ \cite{Baxt60,Rota69I,Rota+Smit72,Rota95} on the algebra
(see also appendix~A).
A few years ago it was shown that the choice of a renormalization scheme in
perturbative quantum field theory corresponds to the choice of a Rota-Baxter operator
\cite{Conn+Krei99,Krei03,Figu+Grac04,Manc04}.
In \cite{EGK04int1,EGK04int2} it has been pointed out that this setting resembles the
loop algebra framework of integrable systems. The antisymmetric part of the
bilinear Rota-Baxter relation (of weight 1) is the famous \emph{classical Yang-Baxter relation},
which plays an important role in integrable system theory \cite{Seme84,Seme02,Fadd+Takh87}.
It should not come as a surprise that various Rota-Baxter relations also appear
in the present work.
\vskip.1cm

Section~\ref{section:basic} introduces the algebra $\A$ which plays a basic role
in this work. Section~\ref{section:ps} then provides a realization in terms of
partial sum calculus. Some other realizations of the algebra $\A$ are briefly
described in appendix~B. Section~\ref{section:A(P)} treats the case of the subalgebra
$\A(P)$ of $\A$ generated by a single element $P$. This plays a central role in
the subsequent sections. Section \ref{section:A(P,Q)} addresses the case of a subalgebra
of $\A$ generated by two commuting elements and an embedding of $\A(P)$. Although this
section is important in order to make contact with the aforementioned algebraic sum
identities, it may be skipped on first reading.
Sections~\ref{section:PsiDOs} and \ref{section:xncKP} relate the algebraic
framework with the ncKP hierarchy and (in the case where $\ast$
is the Moyal product) its xncKP extension.
A more general extension, corresponding to a more general $\ast$-product
(see appendix~C), is studied in section~\ref{section:XncKP}.
Appendix~D sketches a certain generalization of the algebraic
framework which, in particular, allows to introduce an algebraic counterpart of
a Baker-Akhiezer function (formal eigenfunction of a Lax operator like $L$).
Section~\ref{section:concl} contains some conclusions and further remarks.

\section{The basic algebraic structure}
\label{section:basic}
\setcounter{equation}{0}
Let $\A = \bigoplus_{r \geq 1} \A^r$ be a graded linear space over a
field $\mathbb{K}$ of characteristic zero, which becomes an associative algebra
with respect to two products $\prec$ and $\bullet$, which are
bilinear maps $\A^r \times \A^s \rightarrow \A^{r+s}$ and $\A^r
\times \A^s \rightarrow \A^{r+s-1}$, respectively.\footnote{The grading
basically accounts for the notion of `length' mentioned in a previous footnote.}
Furthermore, we require that the two products satisfy the mutual associativity
conditions
\be
    ( \alpha \prec \beta ) \bullet \gamma
  = \alpha \prec ( \beta \bullet \gamma )    \qquad\quad
    ( \alpha \bullet \beta ) \prec \gamma
  = \alpha \bullet ( \beta \prec \gamma )
\ee
for all $\alpha, \beta, \gamma \in \A$.
It is convenient to introduce the notation
\be
   \alpha \succ \beta := \alpha \prec \beta + \alpha \bullet \beta  \label{succ}
\ee
for the combined product which is clearly also associative. This new product induces
a different grading of the algebra: $\A = \bigoplus_{r \geq 1} \A_r$, where $\A_1 = \A^1$
and $\A_r \succ \A_s \subseteq \A_{r+s}$. We also have
$\A_r \bullet \A_s \subseteq \A_{r+s-1}$ and
$\A_r \prec \A_s \subseteq \A_{r+s-1} \bigoplus \A_{r+s}$.
\vskip.1cm

Let $\mathrm{Shuff}(m,n)$ denote the set of $(m,n)$-\emph{shuffles}, i.e.
\be
    \mathrm{Shuff}(m,n) := \{\sigma \in {\cal S}_{m+n} \, | \, \sigma^{-1}(1) < \ldots
    < \sigma^{-1}(m), \; \sigma^{-1}(m+1) < \ldots < \sigma^{-1}(m+n) \}
\ee
where ${\cal S}_n$ is the symmetric group acting on $n$ letters. For example,
\bez
  \mathrm{Shuff}(1,n) &=& \{ \, \{ 1,2,\ldots,n+1 \}, \, \{ 2,1,3,\ldots,n+1 \}, \ldots, \,
  \{ 2,3,\ldots,n+1,1 \} \, \}  \\
  \mathrm{Shuff}(2,2) &=& \{ \, \{ 1,2,3,4 \}, \, \{ 1,3,2,4 \}, \, \{ 1,3,4,2 \},
  \, \{ 3,1,2,4 \}, \, \{ 3,1,4,2 \}, \, \{ 3,4,1,2 \} \, \}
\eez
where a permutation $\sigma$ is described by the ordered set $\{ \sigma(1), \ldots, \sigma(m+n) \}$.
Taking a deck of $m$ cards and another one of $n$ cards, $\mathrm{Shuff}(m,n)$
describes all possible shuffles of the two decks. It has $(m+n)!/(m! \, n!)$ elements.
Clearly, $\mathrm{Shuff}(m,n) = \mathrm{Shuff}(n,m)$.
\vskip.1cm

We define the \emph{main product} $\circ$ in $\A$ by
\be
 \lefteqn{(A_1 \curlywedge_1 \ldots \curlywedge_{m-1} A_m) \circ (A_{m+1} \curlywedge_{m+1}
 \ldots \curlywedge_{m+n-1} A_{m+n})}
 \hspace*{5cm} && \nonumber \\
 & & := \sum_{\sigma \in \mathrm{Shuff}(m,n)} A_{\sigma(1)} \curlywedge'_{\sigma(1)}
     \ldots \curlywedge'_{\sigma(m+n-1)} A_{\sigma(m+n)}
\ee
for $A_1, \ldots, A_{m+n} \in \A^1$. Each $\curlywedge_i$, $1 \leq i \leq m+n-1$,
stands for one of the choices $\prec$ or $\succ$, and
\be
 \curlywedge'_{\sigma(i)} :=
  \left\{ \begin{array}{ll}
  \succ   & \mbox{if } \sigma(i) \leq m < \sigma(i+1) \\
  \prec   & \mbox{if } \sigma(i+1) \leq m < \sigma(i) \\
  \curlywedge_i & \mbox{otherwise}
  \end{array} \right.
\ee
This defines another associative product in $\A$. It is a \emph{mixable shuffle product}
\cite{Guo+Keig00shuffle,Guo00diff} with respect to the product pair $(\prec,\bullet)$,
respectively $(\succ,\bullet)$. In particular, we find
\be
 & & ( A_1 \curlywedge_1 A_2 ) \circ ( A_3 \curlywedge_3 A_4 )
  = \sum_{\sigma \in \mathrm{Shuff}(2,2)} A_{\sigma(1)} \curlywedge'_{\sigma(1)} \ldots
     \curlywedge'_{\sigma(3)} A_{\sigma(4)} \nonumber \\
 &=& A_1 \curlywedge_1 A_2 \succ A_3 \curlywedge_3 A_4 + A_1 \succ A_3 \prec A_2 \succ A_4
    + A_1 \succ A_3 \curlywedge_3 A_4 \prec A_2     \nonumber \\
 & & + A_3 \prec A_1 \curlywedge_1 A_2 \succ A_4 + A_3 \prec A_1 \succ A_4 \prec A_2
     + A_3 \curlywedge_3 A_4 \prec A_1 \curlywedge_1 A_2  \; .
\ee
Furthermore,
\be
   A_1 \circ A_2
 = \sum_{\sigma \in \mathrm{Shuff}(1,1)} A_{\sigma(1)} \curlywedge'_{\sigma(1)} A_{\sigma(2)}
 = A_1 \succ A_2 + A_2 \prec A_1    \label{A1cA2}
\ee
and, more generally,
\be
  & & A_1 \circ ( A_2 \curlywedge_2 A_3 \curlywedge_3 \ldots \curlywedge_n A_{n+1} )
   = \sum_{\sigma \in \mathrm{Shuff}(1,n)} A_{\sigma(1)} \curlywedge'_{\sigma(1)} \ldots
     \curlywedge'_{\sigma(n)} A_{\sigma(n+1)}  \nonumber \\
   &=& A_1 \succ A_2 \curlywedge_2 \ldots \curlywedge_n A_{n+1}
       + A_2 \prec A_1 \succ A_3 \curlywedge_3 \ldots \curlywedge_n A_{n+1}  \nonumber \\
   & & + A_2 \curlywedge_2 A_3 \prec A_1 \succ A_4 \curlywedge_4 \ldots \curlywedge_n A_{n+1}
       + \ldots
       + A_2 \curlywedge_2 A_3 \curlywedge_3 \ldots \curlywedge_n A_{n+1} \prec A_1
\ee
where we can substitute either $\prec$ or $\succ$ for $\curlywedge_2, \ldots, \curlywedge_n$.
Let $\beta = B_1 \curlywedge_1 B_2 \curlywedge_2 \ldots \curlywedge_{n-1} B_n$ with $B_i \in \A^1$ and
$\beta_{[r,s]} := B_r \curlywedge_r \ldots \curlywedge_{s-1} B_s$ for $r \leq s$. The last
formula can then be written more concisely as
\be
   A \circ \beta = A \succ \beta + \sum_{r=1}^{n-1} \beta_{[1,r]} \prec A \succ \beta_{[r+1,n]}
   + \beta \prec A \; .    \label{Acircb}
\ee
It is convenient to introduce the `Sweedler notation' \cite{Swee69}
\be
    A \circ \beta = A \succ \beta + \sum \beta_{(1)} \prec A \succ \beta_{(2)}
                    + \beta \prec A \; .    \label{Acircbeta}
\ee
In a similar way, we obtain
\be
   \beta \circ A = \beta \succ A + \sum \beta_{(1)} \succ A \prec \beta_{(2)} + A \prec \beta \; .
     \label{betacircA}
\ee
\vskip.1cm
\noindent
{\it Remark.} If $(\A^1 , \bullet)$ is unital with a \emph{unit element} $E$, this
extends to $\A$ such that $E \bullet \alpha = \alpha = \alpha \bullet E$.
Note that no rules are specified to resolve expressions like
$E \prec \alpha$ or $\alpha \prec E$.
\hfill $\blacksquare$

\subsection{Some properties of the algebra $\A$}

\begin{lemma} Let $A \in \A^1$ and $\alpha, \beta \in \A$. Then
\be
      A \circ (\alpha \curlywedge \beta)
 &=& (A \circ \alpha) \curlywedge \beta + \alpha \curlywedge (A \circ \beta)
        - \alpha \curlywedge A \curlywedge \beta   \label{A-c-a-cw-b} \\
     (\alpha \curlywedge \beta) \circ A
 &=& (\alpha \circ A) \curlywedge \beta + \alpha \curlywedge (\beta \circ A)
        - \alpha \curlywedge A \curlywedge \beta   \label{a-cw-b-c-A} \\
      \lbrack A , \alpha \curlywedge \beta \rbrack_{\circ}
 &=& [A , \alpha]_{\circ} \curlywedge \beta + \alpha \curlywedge [ A , \beta ]_{\circ}
\ee
where $[ \, , \, ]_\circ$ denotes the commutator with respect to the product $\circ$.
\end{lemma}
{\it Proof:} Because of linearity, it is sufficient to consider the case where
$\alpha \in \A^m$ and $\beta \in \A^n$ for $m,n \in \mathbb{N}$.
Using (\ref{Acircbeta}), we find
\bez
       A \circ (\alpha \curlywedge \beta)
   &=& A \succ (\alpha \curlywedge \beta)
    + \sum (\alpha \curlywedge \beta)_{(1)} \prec A \succ (\alpha \curlywedge \beta)_{(2)}
    + (\alpha \curlywedge \beta) \prec A \\
   &=& (A \succ \alpha) \curlywedge \beta
    + \sum \alpha_{(1)} \prec A \succ \alpha_{(2)} \curlywedge \beta
    + \alpha \prec A \succ \beta  \\
   & & + \alpha \curlywedge \sum \beta_{(1)} \prec A \succ \beta_{(2)}
       + \alpha \curlywedge \beta \prec A \\
   &=& (A \circ \alpha) \curlywedge \beta + \alpha \curlywedge (A \circ \beta)
       + \alpha \prec A \succ \beta - \alpha \prec A \curlywedge \beta
       - \alpha \curlywedge A \succ \beta \; .
\eez
For both choices $\prec$ and $\succ$ for $\curlywedge$ this yields the first identity
of the lemma. The second is obtained in the same way using (\ref{betacircA}).
The third identity is an immediate consequence of the first two.
\hfill $\blacksquare$
\vskip.2cm

In the following we will adopt the convention that the product $\circ$,
which does not satisfy mutual associativity relations with the other products,
takes precedence over the other products. This means that it has to be evaluated first
in expressions containing also other products. For example,
\be
    \alpha \circ \alpha' \curlywedge \beta \circ \beta'
  := (\alpha \circ \alpha') \curlywedge (\beta \circ \beta') \; .
\ee

\begin{lemma}
\be
      (\alpha \prec A) \circ \beta
  &=&  \alpha \prec A \succ \beta
      + \sum \alpha \circ \beta_{(1)} \prec A \succ \beta_{(2)}
      + \alpha \circ \beta \prec A   \label{a-pre-A-c-b}  \\
     (A \succ \alpha) \circ \beta
  &=& A \succ \alpha \circ \beta + \sum \beta_{(1)} \prec A \succ \alpha \circ \beta_{(2)}
      + \beta \prec A \succ \alpha  \label{A-succ-a-c-b}  \\
     \beta \circ (A \prec \alpha)
  &=& \beta \succ A \prec \alpha + \sum \beta_{(1)} \succ A \prec \beta_{(2)} \circ \alpha
      + A \prec \beta \circ \alpha  \label{b-c-A-prec} \\
     \beta \circ (\alpha \succ A)
  &=& \beta \circ \alpha \succ A + \sum \beta_{(1)} \circ \alpha \succ A \prec \beta_{(2)}
      + \alpha \succ A \prec \beta \; .    \label{b-c-a-succ-A}
\ee
\end{lemma}
{\it Proof:} According to the definition of the shuffle product $\circ$, which preserves
the order of the components of each factor (and the product symbols between them),
an expression like $(\alpha \prec A) \circ \beta$ means that we first have to shuffle
$A$ into $\beta$ and afterwards shuffle $\alpha$ into the resulting expression,
but now with the restriction that all components of $\alpha$ have to precede $A$.
For example, in order to evaluate $(A_1 \prec A_2) \circ \beta$, we first compute
\bez
 A_2 \circ \beta = A_2 \succ \beta + \sum \beta_{(1)} \prec A_2 \succ \beta_{(2)}
                   + \beta \prec A_2  \; .
\eez
Then we shuffle $A_1$ into this expression as follows,
\bez
   (A_1 \prec A_2) \circ \beta = A_1 \prec A_2 \succ \beta
        + \sum (A_1 \circ \beta_{(1)}) \prec A_2 \succ \beta_{(2)}
        + (A_1 \circ \beta) \prec A_2  \; .
\eez
This obviously generalizes to
\bez
   (\alpha \prec A) \circ \beta = \alpha \prec A \succ \beta
        + \sum (\alpha \circ \beta_{(1)}) \prec A \succ \beta_{(2)}
        + (\alpha \circ \beta) \prec A
\eez
which is the first identity of this lemma. The others are obtained by similar
considerations.
\hfill $\blacksquare$
\vskip.2cm

The following identity characterizes the main product as a `quasi-shuffle product' \cite{Hoff00}.

\begin{proposition}
\be
       (A \prec \alpha) \circ (B \prec \beta)
 &=& A \prec \alpha \circ (B \prec \beta)
     + B \prec (A \prec \alpha) \circ \beta
     + (A \bullet B) \prec \alpha \circ \beta  \; .   \label{shuffle}
\ee
\end{proposition}
{\it Proof:}
Using (\ref{b-c-A-prec}) and (\ref{succ}), we obtain
\bez
     (A \prec \alpha) \circ (B \prec \beta)
 &=& (A \prec \alpha) \succ B \prec \beta + (A \prec B + A \bullet B) \prec (\alpha \circ \beta) \\
 & & + A \prec \sum \alpha_{(1)} \succ B \prec (\alpha_{(2)} \circ \beta)
     + B \prec (A \prec \alpha) \circ \beta \; .
\eez
The formula (\ref{shuffle}) is now obtained by rewriting the first term on the
right hand side as follows, again with the help of (\ref{b-c-A-prec}),
\bez
     A \prec \alpha \succ B \prec \beta
 &=& A \prec \Big( \alpha \circ (B \prec \beta)
       - \sum \alpha_{(1)} \succ B \prec \alpha_{(2)} \circ \beta
       - B \prec \alpha \circ \beta \Big) \; .
\eez
\hfill $\blacksquare$
\vskip.2cm

In a similar way, one can prove the following identity,
\be
    (A \succ \alpha) \circ (B \prec \beta)
  = A \succ \alpha \circ (B \prec \beta)
    + B \prec (A \succ \alpha) \circ \beta \; .
\ee
\vskip.1cm

\noindent
{\em Remark.} With $A \in \A^1$ let us associate a map $R_A : \A \to \A$ via
$R_A(\alpha) = A \prec \alpha$. Then (\ref{shuffle}) reads
\be
    R_A(\alpha) \circ R_B(\beta) = R_A(\alpha \circ R_B(\beta))
    + R_B(R_A(\alpha) \circ \beta) + R_{A \bullet B}(\alpha \circ \beta) \; .
\ee
In particular, if $A \in \A^1$ satisfies $A \bullet A = - \mathrm{q} \, A$
with $\mathrm{q} \in \mathbb{K}$, then $R_A$ defines a \emph{Rota-Baxter operator} of weight
$\mathrm{q}$ on $(\A,\circ)$ \cite{Baxt60,Rota69I,Rota+Smit72,Rota95} (see also appendix~A
and \cite{Guo00free,Guo+Keig00shuffle,Guo+Keig00Bax-compl,Ebra+Guo04}
for relations with shuffle algebras). Associated with a unit element $E$ is thus a
Rota-Baxter operator of weight $-1$.
If $\mathrm{q} = 0$ and $\alpha = \sum_{n \geq 1} a_n \, A^{\prec \, n}$,
$\beta = \sum_{n \geq 1} b_n \, A^{\prec \, n}$, we obtain
$\alpha \circ \beta = \sum_{n \geq 1} c_n \, P^{\prec \, n}$
with $c_n = \sum_{k=0}^n {n \choose k} \, a_k \, b_{n-k}$, from which we recover
the ring of \emph{Hurwitz series} (divided power series) \cite{Carl49}.
\hfill $\blacksquare$

\begin{theorem}
\label{theorem:circ-comm}
If $[A,B]_\bullet := A \bullet B - B \bullet A$ vanishes for all $A,B \in \A^1$, then
$(\A, \circ)$ is a commutative algebra.
\end{theorem}
{\it Proof:} First we note that $[A,B]_\circ = [A,B]_\bullet$.
(\ref{Acircbeta}) and (\ref{betacircA}) lead to
\bez
      [ A , \beta ]_\circ
  &=& [A , \beta ]_\bullet
      + \sum \Big( \beta_{(1)} \prec A \bullet \beta_{(2)}
         - \beta_{(1)} \bullet A \prec \beta_{(2)} \Big) \\
  &=& \sum_{r=1}^n B_1 \curlywedge_1 \ldots \curlywedge_{r-1} [A,B_r]_\bullet
      \curlywedge_r \ldots \curlywedge_{n-1} B_n
\eez
for $\beta = B_1 \curlywedge_1 \ldots \curlywedge_{n-1} B_n$. This vanishes indeed
as a consequence of our assumption. Furthermore, from (\ref{shuffle}) we obtain
\bez
     [A \prec \alpha , B \prec \beta]_\circ
 &=& A \prec [\alpha , B \prec \beta]_\circ
     + B \prec [A \prec \alpha , \beta]_\circ  \\
 & & + (A \bullet B) \prec \alpha \circ \beta
     - (B \bullet A) \prec \beta \circ \alpha  \; .
\eez
Using our assumption, the last two terms combine to
$(A \bullet B) \prec [\alpha , \beta]_\circ$. Hence this formula can be used
to prove our general statement by induction on the grades of $\alpha$
and $\beta$.
\hfill $\blacksquare$

\subsection{Involutions interchanging $\prec$ and $\succ$}
\label{sec:inv}
There is a fundamental duality in the algebra $\A$ concerning the two
operations $\prec$ and $\succ$. It is convenient to encode this
duality in two involutions which exchange the two products and their
gradings:
\be
    (\alpha \prec \beta)^\psi = \alpha^\psi \succ \beta^\psi
    \qquad\quad
    (\alpha \prec \beta)^\omega = \beta^\omega \succ \alpha^\omega
\ee
where for all $A \in \A^1$ also $A^\psi, \, A^\omega \in \A^1$.
Using the involution property $\gamma^{\psi \psi} = \gamma$,
respectively $\gamma^{\omega \omega} = \gamma$, for all $\gamma \in \A$,
this implies
\be
    (\alpha \bullet \beta)^\psi = - \alpha^\psi \bullet \beta^\psi
    \qquad\quad
    (\alpha \bullet \beta)^\omega = - \beta^\omega \bullet \alpha^\omega \; .
\ee
As a consequence,
\be
    (\alpha \succ \beta)^\psi = \alpha^\psi \prec \beta^\psi
    \qquad\quad
    (\alpha \succ \beta)^\omega = \beta^\omega \prec \alpha^\omega \; .
\ee
We still have the freedom to define the action of the two involutions on the
generators of $\A$.

\begin{proposition}
\be
    (\alpha \circ \beta)^\psi = \beta^\psi \circ \alpha^\psi
    \qquad\quad
    (\alpha \circ \beta)^\omega = \alpha^\omega \circ \beta^\omega \; .
\ee
\end{proposition}
{\it Proof:} by induction with respect to the grade of $\alpha$. For
$\alpha \in \A^1$ the identities easily follow from (\ref{Acircbeta}) and
(\ref{betacircA}). If the identities hold for $\alpha \in \A^n$, they also
hold for $\alpha \in \A^{n+1}$ by use of the identities (\ref{a-pre-A-c-b}) and
(\ref{b-c-a-succ-A}).
\hfill $\blacksquare$
\vskip.2cm

Applying the above involutions to identities in $\A$ generates further identities.
This often provides us with a quick way of proving required relations.

\begin{proposition}
\be
    (A \succ \alpha) \circ (B \succ \beta) &=& A \succ \alpha \circ (B \succ \beta)
     + B \succ (A \succ \alpha) \circ \beta - (B \bullet A) \succ \alpha \circ \beta
     \qquad       \label{shuffle2} \\
    (\alpha \prec A) \circ (\beta \prec B) &=& \alpha \circ (\beta \prec B) \prec A
     + (\alpha \prec A) \circ \beta \prec B + \alpha \circ \beta \prec (A \bullet B)
           \label{shuffle3} \\
    (\alpha \succ A) \circ (\beta \succ B) &=& \alpha \circ (\beta \succ B) \succ A
     + (\alpha \succ A) \circ \beta \succ B - \alpha \circ \beta \succ (B \bullet A)
           \label{shuffle4}  \; .
\ee
\end{proposition}
{\it Proof:} (\ref{shuffle2}) and (\ref{shuffle4}) are obtained by applying ${}^\psi$,
respectively ${}^\omega$, to (\ref{shuffle}). (\ref{shuffle3}) in turn results from
(\ref{shuffle2}) by application of ${}^\omega$ (or from (\ref{shuffle4}) via
${}^\psi$).
\hfill $\blacksquare$

\subsection{Associative products determined by elements of $\A^1$}
\label{subsection:products}
With each $A \in \A^1$ we associate two bilinear maps $\hA, \cA : \A \times \A \to \A$
via
\be
    \hA(\alpha,\beta) &:=& \alpha \, \hA \, \beta := \alpha \prec A \succ \beta
      \label{hA}  \\
    \cA(\alpha,\beta) &:=& \alpha \, \cA \, \beta := \alpha \succ A \prec \beta \; .
      \label{cA}
\ee
The `product notation' is justified since the expressions on the right hand sides
are combined associative with all products defined so far, with the exception of the
main product, and thus also among themselves. In particular,
$(\alpha \, \cA \, \beta) \, \hB \, \gamma = \alpha \, \cA \, (\beta \, \hB \, \gamma)$
so that we are allowed to drop the brackets.

\begin{lemma}
\label{lemma:hAinv}
\be
    (\alpha \, \hA \, \beta)^\psi = \alpha^\psi \, \cA^\psi \, \beta^\psi  \qquad\quad
    (\alpha \, \hA \, \beta)^\omega = \beta^\omega \, \hA^\omega \, \alpha^\omega \; .
\ee
\end{lemma}
{\it Proof:} These are immediate consequences of the definitions (\ref{hA}) and (\ref{cA}),
and the properties of the involutions ${}^\psi$ and ${}^\omega$ (see section~\ref{sec:inv}).
With $B := A^\psi$, $\cA^\psi$ means $\cB$.
\hfill $\blacksquare$

\begin{proposition}
The following derivation properties of $\circ$-multiplication by an
element $B \in \A^1$ hold:
\be
       B \circ (\alpha \, \cA \, \beta)
 &=& (B \circ \alpha) \, \cA \, \beta + \alpha \, \cA \, (B \circ \beta)    \\
       (\alpha \, \hA \, \beta) \circ B
 &=& \alpha \, \hA \, (\beta \circ B) + (\alpha \circ B) \, \hA \, \beta \; .
         \label{hA-circB-deriv}
\ee
\end{proposition}
{\it Proof:} This is easily verified with the help of (\ref{A-c-a-cw-b})
and (\ref{a-cw-b-c-A}).
Note also that the two identities are mapped to each other by application of the
involution ${}^\psi$ (with $A^\psi = A$ for all $A \in \A^1$) and use of
lemma~\ref{lemma:hAinv}.
\hfill $\blacksquare$
\vskip.2cm

The next result is a generalization of the previous proposition.

\begin{proposition}
\be
    \gamma \circ (\alpha \, \cA \, \beta) &=& (\gamma \circ \alpha) \, \cA \, \beta
    + \sum (\gamma_{(1)} \circ \alpha) \, \cA \, (\gamma_{(2)} \circ \beta)
    + \alpha \, \cA \, (\gamma \circ \beta)  \label{gamma_circ_cA}   \\
   (\alpha \, \hA \, \beta) \circ \gamma &=& \alpha \, \hA \, (\beta \circ \gamma)
    + \sum (\alpha \circ \gamma_{(1)}) \, \hA \, (\beta \circ \gamma_{(2)})
    + (\alpha \circ \gamma) \, \hA \, \beta  \; .  \label{hA_circ_gamma}
\ee
\end{proposition}
{\it Proof:} According to the definition of the shuffle product,
$\gamma \circ (\alpha \succ A \prec \beta)$ consists of a sum of terms, two of
which correspond to shuffling of $\gamma$ into $\alpha$, respectively $\beta$.
In addition, we have all possible terms obtained by splitting $\gamma$ into two ordered
parts and shuffling the first into $\alpha$ and the second into $\beta$. The result
is precisely our first formula.
The second formula is obtained in the same way.\footnote{The formulae of this
proposition do \emph{not} hold with $\hA$ and $\cA$ exchanged, if $\circ$ is not commutative.
For example, $B \circ (\alpha \, \hA \, \beta) = (B \circ \alpha) \prec A \succ \beta
+ \alpha \prec B \circ (A \succ \beta) - \alpha \prec B \prec A \succ \beta$
where the last term corrects a double counting of the first two. By use of
(\ref{A-c-a-cw-b}), we find
$B \circ (\alpha \, \hA \, \beta) = (B \circ \alpha) \prec A \succ \beta
 + \alpha \prec A \succ (B \circ \beta) + \alpha \prec [B,A]_\circ \succ \beta$. }
\hfill $\blacksquare$

\section{Realization by partial sum calculus}
\label{section:ps}
\setcounter{equation}{0}
Let $\mathcal{N} := \{I \subset \mathbb{N} \, | \, I \neq \emptyset, \;
|I|<\infty \}$. This is the set of nonempty finite subsets of the set of
natural numbers.
Let $\A$ be the freely generated linear space (over $\mathbb{K}$) with basis
$\{ e_I \, | \; I \in \mathcal{N} \}$.
For $I,J \in \mathcal{N}$ we define the following associative products:
\be
    e_I \prec e_J :=\left\{\begin{array}{ll} e_{I\cup J} & \hbox{if $\max(I)<\min(J)$} \\
     0 & \hbox{otherwise}
    \end{array}\right.
\ee
\be
    e_I \bullet e_J := \left\{\begin{array}{ll} e_{I\cup J} & \hbox{if $\max(I)=\min(J)$} \\
     0 & \hbox{otherwise}
    \end{array}\right.
\ee
and thus
\be
    e_I \succ e_J = \left\{\begin{array}{ll} e_{I\cup J} & \hbox{if $\max(I)\leq\min(J)$}\\
     0 & \hbox{otherwise}
    \end{array}\right.
\ee
For example, $e_{\{2,4,5\}} = e_2 \prec e_4 \prec e_5 \in \A^3$, where we simply write $e_n$
instead of $e_{\{ n \}}$.
Any element $A$ of $\A^1$ can be written as $A = \sum_{n \geq 1} a_n \, e_n$
with $a_n \in \mathbb{K}$.
The $\bullet$-product with another element $B = \sum_{n \geq 1} b_n \, e_n$ of $\A^1$
is then given by
\be
    A \bullet B = \sum_{n \geq 1} a_n \, b_n \, e_n \; .
\ee
There is a formal\footnote{Proper elements of $\A$ are \emph{finite} sums.}
unit element, $E := \sum_{n \geq 1} e_n$.
With $A_i = \sum_{n \geq 1} a_{i,n} \, e_n$, $i=1,\ldots,r$, we obtain
\be
    A_1 \prec \ldots \prec A_r = \sum_{1 \leq n_1 < \ldots <n_r}
    a_{1,n_1} \cdots a_{r,n_r} \; e_{\{ n_1, \ldots, n_r \}} \; .
\ee
 For the main product, we find the simple formula
\be
    e_I \circ e_J = e_{I \cup J} \; .
\ee
\vskip.1cm

The linear map $\Sigma_N : \A \to \mathbb{K}$ defined by
\be
    \Sigma_N(e_I) = \left\{\begin{array}{ll} 1 & \hbox{if $I \subset \{1,2,\ldots,N\}$} \\
     0 & \hbox{otherwise}
    \end{array}\right.
\ee
has the properties
\be
    \Sigma_N(A_1 \prec \ldots \prec A_r)
 &=& \sum_{1 \leq n_1 < \ldots < n_r \leq N}
    a_{1,n_1} \cdots a_{r,n_r}   \label{Sigma(A<...<A)}   \\
    \Sigma_N(A_1 \circ \ldots \circ A_r)
 &=& \Big(\sum_{n_1=1}^N a_{1,n_1} \Big) \cdots \Big( \sum_{n_r=1}^N a_{1,n_r} \Big) \; .
    \label{Sigma-circ}
\ee
By application of $\Sigma_N$ to identities in (the partial sum realization of)
the algebra $\A$, we obtain sum identities of the kind considered
in the introduction, which hold for all $N$. But which identities in $\A$
correspond to the equations of the (x)ncKP hierarchy?
The answer will be given in section~\ref{section:xncKP}.
\vskip.1cm

\noindent
{\em Remark.} The calculus of partial sums is known to carry the
structure of a Rota-Baxter algebra \cite{Baxt60,Rota69I} (see also appendix~A).
We define a map $R$ from $\A$ to a completion (as a projective limit)
$\bar{\A}^1$ of $\A^1$ by
\be
    R(\alpha) = \sum_{N \geq 1} \Sigma_{N-1}(\alpha) \, e_N \qquad \forall \alpha \in \A
       \label{R(alpha)-ps-calc}
\ee
where $\Sigma_0(\alpha) := 0$. It satisfies
\be
    R(\alpha \prec A) = R(R(\alpha) \bullet A)    \label{R(a<A)_ps}
\ee
and therefore
\be
    R(A_1 \prec \ldots \prec A_r)
  = R(R( \ldots R(R(A_1) \bullet A_2) \bullet \ldots) \bullet A_r)   \label{psc_R(prec)->bull}
\ee
for $A_1,\ldots,A_r \in \A^1$. Another simple consequence of (\ref{R(a<A)_ps}) is
\be
    R(\alpha \succ A) = R(R(\alpha) \bullet A + \alpha \bullet A) \; .
\ee
Furthermore, for all $\alpha, \beta \in \A$ the following identity holds,
\be
    R(\alpha \circ \beta) = R(\alpha) \bullet R(\beta) \qquad \forall \alpha,\beta \in \A \; .
    \label{psc_Rcirc_bull}
\ee
Applying $R$ to $A \circ B = A \succ B + B \prec A$ thus leads to
\be
    R(A) \bullet R(B) = R( R(A) \bullet B + A \bullet R(B) + A \bullet B)
    \qquad \forall A,B \in \A^1  \; .
\ee
With obvious extensions of $\bullet$ and $R$, $(\bar{\A}^1,\bullet, R|_{\bar{\A}^1})$
becomes a Rota-Baxter algebra of weight $-1$.
\hfill $\blacksquare$

\section{The subalgebra of $\A$ generated by a single element $P$}
\label{section:A(P)}
\setcounter{equation}{0}
Let $\A(P)$ be the subalgebra of $\A$ generated by an element $P \in \A^1$.
More precisely, if $(\A(P), \bullet)$ has a unit element $E$, then
$\A^1(P)$ is spanned by
\be
     P_n := P^{\bullet \, n}   \qquad \quad   n = 0,1,2, \ldots
\ee
where $P_0 := E$. If $(\A(P), \bullet)$ is not unital, we have to
disregard expressions containing $P_0$ in the following.
Clearly, $(\A^1(P),\bullet)$ is commutative, and thus also $(\A(P),\circ)$
by theorem~\ref{theorem:circ-comm}.
According to section~\ref{subsection:products}, $P$ determines an associative product,
\be
  \alpha \htimes \beta := - \alpha \, \hP \, \beta = - \alpha \prec P \succ \beta
  \qquad \forall \alpha, \beta \in \A(P)
\ee
which will play an important role in our subsequent considerations.

\begin{proposition}
Via the main product, each $A \in \A^1(P)$ acts on a $\htimes$-product according to
the derivation rule
\be
   A \circ (\alpha \htimes \beta)
 = (A \circ \alpha) \htimes \beta + \alpha \htimes (A \circ \beta) \; .
    \label{Acirc-htimes-deriv}
\ee
\end{proposition}
{\it Proof:} by use of (\ref{hA-circB-deriv}), taking the commutativity of $(\A(P),\circ)$
into account.
\hfill $\blacksquare$
\vskip.2cm

It is convenient to introduce the following objects which form a basis of $\A(P)$,
\be
    P_{m_1 \ldots m_k}
 := P_{m_1} \prec \ldots \prec P_{m_k} \; . \label{Pm1...mk}
\ee

\begin{theorem}
\label{theorem:P...-circ-(a-htimes-b)}
\be
    P_{m_1 \ldots m_k} \circ (\alpha \htimes \beta)
  = \sum_{j=0}^k (P_{m_1 \ldots m_j} \circ \alpha) \htimes
      (P_{m_{j+1} \ldots m_k} \circ \beta) \; .
\ee
\end{theorem}
{\it Proof:} Since $\circ$ is commutative in the case under consideration,
(\ref{hA_circ_gamma}) implies
\bez
 & & (A_1 \prec A_2 \prec \ldots \prec A_k) \circ (\alpha \htimes \beta)
  = (A_1 \prec \ldots \prec A_k) \circ \alpha \htimes \beta  \\
 & & + \sum_{l=1}^{k-1} (A_1 \prec \ldots \prec A_l ) \circ \alpha
           \htimes (A_{l+1} \prec \ldots \prec A_k) \circ \beta
     + \alpha \htimes (A_1 \prec \ldots \prec A_k) \circ \beta
\eez
for arbitrary $A_l \in \A^1$. Setting $A_l = P_{m_l}$ completes the proof.
\hfill $\blacksquare$
\vskip.2cm

\noindent
{\it Remark.} It looks natural to consider still another product:
$\alpha \ctimes \beta := \alpha \, \cP \, \beta := \alpha \succ P \prec \beta$.
Choosing the involution ${}^\psi$ in such a way that $P^\psi = P$, lemma (\ref{lemma:hAinv})
implies $(\alpha \ctimes \beta)^\psi = - \alpha^\psi \htimes \beta^\psi$. The
product $\ctimes$ is thus equivalent to the product $\htimes$ and it is sufficient
to deal with the latter, as long as we restrict our considerations to the algebra
$\A(P)$.
\hfill $\blacksquare$

\subsection{Special relations in $\A(P)$ and reminiscences of (x)ncKP}
\label{section:special-rels}
The aim of this section is to derive algebraic identities in $\A(P)$ which
mirror algebraic properties of the (x)ncKP hierarchy, as derived in \cite{DMH04ncKP}.
The results will be important in later sections, where the relation between
identities in $\A(P)$ and the ncKP hierarchy (and extensions) is put
on firmer grounds.

\begin{lemma}
\be
 & & P^{\circ \, n}
  = P \succ P^{\circ \, n-1} - \sum_{r=1}^{n-2}{n-1 \choose r} \,
       P^{\circ \, n-r-1} \htimes P^{\circ r} + P^{\circ \, n-1} \prec P
          \qquad n = 2,3, \ldots \qquad     \label{pn1}   \\
 & & P^{\circ \, n-2} \circ (P \prec P)
  = P^{\circ \, n-1} \prec P - \sum_{r=1}^{n-2} {n-2 \choose r}
       P^{\circ \, n-r-1} \htimes P^{\circ \, r}
          \qquad \; \,   n=3,4, \ldots  \label{plp}
\ee
\end{lemma}
{\it Proof:} For $n=2$, the first relation obviously holds. Let us assume that the formula
holds for some integer $n \geq 2$. Then
\bez
   P^{\circ \, n+1} = P^{\circ \, n} \circ P
 = \Big( P \succ P^{\circ \, n-1} - \sum_{r=1}^{n-2}{n-1 \choose r} \,
    P^{\circ \, n-r-1} \htimes P^{\circ r} + P^{\circ \, n-1} \prec P \Big) \circ P \; .
\eez
Next we use (\ref{a-cw-b-c-A}), $P^{\circ \, 2} = P \succ P + P \prec P$,
and (\ref{Acirc-htimes-deriv}) to obtain
\bez
     P^{\circ \, n+1}
 &=& P \succ P^{\circ \, n} + P^{\circ \, n} \prec P
     - P \htimes P^{\circ \, n-1} - P^{\circ \, n-1} \htimes P  \\
 & & - \sum_{r=1}^{n-2}{n-1 \choose r} \, \Big( P^{\circ \, n-r-1} \htimes P^{\circ r+1}
       + P^{\circ \, n-r} \htimes P^{\circ r} \Big) \; .
\eez
With the help of the combinatorial identity
\be
    {n \choose r} = {n-1 \choose r} + {n-1 \choose r-1}   \label{comb-identity}
\ee
and some simple manipulations, this becomes
\bez
   P^{\circ \, n+1}
 = P \succ P^{\circ \, n} + P^{\circ \, n} \prec P
   - \sum_{r=1}^{n-1}{n \choose r} \, P^{\circ \, n-r} \htimes P^{\circ r}
\eez
so that the first formula of the lemma also holds for $n+1$. The proof of the
second formula can be carried out in a very similar way.
\hfill $\blacksquare$
\vskip.2cm

Let us introduce $U_2 := P$ and
\be
    U_n := (-1)^n \, P \prec P^{\circ \, n-2} \qquad n=3,4, \ldots
    \label{U_n}
\ee

\begin{proposition}
\be
    P \circ U_{n+1}
 = {1 \over 2} (P_2 - P^{\circ \, 2}) \circ U_n - [U_2,U_n]_{\htimes}
   + \sum_{r=1}^{n-2} {n-2 \choose r} \, (-1)^r \,
    U_{n-r} \htimes P^{\circ \, r} \circ U_2  \quad   \label{uit2}
\ee
where $[\alpha,\beta]_{\htimes} := \alpha \htimes \beta - \beta \htimes \alpha$.
\end{proposition}
{\it Proof:} First we note that, by use of (\ref{A-c-a-cw-b}), the definition
(\ref{U_n}) implies $P \circ U_n = - U_{n+1} + P \succ U_n$ and, by multiple use
of this equation,
\bez
    P^{\circ \, 2} \circ U_n
  = P \circ ( P \circ U_n )
  = - U_{n+2} - 2 \, P \circ U_{n+1} - 2 \, P \htimes U_n + P_2 \succ U_n \; .
\eez
Furthermore, with the help of (\ref{A-c-a-cw-b}), $P_2 = P^{\circ \, 2} - 2 \, P \prec P$,
and (\ref{plp}), we obtain
\bez
     P_2 \circ U_n
 &=& (-1)^n \, P_2 \circ ( P \prec P^{\circ \, n-2} )
  =  P_2 \succ U_n + (-1)^n \, P \prec (P_2 \circ P^{\circ \, n-2}) \\
 &=& P_2 \succ U_n + U_{n+2} - 2 \, (-1)^n \, P \prec ((P \prec P) \circ P^{\circ \, n-2}) \\
 &=& P_2 \succ U_n + U_{n+2} + 2 \, U_{n+1} \prec P
     + 2 \, (-1)^n \, P \prec \sum_{r=1}^{n-2} {n-2 \choose r} P^{\circ \, n-r-1} \htimes
      P^{\circ \, r} \\
 &=& P_2 \succ U_n + U_{n+2} + 2 \, U_{n+1} \prec P
     - 2 \sum_{r=1}^{n-2} {n-2 \choose r} (-1)^r \, U_{n-r+1} \htimes P^{\circ \, r} \; .
\eez
Now we can eliminate the products $\prec$ and $\succ$ from this expression
with the help of our first result and
\bez
    U_{n+1} \prec P
 = - U_{n+2} + \sum_{r=1}^{n-1} {n-1 \choose r} (-1)^r \, U_{n-r+1} \htimes P^{\circ \, r}
\eez
which is obtained by applying $P\prec \,$ to (\ref{pn1}). After simple manipulations and use
of (\ref{comb-identity}), this results in the desired formula.
\hfill $\blacksquare$
\vskip.2cm

Next we introduce $H^{(m_1,\ldots,m_r)}_1 := P_{m_r} \succ \ldots \succ P_{m_1}$ and
\be
    H^{(m_1, \ldots, m_r)}_{n+1} := H_n \succ P_{m_r} \succ \ldots \succ P_{m_1}
    \qquad \quad  n \in \mathbb{N}   \label{H^m1...mr_n}
\ee
where
\be
   H_n := H^{(1)}_n := P^{\succ n} \qquad \quad  n \in \mathbb{N}  \; .
\ee

\begin{proposition}
\be
    P_n \circ H^{(m)}_k - P_m \circ H^{(n)}_k
    - \sum_{j=1}^{k-1} [H^{(m)}_j , H^{(n)}_{k-j}]_{\htimes} = 0 \; .  \label{Pn-circ-H^m_k}
\ee
\end{proposition}
{\it Proof:} Using (\ref{betacircA}) and (\ref{H^m1...mr_n}), we obtain
\bez
     H_{k-1} \circ P_m
 &=& H_{k-1} \succ P_m + \sum_{j=1}^{k-2} H_j \succ P_m \prec H_{k-j-1} + P_m \prec H_{k-1} \\
 &=& H^{(m)}_k + \sum_{j=1}^{k-3} H^{(m)}_{j+1} \prec P \succ H_{k-j-2} + H^{(m)}_{k-1} \prec P
     + P_m \prec P \succ H_{k-2} \\
 &=& H^{(m)}_k - \sum_{j=1}^{k-2} H^{(m)}_j \htimes H_{k-j-1} + H^{(m)}_{k-1} \prec P
\eez
and thus
\bez
    H_{k-1} \circ P_m \succ P_n
  = H^{(m)}_k \succ P_n - \sum_{j=1}^{k-1} H^{(m)}_j \htimes H^{(n)}_{k-j} \; .
\eez
This is used to derive
\bez
     P_m \circ H^{(n)}_k
 &=& P_m \circ ( P^{\succ k-1} \succ P_n ) \\
 &=& (P_m \circ P^{\succ k-1}) \succ P_n + P^{\succ k-1} \succ (P_m \circ P_n)
     - H^{(m)}_k \succ P_n \\
 &=& H_{k-1} \succ (P_m \circ P_n) - \sum_{j=1}^{k-1} H^{(m)}_j \htimes H^{(n)}_{k-j}
\eez
from which (\ref{Pn-circ-H^m_k}) follows by anti-symmetrization with respect to $m,n$.
\hfill $\blacksquare$

\begin{proposition}
\label{proposition:H^m_n-recursion}
\be
    H^{(m+1)}_n
 = - P_m \circ H^{(1)}_n + H^{(m)}_{n+1} + H^{(1)}_{m+n}
   - \sum_{r=1}^{n-1} H^{(m)}_{n-r} \htimes H^{(1)}_r
   + \sum_{r=1}^{m-1} H^{(m-r)}_n \htimes H^{(1)}_r \;. \label{hnm}
\ee
\end{proposition}
{\it Proof:} First we obtain
\bez
    P_m \succ P = H_{m+1} - \sum_{r=1}^{m-1} P_{m-r} \prec H_{r+1}
\eez
by induction on $m$. This shows that
\bez
    P_m \succ H_n = H_{m+n} + \sum_{r=1}^{m-1} P_{m-r} \htimes H_{n+r-1}
\eez
holds for $n=1$, and the general formula is easily verified by induction on $n$.
According to (\ref{Acircbeta}),
\bez
   P_m \circ H_n
 = P_m \succ H_n + \sum_{r=1}^{n-1} H_r \prec P_m \succ H_{n-r} + H_n \prec P_m \; .
\eez
Using $H_n \prec P_m = H^{(m)}_{n+1} - H^{(m+1)}_n$, which is easily verified, this becomes
\bez
   H^{(m+1)}_n - H^{(m)}_{n+1} + P_m \circ H_n
 = P_m \succ H_n + \sum_{r=1}^{n-1} H_r \prec P_m \succ H_{n-r} \; .
\eez
Now we eliminate all expressions $P_m \succ H_l$ by means of the corresponding
formula above to get
\bez
     H^{(m+1)}_n - H^{(m)}_{n+1} + P_m \circ H_n
 &=& H_{m+n} + \sum_{r=1}^{m-1} P_{m-r} \htimes H_{n+r-1} \\
 & & + \sum_{r=1}^{n-1} H_r \prec \Big( H_{m+n-r}
     + \sum_{k=1}^{m-1} P_{m-k} \htimes H_{n-r+k-1} \Big) \; .
\eez
Next we use $H_r \prec P_{m-k} = H^{(m-k)}_{r+1} - H^{(m-k+1)}_r$.
Some rearrangements then lead to (\ref{hnm}).
\hfill $\blacksquare$

\begin{proposition}
\label{proposition:H^m..._n-recursion}
\be
    H^{(m_1, \ldots, m_{r+1})}_n = H^{(m_1, \ldots, m_r)}_{n+m_{r+1}}
    + \sum_{k=1}^{m_{r+1}-1} H^{(m_{r+1}-k)}_n
    \htimes H^{(m_1, \ldots, m_r)}_k    \qquad  r=1,2,\ldots   \label{H-recursion}
\ee
\end{proposition}
{\it Proof:}
By induction one easily verifies that
\bez
    P_n = H_n - \sum_{k=1}^{n-1} P_{n-k} \prec H_k \; .
\eez
Using this in the definition (\ref{H^m1...mr_n}), we find
\bez
     H^{(m_1, \ldots, m_{r+1})}_n
 &=& H_{n-1} \succ (H_{m_{r+1}} - \sum_{k=1}^{m_{r+1}-1} P_{m_{r+1}-k} \prec H_k )
     \succ P_{m_r} \succ \ldots \succ P_{m_1}  \\
 &=& H^{(m_1, \ldots, m_r)}_{n+m_{r+1}-1}
     - \sum_{k=1}^{m_{r+1}-1} H^{(m_{r+1}-k)}_n \prec P \succ H^{(m_1, \ldots, m_r)}_k
\eez
which is (\ref{H-recursion}).
\hfill $\blacksquare$
\vskip.2cm

Let $C^{(m_1, \ldots, m_r)}_1 := (-1)^r \, P_{m_1 \ldots m_r}$ and
\be
  C^{(m_1, \ldots, m_r)}_{n+1} := (-1)^{n+r} P_{m_1 \ldots m_r} \prec P^{\prec n}
    \qquad \quad  n \in \mathbb{N} \; .   \label{C^(m...m)_n}
\ee

\begin{proposition}
\label{proposition:C^m..._n-recursion}
\be
  C^{(m+1)}_n &=& P_m \circ C^{(1)}_n + C^{(m)}_{n+1} + C^{(1)}_{m+n}
      + \sum_{r=1}^{n-1} C^{(1)}_r \htimes C^{(m)}_{n-r}
      - \sum_{r=1}^{m-1} C^{(1)}_r \htimes C^{(m-r)}_n  \qquad  \\
  C^{(m_1, \ldots, m_{r+1})}_n &=& C^{(m_1, \ldots, m_r)}_{n+m_{r+1}}
    - \sum_{k=1}^{m_{r+1}-1} C^{(m_1, \ldots, m_r)}_k \htimes C^{(m_{r+1}-k)}_n \; .
      \label{cmm}
\ee
\end{proposition}
{\it Proof:} Choose the involution ${}^\omega$ such that $P^\omega = -P$. Then
$P_r{}^\omega = - P_r$, $(\alpha \htimes \beta)^\omega = -\beta^\omega \htimes \alpha^\omega$,
and $C^{(m_1, \ldots, m_r)}_n = (H^{(m_1, \ldots, m_r)}_n)^\omega$. Now our statements
follow by application of ${}^\omega$ to (\ref{hnm}) and (\ref{H-recursion}).
\hfill $\blacksquare$
\vskip.2cm

Let
\be
    A_{mn} := {1\over 2} (P_{m n} - P_{n m} )
            = {1\over 2} (P_m \prec P_n - P_n \prec P_m)
            = {1 \over 2} (P_m \succ P_n - P_n \succ P_m) \; .  \label{A_mn}
\ee

\begin{proposition}
\be
    A_{mn} \circ (\alpha \htimes \beta) = A_{mn} \circ \alpha \htimes \beta
    + \alpha \htimes A_{mn} \circ \beta + {1 \over 2} (P_m \circ \alpha \htimes P_n \circ \beta
    - P_n \circ \alpha \htimes P_m \circ \beta) \; .
\ee
\end{proposition}
{\it Proof:} This follows directly from theorem~\ref{theorem:P...-circ-(a-htimes-b)}.
\hfill $\blacksquare$

\begin{proposition}
\be
     A_{mn}
 &=& - {1 \over 2} (P_{m+n} + P_m \circ P_n) + H^{(n)}_{m+1}
     + \sum_{r=1}^{m-1} P_r \htimes H^{(n)}_{m-r} \nonumber \\
 &=& - {1 \over 2} (P_{m+n} - P_m \circ P_n) - C^{(n)}_{m+1}
     - \sum_{r=1}^{m-1} C^{(n)}_{m-r} \htimes P_r \; .
\ee
\end{proposition}
{\it Proof:} Using $P_m \circ P_n = P_m \succ P_n + P_n \prec P_m$
and $P_{m+n} = P_m \succ P_n - P_m \prec P_n$ we find
\bez
  A_{mn} = P_m \succ P_n - \frac{1}{2} ( P_m \circ P_n + P_{m+n} ) \; .
\eez
The first equality of the proposition now follows with the
help of
\bez
    P_m \succ P_n = H^{(n)}_{m+1} + \sum_{r=1}^{m-1} P_r \htimes H^{(n)}_{m-r}
\eez
which is a special case of (\ref{H-recursion}). The second equality is
obtained by application of ${}^\omega$ to the first.
\hfill
$\blacksquare$
\vskip.2cm

Adding the two expressions for $A_{mn}$ derived in the last proposition, leads to
\be
    A_{mn} = - {1 \over 2} \Big(P_{m+n} + C^{(n)}_{m+1} - H^{(n)}_{m+1}
     + \sum_{r=1}^{m-1} (C^{(n)}_{m-r} \htimes P_r - P_r \htimes H^{(n)}_{m-r}) \Big)
     \label{Amn-id}
\ee
and subtraction yields
\be
    P_m \circ P_n = C^{(n)}_{m+1} + H^{(n)}_{m+1}
     + \sum_{r=1}^{m-1} (C^{(n)}_{m-r} \htimes P_r + P_r \htimes H^{(n)}_{m-r}) \; .
     \label{PcircP-id}
\ee
\vskip.1cm

As a consequence of propositions~\ref{proposition:H^m_n-recursion},
\ref{proposition:H^m..._n-recursion} and \ref{proposition:C^m..._n-recursion}
and some results of the following subsection (see (\ref{Hn_Schur}) and (\ref{Cn_Schur})),
the expressions $C^{(m_1, \ldots, m_r)}_n$ and $H^{(m_1, \ldots, m_r)}_n$ can be
iteratively expressed completely in terms of only $P_m$, $m = 1,2,\ldots$, the
main product $\circ$ and the $\htimes$-product. We will refer to this result in
sections~\ref{section:xncKP} and \ref{section:XncKP}.
\vskip.1cm

Equation (\ref{PcircP-id}) shows that the expressions constructed in this way
are not all independent, but satisfy certain identities, and these actually
correspond to ncKP equations.
This correspondence will be firmly established in section~\ref{section:xncKP}.
At this stage we already recognize it by comparing identities derived above
with corresponding formulae in section 5 of \cite{DMH04ncKP}, keeping the relations
in the introduction and (\ref{Sigma-circ}) in mind. In this way, the ncKP expression
(5.31) in \cite{DMH04ncKP} for $\phi_{t_m t_n}$ finds its algebraic counterpart
in (\ref{PcircP-id}), provided that $\ast$ corresponds to $\htimes$.
Such a (at this point still somewhat vague) correspondence is indeed observed
between further (x)ncKP relations in \cite{DMH04ncKP} and algebraic identities
in this section. The first non-trivial equation which arises from (\ref{PcircP-id})
is the one with $m=n=2$ and yields
\be
   4 \, P \circ P_3 - P^{\circ 4} - 3 \, P_2 \circ P_2
   = 6 \, P \circ (P \htimes P) - 6 \, ( P \htimes P_2 - P_2 \htimes P )
   \label{KP-Pid}
\ee
which should be compared with (\ref{ncKP-T-id}) (see also the end of
section~\ref{subsection:embedding}).
\vskip.1cm

Taking further algebraic objects built with $\prec$ into consideration,
we obtain additional identities. With the choice
$\{ P_n , A_{mn} \}$ we have the identities (\ref{Amn-id}) and a correspondence
with xncKP equations is achieved (cf (5.30) in \cite{DMH04ncKP}). This will be made
precise in section~\ref{subsection:xncKP}.
Since the basis $\{ P_{m_1 \ldots m_k} \}$ of $\A(P)$ contains more
objects, one should expect that an extension of the ncKP
hierarchy exists which contains counterparts of all of them.
This expectation will be confirmed in section~\ref{section:XncKP}.
\vskip.1cm

Let us recall the underlying idea which might have gone lost during the development of
so much formalism. In the partial sum calculus realization, identities like
(\ref{PcircP-id}) become relations between sums where the summations run from
$1$ to some number $N \in \mathbb{N}$. The latter number is completely arbitrary,
however. Hence we obtain families of sum identities if we let $N$ run through the
natural numbers. Mapping the original identities in $\A(P)$ properly to partial differential
equations, as sketched in the introduction, the resulting differential equations will
be solvable by the ansatz (\ref{eps-expansion}) and thus admit KP-like soliton solutions.

\subsection{Symmetric functions}
A simple calculation yields
\be
     P_m \circ H_n
 &=& P_m \succ H_n + \sum_{r=1}^{n-1} H_r \prec P_m \succ H_{n-r} + H_n \prec P_m \nonumber \\
 &=& P_m \succ H_n + \sum_{r=2}^{n-1} H_{r-1} \succ (P \succ P_m - P_{m+1}) \succ H_{n-r}
      \nonumber \\
 & & + (P \succ P_m - P_{m+1}) \succ H_{n-1} + H_n \prec P_m   \nonumber \\
 &=& P_m \succ H_n + \sum_{r=1}^{n-1} H_r \succ P_m \succ H_{n-r}
     - \sum_{r=1}^{n-2} H_r \succ P_{m+1} \succ H_{n-r-1}  \nonumber \\
 & & - P_{m+1} \succ H_{n-1} + H_n \succ P_m - H_{n-1} \succ P_{m+1}  \; .
\ee
Summing this relation properly, we obtain
\be
  n H_n = \sum_{r=1}^n P_r \circ H_{n-r} \qquad \quad  n \in \mathbb{N} \; .  \label{nhn}
\ee
A similar calculation, or a simple application of the involution ${}^\psi$ to the last
formula\footnote{If we choose the involution ${}^\psi$ such that $P^\psi = P$, then
$P_n{}^\psi= (-1)^{n-1}P_n$ and $C_n{}^\psi = H_n$.},
leads to
\be
    n C_n = \sum_{r=1}^n(-1)^{r-1} C_{n-r} \circ P_r \qquad \quad  n \in \mathbb{N}
               \label{nen}
\ee
where
\be
  C_n := P^{\prec n} = (-1)^n \, C^{(1)}_n  \qquad \quad  n \in \mathbb{N} \; . \label{C_n}
\ee
Defining generating functions (with an indeterminate $\la$) by
\be
    H(\la) := \sum_{n \geq 0} H_n \, \la^n \qquad
    C(\la) := \sum_{n \geq 0} C_n \, \la^n \qquad
    P(\la) := \sum_{n \geq 1} P_n \, \la^{n-1}
    \label{HCP(lambda)}
\ee
where $H_0 = C_0 = I$ with a unit\footnote{Here the unit element $I$ is only introduced
temporarily in order to achieve compact expressions in terms of the exponential function.}
$I$ of the $\circ$-product, allows us to express (\ref{nhn}) and (\ref{nen}) in the form
\be
    {d \over d \la} H(\la) = P(\la) \circ H(\la)  \qquad \quad
    {d \over d \la} C(\la) = P(-\la) \circ C(\la) \; .
\ee
Setting
\be
    \tilde{P}(\la) := \int P(\la) \, d \la = \sum_{n \geq 1} {P_n \over n} \la^n
    \label{tP(lambda)}
\ee
we find
\be
    H(\la) = e_\circ^{\tilde{P}(\la)}  \qquad \quad
    C(\la) = e_\circ^{-\tilde{P}(-\la)}   \label{H(lambda),C(lambda)}
\ee
where the exponentials are built with the $\circ$-product.
This implies $C(-\la) \circ H(\la) = I$ and thus
\be
    \sum_{r=0}^n (-1)^r C_r \circ H_{n-r} =0 \; .
\ee
Moreover, recalling the definition
\be
    e^{\sum_{n \geq 1} x_n \la^n} = \sum_{n \geq 0} \chi_n(x_1,x_2,x_3,\ldots) \, \la^n
\ee
(with commuting variables $x_k$, $k=1,2, \ldots$) of the \emph{Schur polynomials},
we obtain
\be
  H_n &=& \chi_n(P,P_2/2,P_3/3,\ldots)
       = \sum_{|\mu|=n} z_\mu^{-1} \, P_1^{\circ m_1} \circ \ldots \circ P_n^{\circ m_n}
         \label{Hn_Schur} \\
  C_n &=& (-1)^n \, \chi_n(-P,-P_2/2,-P_3/3,\ldots) \nonumber \\
      &=& (-1)^n \sum_{|\mu|=n} z_\mu^{-1} \, (-1)^{m_1 + \ldots + m_n} \,
            P_1^{\circ m_1} \circ \ldots \circ P_n^{\circ m_n}
            \label{Cn_Schur}
\ee
where the sum is over all partitions $\mu = (1^{m_1} 2^{m_2} \ldots n^{m_n})$ of $n$
(so that $n = m_1 \, 1 + m_2 \, 2 + \ldots + m_n \, n$ with $m_r \in \mathbb{N} \cup \{ 0 \}$),
and
\be
    z_\mu := \prod_{r=1}^n r^{m_r} \, m_r!  \; .
\ee
\vskip.1cm

Writing $P = \sum_{k \geq 1} p_k \, e_k$ in the case of the partial sum calculus,
\be
   \Sigma_N(P_n) = \sum_{k=1}^N p_k^n
\ee
is the $n$th \emph{power sum},
\be
   \Sigma_N(C_n) = \sum_{1 \leq k_1 < \ldots < k_n \leq N} p_{k_1} \cdots p_{k_n}
\ee
the $n$th \emph{elementary symmetric polynomial}, and
\be
   \Sigma_N(H_n) = \sum_{1 \leq k_1 \leq \ldots \leq k_n \leq N} p_{k_1} \cdots p_{k_n}
\ee
the \emph{complete symmetric polynomial} of degree $n$
in $N$ indeterminates $p_1, \ldots, p_N$ \cite{Macd79}.
\vskip.1cm

\noindent
{\it Remark.} Applying the Rota-Baxter operator $R$ defined in (\ref{R(alpha)-ps-calc})
to $C(\lambda)$, using (\ref{C_n}), (\ref{HCP(lambda)}), (\ref{H(lambda),C(lambda)})
and (\ref{psc_R(prec)->bull}), leads to
\be
    R(e_\circ^{-\tilde{P}(-\lambda)})
  = \sum_{n \geq 0} \lambda^n R(R(\cdots R(R(P) \bullet P) \bullet \cdots) \bullet P)
\ee
On the other hand, according to (\ref{psc_Rcirc_bull}) we have
\be
    R(e_\circ^{-\tilde{P}(-\lambda)})
  = e_\bullet^{- R(\tilde{P}(-\lambda))} \; .
\ee
With the help of $\ln(1+x) = - \sum_{n \geq 1} (-1)^n x^n/n$, we can write
\be
    \tilde{P}(-\lambda)
  = \sum_{n \geq1} (-1)^n P^{\bullet n} \lambda^n /n
  = - \ln_\bullet (1+\lambda P) \; .
\ee
Hence
\be
    \sum_{n \geq 0} \lambda^n R(R(\cdots R(R(P) \bullet P) \bullet \cdots) \bullet P)
  = \exp_\bullet ( -R (\ln_\bullet(1+\lambda P))
\ee
which is the famous \emph{Spitzer's formula} \cite{Spit56,Baxt60,Cart72,Rota+Smit72,EGK04Spitzer}.
\hfill $\blacksquare$

\section{Embedding of $\A(P)$ into an algebra generated by two elements}
\label{section:A(P,Q)}
\setcounter{equation}{0}
In the previous section we suggested a correspondence between identities in
$\A(P)$ and the ncKP hierarchy (and certain extensions). Writing
$P = \sum_{n \geq 1} p_n \, e_n$ in the partial sum realization and taking
a look at the algebraic identities presented in the introduction, one
immediately concludes that a second element $Q = \sum_{n \geq 1} q_n \, e_n$
is required. But in this section we show that it is actually sufficient
to restrict considerations to $\A(P)$. This covers an important aspect of
our framework (see also the conclusions). The material of the present section is,
however, not used in the following sections.
\vskip.1cm

In the following, $(\A(P), \bullet)$ will \emph{not} be regarded as unital, i.e.,
we exclude a possible unit element $E$. It is convenient (though not necessary)
to augment the algebra $\A$ by a new element $I$.
The necessary preparations are presented in the next two subsections.
The third subsection presents the main result, namely the existence of an
`embedding' $\Psi$ of $\A(P)$ into an algebra generated by two elements $P,Q$
such that certain homomorphism properties hold.
The last subsection contains supplementary material (a generalization of
symmetric functions).

\subsection{The augmented algebra $\tA$}
The new element $I$ will be required to satisfy
\be
    I \prec \alpha = \alpha = \alpha \prec I  \qquad\quad
    I \succ \alpha = \alpha = \alpha \succ I  \qquad\quad
    I \circ \alpha = \alpha = \alpha \circ I
\ee
which implies
\be
    \alpha \bullet I = I \bullet \alpha =0 \; .
\ee
A further consequence is
\be
    (\alpha \succ I) \prec \beta = \alpha \prec \beta    \qquad
    \alpha \succ (I \prec \beta) = \alpha \succ \beta
\ee
which shows that we are forced to give up associativity in these particular combinations.
\vskip.1cm

The \emph{augmented algebra} $\tA$ is again a graded algebra, with
$\tA^0 = \tA_0 = \mathbb{K} \, I$ and
$\tA = \bigoplus_{r \geq 0} \tA^r = \bigoplus_{r \geq 0} \tA_r$
where $\tA^r \simeq \A^r$, $\tA_r \simeq \A_r$ for $r \geq 1$.
\vskip.1cm

With each $A \in \tA^1$ we associate products via (\ref{hA}) and (\ref{cA})
which are essentially\footnote{Non-associativity only appears in special expressions
involving $I$.}
combined associative with all other products defined so far, with the exception of the
main product, and thus also among themselves. In particular, we have
\be
   (\alpha \, \cA \, \beta) \, \hB \, \gamma = \alpha \, \cA \, (\beta \, \hB \, \gamma)
   \qquad
   (\alpha \, \hA \, \beta) \, \cB \, \gamma = \alpha \, \hA \, (\beta \, \cB \, \gamma)
   \label{mixed-I-assoc}
\ee
for all $A,B \in \tA^1$, $\alpha, \beta, \gamma \in \tA$, and
\be
  \begin{array}{l@{\, = \,}l}
  (\alpha \, \hA \, \beta) \, \hB \, \gamma & \alpha \, \hA \, (\beta \, \hB \, \gamma) \\
  (\alpha \, \cA \, \beta) \, \cB \, \gamma & \alpha \, \cA \, (\beta \, \cB \, \gamma)
  \end{array}
  \qquad \mbox{if} \quad \beta \neq I
\ee
so that we are allowed to drop the brackets and simply write, e.g.,
$\alpha \, \hA \, \beta \, \hB \, \gamma$ if $\beta \neq I$. Since
\be
  &&I \, \hA \, I = A \qquad\quad
    I \, \hA \, \alpha = A \succ \alpha \qquad\quad
    \alpha \, \hA \, I = \alpha \prec A  \\
  &&I \, \cA \, I = A  \qquad\quad
    I \, \cA \, \alpha = A \prec \alpha \qquad\quad
    \alpha \, \cA \, I = \alpha \succ A
\ee
we can express any element of $\tilde{\A}$ in terms of these operators. For example,
\bez
 & &  A_1 \succ A_2 \prec A_3 \prec A_4 \succ A_5
  = (A_1 \succ A_2 \prec A_3) \, \hA_4 \, A_5 \\
 &=& \Big( (A_1 \succ A_2) \, \hA_3 \, I \Big) \, \hA_4 \, (I \, \hA_5 \, I)
  = \Big( (I \hA_1 \, (I \, \hA_2 \, I) ) \, \hA_3 \, I \Big) \, \hA_4 \, (I \, \hA_5 \, I) \\
 &=& \Big( I \hA_1 \, (I \, \hA_2 \, I) \, \hA_3 \, I \Big) \, \hA_4 \, (I \, \hA_5 \, I)
  =: I \, \hA_1 \, (I \, \hA_2 \,I) \, (\hA_3 \, I) \, \hA_4 \, (I \, \hA_5 \, I)
\eez
where we introduced a simplified notation in the last step. The remaining brackets take
care of the non-associativity of certain products with $I$.
In the same way we get
\bez
     A_1 \succ A_2 \prec A_3 \prec A_4 \succ A_5
   = (I \, \cA_1 \, I) \, \cA_2 \, (I \, \cA_3) \, (I \, \cA_4 \, I) \, \cA_5 \, I \; .
\eez
Eliminating the $I$'s at both ends, we obtain two linear maps,
$\alpha \mapsto \hat{\alpha}$, respectively $\alpha \mapsto \check{\alpha}$.
In particular,
\bez
    A_1 \succ A_2 \prec A_3 \prec A_4 \succ A_5 \; &\stackrel{\hat{ }}{\mapsto}&\;
    \hA_1 \, (I \, \hA_2 \, I) \, (\hA_3 \, I) \, \hA_4 \, (I \, \hA_5) \\
    A_1 \succ A_2 \prec A_3 \prec A_4 \succ A_5 \; &\stackrel{\check{ }}{\mapsto}&\;
    (\cA_1 \, I) \, \cA_2 \, (I \, \cA_3) \, (I \, \cA_4 \, I) \, \cA_5  \; .
\eez
The following properties are quite evident:
\bez
 & & \alpha \prec \beta \stackrel{\hat{ }}{\mapsto} (\hat{\alpha} I) \hat{\beta}
     \qquad \quad
     \alpha \succ \beta \stackrel{\hat{ }}{\mapsto} \hat{\alpha} (I \hat{\beta})  \\
 & & \alpha \prec \beta \stackrel{\check{ }}{\mapsto} \check{\alpha} (I \check{\beta})
     \qquad \quad
     \alpha \succ \beta \stackrel{\check{ }}{\mapsto} (\check{\alpha} I) \check{\beta} \, .
\eez

The identities (\ref{Acircbeta}) and (\ref{betacircA}) can be written as
\be
      A \circ \alpha
  &=& I \, \hA \, \alpha + \sum \alpha_{(1)} \, \hA \, \alpha_{(2)}
      + \alpha \, \hA \, I =: {\sum}' \alpha_{(1)} \, \hA \, \alpha_{(2)} \label{Acirc-hA} \\
      \alpha \circ A
  &=& I \, \cA \, \alpha + \sum \alpha_{(1)} \, \cA \, \alpha_{(2)}
      + \alpha \, \cA \, I =: {\sum}' \alpha_{(1)} \, \cA \, \alpha_{(2)} \; .
      \label{circA-cA}
\ee

\subsection{The augmented subalgebra $\tA(P)$}
Let $\tA(P)$ be the subalgebra of $\tA$ obtained from the algebra $\A(P)$, which is
generated by a single element $P \in \A^1$, by augmenting it with $I$.
Then
\be
   P_n = I \, \hP_n I  \qquad \quad   n \in \mathbb{N}  \; .
\ee
Clearly, $(\tA^1(P),\bullet)$ is commutative, and thus also $(\tA(P),\circ)$
according to theorem~\ref{theorem:circ-comm}.

\begin{lemma}
\label{lemma:hP_n&cP_n}
The following identities hold for all $n \in \mathbb{N}$,
\be
    \hP_{n+1} &=& \hP \, (I \, \hP_n) - (\hP \, I) \, \hP_n  \label{hP_n} \\
    \cP_{n+1} &=& (\cP \, I) \, \cP_n - \cP \, (I \, \cP_n) \; . \label{cP_n}
\ee
\end{lemma}
{\it Proof:}
\bez
     \alpha \, \hP_{n+1} \, \beta
 &=& \alpha \prec ( P \succ P_n - P \prec P_n) \succ \beta
  = \alpha \, \hP \, (P_n \succ \beta) - \alpha \prec ( P \, \hP_n \, \beta ) \\
 &=& \alpha \, \hP \, (I \, \hP_n) \, \beta - (\alpha \, \hP \, I) \, \hP_n \, \beta \; .
\eez
The second identity is verified in the same way.
\hfill $\blacksquare$
\vskip.2cm

The product $\htimes$ introduced in section~\ref{section:A(P)}, extended to $\tA$,
is essentially associative:
\be
    (\alpha \htimes \beta) \htimes \gamma = \alpha \htimes (\beta \htimes \gamma)
    \qquad \quad \forall \, \alpha, \beta, \gamma \in \tA(P) \, , \quad \beta \neq I \; .
    \label{mnonas}
\ee
Note that
\be
    I \htimes I = - P \qquad\quad
    I \htimes \alpha = - P \succ \alpha \qquad\quad
    \alpha \htimes I = - \alpha \prec P  \; .
\ee
By iterative use of (\ref{hP_n}), we can express $\hP_n$ in terms of only $I$
and the product $\htimes$. For example,
\be
      \alpha \, \hP_2 \, \beta
  &=& \alpha \, \hP \, (I \, \hP) \, \beta - \alpha \, (\hP \, I) \, \hP \, \beta
   = - \alpha \htimes ( I \, \hP \, \beta) + (\alpha \, \hP \, I) \htimes \beta \nonumber \\
  &=& \alpha \htimes ( I \htimes \beta) - (\alpha \htimes I) \htimes \beta \; .
\ee
Since we have
\be
    P_m \prec P_n  = P_m \, \hP_n I     \qquad \quad
    P_m \succ P_n = I \, \hP_m \, P_n
\ee
and similar formulae for expressions of higher grade, it follows that the algebraic
structure of $\tA(P)$ can be expressed completely in terms of the element $I$
and the product $\htimes$.
Further examples of expressions in terms of $I$ are $C_n = I \, (\hP \, I)^n$,
$H_n = (I \, \hP)^n I$,
\be
    P_{m_1 \ldots m_k}
  = (I \, \hP_{m_1} I) (\hP_{m_2} \, I) \cdots (\hP_{m_k} I)
\ee
and (cf (\ref{Acirc-hA}))
\be
    P_n \circ P_{m_1 \ldots r_k}
  = \sum_{l=0}^k \Big( (I \, \hP_{m_1} \, I) \cdots (\hP_{m_l} \, I) \Big) \hP_n \,
    \Big( (I \, \hP_{m_{l+1}} \, I) \cdots (\hP_{m_k} \,I) \Big) \; .  \label{P_n-circ}
\ee

\subsection{The embedding}
\label{subsection:embedding}
Let $\tA(P,Q)$ denote the subalgebra of $\tA$ generated by two fixed elements
$P,Q \in \A^1$ with the property $P \bullet Q = Q \bullet P$, so that $(\tA^1,\bullet)$
is commutative and then, by theorem~\ref{theorem:circ-comm}, also $(\tA(P,Q),\circ)$.
Let us introduce the product
\be
      \alpha \times \beta
  := - \alpha \, \bT \, \beta
   = \alpha \succ Q \prec \beta - \alpha \prec P \succ \beta
\ee
where
\be
    \bT := \hP - \cQ  \; .
\ee
The product $\times$ is essentially associative, i.e.,
\be
    (\alpha \times \beta) \times \gamma = \alpha \times (\beta \times \gamma)
\ee
for all $\alpha, \beta, \gamma \in \tA(P,Q)$ with $\beta \neq I$.
\vskip.1cm

Next we define a linear map $\Psi : \tilde{\A}(P) \rightarrow \tilde{\A}(P,Q)$ by
$\Psi(I) = I$ and the homomorphism property
\be
   \Psi(\alpha \htimes \beta) = \Psi(\alpha) \times \Psi(\beta) \qquad \quad
         \forall \alpha, \beta \in \tA(P) \; .  \label{Psi-homom}
\ee
Since $I$ generates $\tA(P)$ using the product $\htimes$, this defines $\Psi$ on
$\tA(P)$. In particular, it leads to
\be
   \Psi(P) = - \Psi(I \htimes I) = - \Psi(I) \times \Psi(I) = - I \times I
           = P - Q = I \, \bT \, I
\ee
and
\be
   \Psi( P \succ \alpha ) = I \, \bT \, \Psi(\alpha)    \qquad\quad
   \Psi( \alpha \prec P ) = \Psi(\alpha) \, \bT \, I \; .
\ee
Resolving the definitions of the two products in (\ref{Psi-homom}), the homomorphism property
of $\Psi$ reads $\Psi(\alpha \hP \beta) = \Psi(\alpha) \bT \Psi(\beta)$, which can be
expressed in the short form $\Psi(\hP) = \bT$.

\begin{proposition}
\label{proposition:Psi(P_n)}
\be
    \Psi(\hP_n) = \hP_n - \cQ_n
      \qquad \quad n=1,2, \ldots    \label{Psi(P_n)}
\ee
where $\hP_n$ and $\cQ_n$ are determined by $P_n = P^{\bullet \, n}$ and $Q_n = Q^{\bullet \, n}$,
respectively.
\end{proposition}
{\it Proof:} By construction of $\Psi$, (\ref{Psi(P_n)}) holds for $n=1$.
Let us now assume that $\Psi(\hP_n) = \hP_n - \cQ_n =: \bT_{n}$ holds for
fixed $n \in \mathbb{N}$. Then $\Psi$ applied to (\ref{hP_n}) yields
\bez
      \Psi( \alpha \, \hP_{n+1} \beta )
  &=& \Psi( \alpha \, \hP \, (I \, \hP_n) \, \beta) - \Psi( \alpha \, (\hP \, I) \, \hP_n \, \beta) \\
  &=& - \Psi( \alpha ) \times \Psi( I \, \hP_n \, \beta)
      - \Psi( \alpha \, \hP \, I) \, \bT_{n} \, \Psi(\beta) \\
  &=& \Psi( \alpha ) \, \bT \, ( \Psi(I) \, \bT_{n} \, \Psi(\beta) )
      - ( \Psi( \alpha ) \, \bT \, I ) \, \bT_{n} \, \Psi(\beta) \\
  &=& \Psi( \alpha ) \Big( \bT \, ( I \, \bT_{n} )
      - ( \bT \, I) \, \bT_{n} \Big) \, \Psi(\beta) \; .
\eez
Making use of (\ref{mixed-I-assoc}), (\ref{hP_n}) and (\ref{cP_n}), we find
\bez
     \bT \, (I \, \bT_{n}) - (\bT \, I) \, \bT_{n}
   = \hP \, (I \, \hP_n) - (\hP \, I) \, \hP_n
     + \cQ \, (I \, \cQ_n) - (\cQ \, I) \, \cQ_n
   = \hP_{n+1} - \cQ_{n+1} \; .
\eez
Hence $\Psi(\hP_{n+1}) = \hP_{n+1} - \cQ_{n+1}$ which completes the induction.
\hfill $\blacksquare$
\vskip.2cm

Resolving the definitions involved, (\ref{Psi(P_n)}) reads
\be
  \Psi(\alpha \prec P_n \succ \beta)
 = \Psi(\alpha) \prec P_n \succ \Psi(\beta) - \Psi(\alpha) \succ Q_n \prec \Psi(\beta) \; .
\ee
In particular, we obtain
\be
        \Psi(\alpha \prec P_n)
  = \Psi(\alpha) \prec P_n - \Psi(\alpha) \succ Q_n  \qquad
        \Psi(P_n \succ \beta)
  = P_n \succ \Psi(\beta) - Q_n \prec \Psi(\beta) \; .
\ee

\begin{theorem}
\label{theorem:Psi-circ-hom}
The map $\Psi$ defined above is a main product homomorphism, i.e.,
\be
     \Psi(\alpha \circ \beta) = \Psi(\alpha) \circ \Psi(\beta) \qquad \quad
     \forall \alpha, \beta \in \tA(P) \; .
\ee
\end{theorem}
{\it Proof:} First we prove this property for $\alpha, \beta \in \tA^1(P)$.
It is sufficient to consider
\bez
     \Psi(P_r \circ P_s)
 &=& \Psi(P_r \succ P_s + P_s \prec P_r)
  =  \Psi(I \hP_r \, P_s + P_s \hP_r \, I)
  = I \bT_{r} \, T_{s} + T_{s} \bT_{r} \, I  \\
 &=& P_r \circ (P_s-Q_s) - (P_s-Q_s) \circ Q_r
  = (P_r-Q_r) \circ (P_s-Q_s)
\eez
where $\bT_{r} := \hP_r - \cQ_r$, $T_{r} := P_r - Q_r$ (in deviation from our
previous notation), and we used the commutativity of the main product in the last step.
Hence $\Psi(P_r \circ P_s) = T_{r} \circ T_{s} = \Psi(P_r) \circ \Psi(P_s)$
and thus $\Psi(A \circ B) = \Psi(A) \circ \Psi(B)$ for all $A,B \in \tA^1(P)$.

Next, we show that $\Psi(A \circ \beta ) = \Psi(A) \circ \Psi(\beta)$,
$\forall \beta \in \tA(P)$. It suffices to consider $A = P_n$ and
\bez
   \beta = P_{m_1} \prec \ldots \prec P_{m_k}
         = (I \, \hP_{m_1} I) \, (\hP_{m_2} \, I) \cdots (\hP_{m_k}I) \; .
\eez
By use of (\ref{P_n-circ}) and proposition~\ref{proposition:Psi(P_n)}, we find
\bez
     \Psi(P_n \circ \beta)
 &=& {\sum}' \Psi(\beta_{(1)} \hP_n \, \beta_{(2)})
  =  {\sum}' \Psi(\beta_{(1)}) \bT_{n} \, \Psi(\beta_{(2)})  \; .
\eez
Iterated application of proposition~\ref{proposition:Psi(P_n)} leads to
\bez
     \Psi(P_{m_1} \prec \ldots \prec P_{m_j})
 &=& \Psi \Big( (I \hP_{m_1} \, I) \cdots (\hP_{m_j} \, I) \Big)
  =  (I \bT_{m_1} \, I) \cdots (\bT_{m_j} \, I)  \\
     \Psi(P_{m_{j+1}} \prec \ldots \prec P_{m_k})
 &=& \Psi \Big( (I \hP_{m_{j+1}} \, I) \cdots (\hP_{m_k} \, I) \Big)
  =  (I \bT_{m_{j+1}} \, I) \cdots (\bT_{m_k} \, I)
\eez
and
\bez
  \Psi(\beta) = (I \bT_{m_1} \, I) \cdots (\bT_{m_j} \, I) (\bT_{m_{j+1}} \, I)
                  \cdots (\bT_{m_k} \, I)
\eez
so that
\bez
    \Psi(\beta_{(1)}) = \Psi(\beta)_{(1)} \qquad\quad
    \Psi(\beta_{(2)}) = \Psi(\beta)_{(2)} \; .
\eez
It follows that
\bez
     \Psi(P_n \circ \beta)
 &=& {\sum}' \Psi(\beta)_{(1)} \, \bT_n \, \Psi(\beta)_{(2)} \\
 &=& {\sum}' \Psi(\beta)_{(1)} \, \hP_n \, \Psi(\beta)_{(2)}
     - {\sum}' \Psi(\beta)_{(1)} \, \cQ_n \, \Psi(\beta)_{(2)} \\
 &=& P_n \circ \Psi(\beta) - \Psi(\beta) \circ Q_n
  = T_n \circ \Psi(\beta)
\eez
where we used (\ref{gamma_circ_cA}), (\ref{hA_circ_gamma}), and again the commutativity
of the main product in the last two steps. This implies
$\Psi(A \circ \beta ) = \Psi(A) \circ \Psi(\beta)$.

Finally, we prove our assertion in the general case by induction. We assume that it
holds for all $\alpha \in \tA^m(P)$ where $1 \leq m \leq n$ for fixed $n$, and all
$\beta \in \tA(P)$. The induction step is then carried out with the help of
(\ref{A-succ-a-c-b}), i.e.,
\bez
   (P_r \succ \alpha) \circ \beta
 = I \, \hP_r \, (\alpha \circ \beta)
   + \sum \beta_{(1)} \, \hP_r \, (\alpha \circ \beta_{(2)}) + \beta \, \hP_r \, \alpha
 = {\sum}' \beta_{(1)} \, \hP_r \, (\alpha \circ \beta_{(2)}) \; .
\eez
Applying $\Psi$ and using proposition~\ref{proposition:Psi(P_n)}, leads to
\bez
     \Psi((P_r \succ \alpha) \circ \beta)
 &=& {\sum}' \Psi(\beta_{(1)}) \, \bT_{r} \, \Psi(\alpha \circ \beta_{(2)})
  = {\sum}' \Psi(\beta)_{(1)} \, \bT_{r} \, \Psi(\alpha) \circ \Psi(\beta)_{(2)} \\
 &=& {\sum}' \Psi(\beta)_{(1)} \, \hP_r \, \Psi(\alpha) \circ \Psi(\beta)_{(2)}
    - {\sum}' \Psi(\beta)_{(1)} \, \cQ_r \, \Psi(\alpha) \circ \Psi(\beta)_{(2)} \\
 &=& (P_r \succ \Psi(\alpha)) \circ \Psi(\beta) - \Psi(\beta) \circ (Q_r \prec \Psi(\alpha)) \\
 &=& \Psi(P_r \succ \alpha) \circ \Psi(\beta)
\eez
where we made use of (\ref{A-succ-a-c-b}), (\ref{b-c-A-prec}), and the commutativity
of the $\circ$ product. This implies that our assertion also holds for
$\alpha \in \tA^{n+1}(P)$.
\hfill $\blacksquare$
\vskip.2cm

For generic $Q$, the map $\Psi : \tA(P) \rightarrow \tA(P,Q)$ is injective.
It follows that $\Psi$ is an isomorphism of the (double) algebras $(\A(P),\htimes,\circ)$
and $(\A(P/Q),\times,\circ)$ where $\A(P/Q) := \Psi(\A(P))$.\footnote{From the construction
of $\Psi$ it is evident that the elements of $\A(P/Q)$ are invariant under simultaneous
translations $P \mapsto P+A$, $Q \mapsto Q+A$ with any $A \in \tA^1$ such that
$A \bullet P = P \bullet A = 0$ and $A \bullet Q = Q \bullet A = 0$.}
Applying $\Psi$ and afterwards $\Sigma_N$ to the identity (\ref{KP-Pid}), for example,
we recover the algebraic sum identity (\ref{ncKP-T-id}).

\subsection{`Supersymmetric' functions}
Let us introduce
\be
  \tilde{T}(\la) := \sum_{n \geq 1} {T_n \over n} \, \la^n := \tilde{P}(\la) - \tilde{Q}(\la)
\ee
where $\tilde{P}(\la)$ is given by (\ref{tP(lambda)}) and $\tilde{Q}(\la)$ is defined
in the same way (with $P$ replaced by $Q$). Using the commutativity of $\circ$,
we obtain
\be
   H^{P/Q}(\la) := \sum_{n \geq 0} H^{P/Q}_n \, \la^n
                := e_\circ^{\tilde{T}(\la)}
                 = e_\circ^{\tilde{P}(\la)} \circ e_\circ^{-\tilde{Q}(\la)}
                 = H^P(\la) \circ C^Q(-\la)
\ee
where $H^P(\la)$ is given by the first of relations (\ref{H(lambda),C(lambda)}),
and $C^Q(\la)$ by the second with $P$ replaced by $Q$. Hence
\be
    H^{P/Q}_n = \sum_{r=0}^n (-1)^{n-r} \, H^P_r \circ C^Q_{n-r} \; .
\ee
In the same way, we obtain
\be
    C^{P/Q}(\la) := \sum_{n \geq 0} C^{P/Q}_n \, \la^n := e_\circ^{-\tilde{T}(-\la)}
    \qquad \quad
    C^{P/Q}_n = (-1)^n \, H^{Q/P}_n \; .
\ee
As a consequence of theorem~\ref{theorem:Psi-circ-hom},
\be
    H^{P/Q}_n = \Psi(H^P_n) \qquad \qquad
    C^{P/Q}_n = \Psi(C^P_n) \; .
\ee

Using $P = \sum_{n \geq 1} p_n \, e_n$ and $Q = \sum_{n \geq 1} q_n \, e_n$
in the partial sum calculus, we obtain
\be
    \Sigma_N(T_r) = \sum_{k=1}^N (p_k^r - q_k^r) \; .  \label{Sigma(Tr)}
\ee
A function $f(p_1,\ldots,p_N, q_1,\ldots,q_N)$ is called \emph{doubly symmetric}
if it is invariant under permutations of $p_1,\ldots,p_N$,
as well as permutations of $q_1,\ldots,q_N$.\footnote{Generalizations are
sometimes called `multi-symmetric functions', see \cite{Olve+Shak92} and the
references cited therein.}
A doubly symmetric function is called \emph{supersymmetric} if the substitution
$q_1 = p_1$ results in a function which is independent of $p_1$ \cite{Stem85}.\footnote{Such
functions have been called \emph{bisymmetric} in \cite{MNR81}.}
Together with $1$, the sums (\ref{Sigma(Tr)}) actually generate the algebra of
supersymmetric polynomials of $N+N$ indeterminates \cite{Stem85}.
Then $\Sigma_N(C^{P/Q}_n)$ and $\Sigma_N(H^{P/Q}_n)$ are the \emph{elementary},
respectively the \emph{complete supersymmetric polynomials} (see \cite{Mole+Reta04}).

\section{From $\A(P)$ to the algebra of $\Psi$DOs}
\label{section:PsiDOs}
\setcounter{equation}{0}
In the following, $\mathcal{R}$ denotes the $\mathbb{K}$-algebra of formal
pseudo-differential operators generated by a generic\footnote{In the sense that no
non-trivial identities should hold in $\mathcal{R}$.}
$L$ of the form (\ref{L}) with the product $\ast$ and the projection $( \, )_{\geq 0}$.
For $X,Y \in \mathcal{R}$,
\be
       X \vtr Y
  &:=& X_{\geq 0} \ast Y_{\geq 0} - X_{<0} \ast Y_{<0} \label{vtr} \\
   &=& X_{\geq 0} \ast Y - X \ast Y_{<0}
    = X \ast Y_{\geq 0} - X_{<0} \ast Y  \nonumber
\ee
defines an associative product\footnote{This product already appeared in
\cite{Seme84}. It is an example of an associative product defined more generally
in the framework of Rota-Baxter algebras, see appendix~A. Indeed, $R(X) := X_{\geq 0}$
defines a Rota-Baxter operator on the algebra $(\mathcal{R},\ast)$. Then
$X \vtr Y = R(X) \ast Y + X \ast R(Y) - X \ast Y$.}
$\mathcal{R} \times \mathcal{R} \rightarrow \mathcal{R}$.
As an immediate consequence of its definition, $\vartriangle$ has the property
\be
       \res (X \vtr Y) = 0  \qquad \quad \forall \, X,Y \in \mathcal{R} \; .
       \label{res-vtr}
\ee
\vskip.1cm

In this section we consider $\A(P)$ \emph{not} as unital, i.e., we exclude
a possible unit element from the algebra. A corresponding extension is certainly
possible, but not needed for our purposes.
Let $\ell,r : \A(P) \to \mathcal{R}$ be the two linear maps defined
iteratively by $\ell(P) = r(P) = L$ and
\be
    \ell(\alpha \prec P) &=& - \ell(\alpha)_{<0} \ast L  \qquad \quad
    \ell(\alpha \succ P) = \ell(\alpha)_{\geq 0} \ast L  \label{ell(a<>P)} \\
    r(P \prec \alpha) &=& - L \ast r(\alpha)_{\geq 0}  \qquad \quad
    r(P \succ \alpha) = L \ast r(\alpha)_{<0} \, .   \label{r(P<>a)}
\ee
The pseudo-differential operators defined by
\be
     L^{m_1, \ldots, m_k} := \ell ( P_{m_1 \ldots m_k} )  \label{L^m...m-def}
\ee
will be important in the following.
In particular, they are used to define operators $\delta_{m_1 \ldots m_k}$
in $\mathcal{R}$ iteratively by
\be
     \delta_{m_1 \ldots m_k} L
 &=& - [L^{m_1, \ldots, m_k}_{<0} , L]
     + \sum_{j=1}^{k-1} ( \delta_{m_1 \ldots m_j} L )
       \ast L^{m_{j+1}, \ldots, m_k}_{<0}  \\
    \delta_{m_1 \ldots m_k}( X_{\geq 0} )
 &=& (\delta_{m_1 \ldots m_k} X)_{\geq 0}   \label{delta-geqX}
\ee
and the generalized derivation rule
\be
    \delta_{m_1 \ldots m_k} (X \ast Y)
  = \sum_{j=0}^k (\delta_{m_1 \ldots m_j} X) \ast (\delta_{m_{j+1} \ldots m_k} Y)
    \label{delta-genderiv}
\ee
where $\delta_{m_1 \ldots m_j} = \mathrm{id}$ if $j=0$ and $\delta_{m_{j+1} \ldots m_k}
= \mathrm{id}$ if $j=k$.
We already met the simplest members $\delta_n$ of this family in the introduction, for
which the last formula reduces to the ordinary derivation rule.
After some preparations in the first two subsections, the third demonstrates that
the $\delta_{m_1 \ldots m_k}$ commute with each other.
In the last subsection we explore properties of the linear map
$\Phi : \A(P) \rightarrow \mathcal{R}_0$ defined by
\be
    \Phi(\alpha) := \res(\ell(\alpha))  \qquad \forall \alpha \in \A(P) \; . \label{Phi}
\ee
This map will play a crucial role in the following sections. The reader may jump
from here directly to section~\ref{section:xncKP} and skip the more technical subsections
on first reading.

\subsection{Properties of the maps $\ell$ and $r$}

\begin{lemma}
\be
    \ell(P_n) = L^n = r(P_n) \qquad \quad  n = 1,2, \ldots   \label{ell(P_n)}
\ee
\end{lemma}
{\it Proof:} Using the definition of $\ell$, we find
\bez
  \ell(P_{n+1}) = \ell(P_n \bullet P)
                = \ell(P_n \succ P) - \ell(P_n \prec P)
                = \ell(P_n)_{\geq 0} \ast L + \ell(P_n)_{<0} \ast L
                = \ell(P_n) \ast L \; .
\eez
Now the statement for $\ell$ follows by induction. The corresponding statement
for the map $r$ is obtained in the same way.
\hfill $\blacksquare$

\begin{lemma}
 For $n \in \mathbb{N}$ we have
\be
    \ell(\alpha \prec P_n) &=& - \ell(\alpha)_{<0} \ast L^n  \label{ell(a<P_n)} \\
    \ell(\alpha \succ P_n) &=& \ell(\alpha)_{\geq 0} \ast L^n  \\
       r(P_n \prec \alpha) &=& - L^n \ast r(\alpha)_{\geq 0}  \label{r(P_n<a)} \\
       r(P_n \succ \alpha) &=& L^n \ast r(\alpha)_{<0} \; .
\ee
\end{lemma}
{\it Proof:}
\bez
     \ell(\alpha \prec P_{n+1})
  &=& \ell( \alpha \prec P_n \bullet P)
   = \ell( (\alpha \prec P_n) \succ P) - \ell( (\alpha \prec P_n) \prec P) \\
  &=& \ell(\alpha \prec P_n)_{\geq 0} \ast L + \ell(\alpha \prec P_n)_{<0} \ast L
   = \ell(\alpha \prec P_n) \ast L \; .
\eez
Together with $\ell(\alpha \prec P) = - \ell(\alpha)_{<0} \ast L$, the
first relation of the lemma follows by induction. The other relations are
obtained in an analogous way.
\hfill $\blacksquare$
\vskip.2cm

By iterated application of the preceding lemma, we obtain
\be
     L^{m_1, \ldots, m_k}
 &=& (-1)^{k-1} \, (( \ldots ( L^{m_1}{}_{<0} \ast L^{m_2})_{<0}
     \ldots )_{<0} \ast L^{m_{k-1}} )_{<0} \ast L^{m_k} \label{L^m...m} \\
     \tilde{L}^{m_1, \ldots, m_k}
 &:=& r( P_{m_1 \ldots m_k} )  \nonumber \\
 &=& (-1)^{k-1} \, L^{m_1} \ast ( L^{m_2} \ast (
     \ldots ( L^{m_{k-1}} \ast L^{m_k}{}_{\geq 0} )_{\geq 0}
     \ldots )_{\geq 0} )_{\geq 0}  \, .  \label{tildeL^m...m}
\ee
Since the elements $P_{m_1 \ldots m_k}$ defined in (\ref{Pm1...mk}) span
$\A(P)$, this allows to compute $\ell(\alpha)$ and $r(\alpha)$ for all
$\alpha \in \A(P)$.

\begin{proposition}
\label{proposition:rightR-to-leftR}
In terms of \footnote{This notation avoids complex nested expressions like those
in (\ref{L^m...m}) and (\ref{tildeL^m...m}). For example,
$X_1 \ast \overrightarrow{R} X_2 \ast \ldots \ast \overrightarrow{R} X_k
= X_1 \ast (X_2 \ast( \ldots  (X_k)_{\geq 0} )_{\geq 0} \ldots )_{\geq 0}$.}
$\overrightarrow{R} X := X_{\geq 0}$ and $X \overleftarrow{R}:= X_{<0}$
the following identity holds in $\mathcal{R}$,
\be
 & & X_1 \ast \overrightarrow{R} X_2 \ast \ldots \ast \overrightarrow{R} X_k
  = X_1 \overleftarrow{R} \ast \ldots \ast X_{k-1} \overleftarrow{R} \ast X_k  \nonumber  \\
 & & \qquad
    + \sum_{j=1}^{k-1} (X_1 \overleftarrow{R} \ast \ldots \ast X_{j-1} \overleftarrow{R} \ast X_j)
      \vtr (X_{j+1} \ast \overrightarrow{R} X_{j+2} \ast \ldots \ast \overrightarrow{R} X_k) \; .
      \label{rightR-to-leftR}
\ee
\end{proposition}
{\it Proof:}
The formula is easily verified for $k=2$. The general formula is proved by induction on $k$.
For $k+1$ we write the left hand side as
\bez
   X_1 \ast \overrightarrow{R} X_2 \ast \ldots \ast \overrightarrow{R} X_{k+1}
 = X_1 \ast \overrightarrow{R} X_2 \ast \ldots \ast \overrightarrow{R} X_{k-1}
   \ast \overrightarrow{R} ( X_k \ast \overrightarrow{R} X_{k+1} )
\eez
to which we can now apply the induction hypothesis. After use of the identities
\bez
    X_k \ast \overrightarrow{R} X_{k+1}
 = (X_k \overleftarrow{R}) \ast X_{k+1} + X_k \vtr X_{k+1}
\eez
and
\bez
     Y \ast ( X_k \vtr X_{k+1} )
   - ( Y \ast (\overrightarrow{R} X_k) )\overleftarrow{R} \ast X_{k+1}
   - ( Y \ast X_k ) \vtr X_{k+1} = 0
\eez
for $Y$ with $Y = Y \overleftarrow{R} = Y_{<0}$, the formula with $k$ replaced by
$k+1$ results.
\hfill $\blacksquare$
\vskip.2cm

\noindent
{\bf Corollary}
\be
    \tilde{L}^{m_1, \ldots, m_k} = L^{m_1, \ldots, m_k}
    - \sum_{j=1}^{k-1} L^{m_1,\ldots,m_j} \vtr \tilde{L}^{m_{j+1}, \ldots, m_k}
    \label{tL=L-vtr}
\ee
and thus
\be
    r(\alpha) = \ell(\alpha) - \sum \ell(\alpha_{(1)}) \vtr r(\alpha_{(2)}) \qquad \quad
    \forall \alpha \in \A(P) \; .    \label{r(a)=l(a)-vtr}
\ee
\hfill $\blacksquare$

\begin{lemma}
\be
     r( P_{m_1 \ldots m_k} \circ P_n )
 &=& - L^{m_1, \ldots, m_k}_{<0} \ast L^n
     - \sum_{j=1}^{k-1} L^{m_1, \ldots, m_j}_{<0} \ast L^n \ast \tilde{L}^{m_{j+1}, \ldots, m_k}_{<0}
     \nonumber \\
 & & + L^n \ast \tilde{L}^{m_1, \ldots, m_k}_{<0}
     - \sum_{j=1}^k L^{m_1, \ldots, m_j} \vtr r(P_{m_{j+1} \ldots m_k} \circ P_n) \; .
     \label{r(Pm1...mk-c-Pn}
\ee
\end{lemma}
{\it Proof:} Using (\ref{a-cw-b-c-A}), the commutativity of $\circ$, and
(\ref{A1cA2}), we find
\bez
     P_{m_1 \ldots m_k} \circ P_n
 &=& (P_{m_1} \prec P_{m_2 \ldots m_k}) \circ P_n \\
 &=& (P_{m_1} \circ P_n - P_{m_1} \prec P_n) \prec P_{m_2 \ldots m_k}
     + P_{m_1} \prec ( P_{m_2 \ldots m_k} \circ P_n ) \\
 &=& P_n \succ P_{m_1 \ldots m_k} + P_{m_1} \prec ( P_{m_2 \ldots m_k} \circ P_n )
\eez
so that
\bez
     r(P_{m_1 \ldots m_k} \circ P_n)
 &=& r( P_{m_1} \prec P_{m_2 \ldots m_k} \circ P_n )
     + r( P_n \succ P_{m_1 \ldots m_k} )  \\
 &=& - \ell(P_{m_1}) \ast r(P_{m_2 \ldots m_k} \circ P_n)_{\geq 0}
     + \ell(P_n) \ast r(P_{m_1 \ldots m_k})_{<0} \\
 &=& - \ell(P_{m_1})_{<0} \ast r(P_{m_2 \ldots m_k} \circ P_n)
     - \ell(P_{m_1}) \vtr r(P_{m_2 \ldots m_k} \circ P_n) \\
 & & + \ell(P_n) \ast r(P_{m_1 \ldots m_k})_{<0}
\eez
and
\bez
 & & \ell(P_{m_1})_{<0} \ast r(P_{m_2 \ldots m_k} \circ P_n) \\
 &=& \ell(P_{m_1})_{<0} \ast
     \Big( r( P_{m_2} \prec P_{m_3 \ldots m_k} \circ P_n )
     + r( P_n \succ P_{m_2 \ldots m_k} ) \Big) \\
 &=& \ell(P_{m_1})_{<0} \ast
     \Big( - L^{m_2} \ast r( P_{m_3 \ldots m_k} \circ P_n )_{\geq 0}
     + r( P_n \succ P_{m_2 \ldots m_k} ) \Big) \\
 &=& \ell(P_{m_1 m_2}) \ast r(P_{m_3 \ldots m_k} \circ P_n )_{\geq 0}
     + \ell(P_{m_1})_{<0} \ast r( P_n \succ P_{m_2 \ldots m_k} ) \\
 &=& \ell(P_{m_1 m_2})_{<0} \ast r(P_{m_3 \ldots m_k} \circ P_n )
     + \ell(P_{m_1 m_2}) \vtr r(P_{m_3 \ldots m_k} \circ P_n ) \\
 & & + \ell(P_{m_1})_{<0} \ast L^n \ast r( P_{m_2 \ldots m_k} )_{<0} \; .
\eez
By iteration, we obtain
\bez
 & & r(P_{m_1 \ldots m_k} \circ P_n)
  = \ell(P_n) \ast r(P_{m_1 \ldots m_k})_{<0}
    - \sum_{j=1}^{k-1} \ell(P_{m_1 \ldots m_j}) \vtr r(P_{m_{j+1} \ldots m_k} \circ P_n) \\
 & & \qquad
    - \sum_{j=1}^{k-2} \ell(P_{m_1 \ldots m_j})_{<0} \ast L^n \ast r(P_{m_{j+1} \ldots m_k})_{<0}
    - \ell(P_{m_1 \ldots m_{k-1}})_{<0} \ast r(P_{m_k} \circ P_n)  \; .
\eez
Next we convert the last term:
\bez
 & & \ell(P_{m_1 \ldots m_{k-1}})_{<0} \ast r(P_{m_k} \circ P_n)
  = \ell(P_{m_1 \ldots m_{k-1}})_{<0} \ast r(P_{m_k} \prec P_n + P_n \succ P_{m_k}) \\
 &=& \ell(P_{m_1 \ldots m_{k-1}})_{<0} \ast ( - L^{m_k} \ast L^n{}_{\geq 0}
     + L^n \ast r(P_{m_k})_{<0} )  \\
 &=& \ell(P_{m_1 \ldots m_k}) \ast L^n{}_{\geq 0}
     + \ell(P_{m_1 \ldots m_{k-1}})_{<0} \ast L^n \ast r(P_{m_k})_{<0} \\
 &=& \ell(P_{m_1 \ldots m_k})_{<0} \ast L^n
     + \ell(P_{m_1 \ldots m_k}) \vtr r(P_n)
     + \ell(P_{m_1 \ldots m_{k-1}})_{<0} \ast L^n \ast r(P_{m_k})_{<0} \; .
\eez
Insertion of this result into the previous formula yields (\ref{r(Pm1...mk-c-Pn}).
\hfill $\blacksquare$

\begin{lemma}
\be
      r( P_{m_1 \ldots m_k} \circ (P_n \prec \alpha) )
 &=& - L^n \ast r(P_{m_1 \ldots m_k} \circ \alpha)_{\geq 0}
     - \sum_{j=1}^{k} L^{m_1, \ldots, m_j}_{\geq 0} \ast L^n
         \ast r( P_{m_{j+1} \ldots m_k} \circ \alpha)_{\geq 0} \nonumber \\
 & & - \sum_{j=1}^{k} L^{m_1, \ldots, m_j} \vtr r(P_{m_{j+1} \ldots m_k} \circ (P_n \prec \alpha))
     \; .  \label{r(Pm1...mk-c-(Pn<a)}  \qquad
\ee
\end{lemma}
{\it Proof:} First we note that (\ref{shuffle}) implies the identity
\bez
 & &  P_{m_1 \ldots m_k} \circ (P_n \prec \alpha)
  =  (P_{m_1} \prec P_{m_2 \ldots m_k}) \circ (P_n \prec \alpha)  \\
 &=& P_{m_1} \prec ( P_{m_2\ldots m_k} \circ (P_n \prec \alpha) )
     + P_n \prec ( P_{m_1 \ldots m_k} \circ \alpha )
     + P_{m_1 + n} \prec ( P_{m_2\ldots m_k} \circ \alpha )
\eez
so that
\bez
     r( P_{m_1 \ldots m_k} \circ (P_n \prec \alpha) )
 &=& - L^{m_1} \ast r( P_{m_2\ldots m_k} \circ (P_n \prec \alpha) )_{\geq 0}
     - L^n \ast r( P_{m_1 \ldots m_k} \circ \alpha )_{\geq 0} \\
 & & - L^{m_1 + n} \ast r( P_{m_2\ldots m_k} \circ \alpha )_{\geq 0} \\
 &=& - L^{m_1}_{<0} \ast r( P_{m_2\ldots m_k} \circ (P_n \prec \alpha) )
     - L^{m_1} \vtr r( P_{m_2\ldots m_k} \circ (P_n \prec \alpha) ) \\
 & & - L^n \ast r( P_{m_1 \ldots m_k} \circ \alpha )_{\geq 0}
     - L^{m_1 + n} \ast r( P_{m_2\ldots m_k} \circ \alpha )_{\geq 0}  \; .
\eez
This is a recursion formula, so we can rewrite the first term on the
right hand side as follows,
\bez
 & & L^{m_1}_{<0} \ast r( P_{m_2\ldots m_k} \circ (P_n \prec \alpha) )  \\
 &=& L^{m_1}_{<0} \ast \Big( - L^{m_2} \ast r( P_{m_3\ldots m_k} \circ (P_n \prec \alpha) )_{\geq 0}
     - L^n \ast r( P_{m_2 \ldots m_k} \circ \alpha )_{\geq 0}  \\
 & & - L^{m_2 + n} \ast r( P_{m_3\ldots m_k} \circ \alpha )_{\geq 0} \Big)  \\
 &=& L^{m_1,m_2} \ast r( P_{m_3\ldots m_k} \circ (P_n \prec \alpha) )_{\geq 0}
     - L^{m_1}_{<0} \ast L^n \ast r( P_{m_2 \ldots m_k} \circ \alpha )_{\geq 0} \\
 & & + L^{m_1,m_2} \ast L^n \ast r( P_{m_3\ldots m_k} \circ \alpha )_{\geq 0} \\
 &=& L^{m_1,m_2}_{<0} \ast r( P_{m_3\ldots m_k} \circ (P_n \prec \alpha) )
     + L^{m_1,m_2} \vtr r( P_{m_3\ldots m_k} \circ (P_n \prec \alpha) ) \\
 & & - L^{m_1}_{<0} \ast L^n \ast r( P_{m_2 \ldots m_k} \circ \alpha )_{\geq 0}
     + L^{m_1,m_2} \ast L^n \ast r( P_{m_3\ldots m_k} \circ \alpha )_{\geq 0} \; .
\eez
Hence we obtain
\bez
 & &   r( P_{m_1 \ldots m_k} \circ (P_n \prec \alpha) )  \\
 &=& - L^{m_1,m_2}_{<0} \ast r( P_{m_3\ldots m_k} \circ (P_n \prec \alpha) )
     - L^n \ast r( P_{m_1 \ldots m_k} \circ \alpha )_{\geq 0} \\
 & & - L^{m_1}_{\geq 0} \ast L^n \ast r( P_{m_2 \ldots m_k} \circ \alpha )_{\geq 0}
     - L^{m_1,m_2} \ast L^n \ast r( P_{m_3\ldots m_k} \circ \alpha )_{\geq 0} \\
 & & - L^{m_1} \vtr r( P_{m_2\ldots m_k} \circ (P_n \prec \alpha) )
     - L^{m_1,m_2} \vtr r( P_{m_3\ldots m_k} \circ (P_n \prec \alpha) )  \; .
\eez
In the next step we proceed as follows,
\bez
 & & L^{m_1,m_2}_{<0} \ast r( P_{m_3\ldots m_k} \circ (P_n \prec \alpha) ) \\
 &=& L^{m_1,m_2,m_3}_{<0} \ast r( P_{m_4\ldots m_k} \circ (P_n \prec \alpha) )
     + L^{m_1,m_2,m_3} \vtr r( P_{m_4\ldots m_k} \circ (P_n \prec \alpha) ) \\
 & & - L^{m_1,m_2}_{<0} \ast L^n \ast r( P_{m_3 \ldots m_k} \circ \alpha )_{\geq 0}
     + L^{m_1,m_2,m_3} \ast L^n \ast r( P_{m_4\ldots m_k} \circ \alpha )_{\geq 0}
\eez
and so forth. In the last step, we have to use
\bez
   P_{m_k} \circ (P_n \prec \alpha) = P_{m_k} \prec P_n \prec \alpha
    + P_n \prec (P_{m_k} \circ \alpha) + P_{m_k+n} \prec \alpha
\eez
which follows from (\ref{A-c-a-cw-b}) and (\ref{A1cA2}), so that
\bez
     L^{m_1,\ldots,m_{k-1}}_{<0} \ast r( P_{m_k} \circ (P_n \prec \alpha) )
 &=& L^{m_1,\ldots,m_k}_{\geq 0} \ast L^n \ast r(\alpha)_{\geq 0}
     + L^{m_1,\ldots,m_k} \vtr r( P_n \prec \alpha )  \\
 & & - L^{m_1,\ldots,m_{k-1}} \ast L^n \ast r( P_{m_k} \circ \alpha )_{\geq 0} \; .
\eez
Finally we obtain (\ref{r(Pm1...mk-c-(Pn<a)}).
\hfill $\blacksquare$

\subsection{Properties of the generalized derivations}

\begin{lemma}
\be
     \delta_{m_1 \ldots m_k} L^n
 &=& - [ L^{m_1, \ldots, m_k}_{<0} , L^n ]
     + \sum_{j=1}^{k-1} ( \delta_{m_1 \ldots m_j} L^n ) \ast L^{m_{j+1}, \ldots, m_k}_{<0}
      \label{delta_m...L^n<}  \\
     \delta_{m_1 \ldots m_k} L^n
 &=& [ L^{m_1, \ldots, m_k}_{\geq 0} , L^n]
    - \sum_{j=1}^{k-1} (\delta_{m_1 \ldots m_j} L^n) \ast L^{m_{j+1}, \ldots, m_k}_{\geq 0}
    \; .  \label{delta_m...L^n>}
\ee
\end{lemma}
{\it Proof:} By definition, the first equality holds for $n=1$ and arbitrary $k \in \mathbb{N}$.
Fix $k$ and $n$ and suppose it holds with $k$ replaced by any $j \in \mathbb{N}$ with $j < k$
and $n$ replaced by any $m \in \mathbb{N}$ with $m \leq n$.
Using this and the generalized derivation property, we find
\bez
 & &  \delta_{m_1 \ldots m_k} L^{n+1}
  = (\delta_{m_1 \ldots m_k} L^n) \ast L + L^n \ast \delta_{m_1 \ldots m_k} L
     + \sum_{j=1}^{k-1} (\delta_{m_1 \ldots m_j} L^n) \ast \delta_{m_{j+1} \ldots m_k} L  \\
 &=& - [L^{m_1, \ldots, m_k}_{<0} , L^{n+1}]
     + \sum_{i=1}^{k-1} (\delta_{m_1 \ldots m_i} L^n)
       \ast \sum_{j=i}^{k-1} (\delta_{m_{i+1} \ldots m_j} L)
     \ast L^{m_{j+1}, \ldots, m_k}_{<0}  \\
 & & + \sum_{j=1}^{k-1} L^n \ast (\delta_{m_1 \ldots m_j} L) \ast L^{m_{j+1}, \ldots, m_k}_{<0} \\
 &=& - [L^{m_1, \ldots, m_k}_{<0} , L^{n+1}]
     + \sum_{j=1}^{k-1} \Big( \sum_{i=1}^j (\delta_{m_1 \ldots m_i} L^n)
       \ast \delta_{m_{i+1} \ldots m_j} L
     + L^n \ast \delta_{m_1 \ldots m_j} L \Big) \ast L^{m_{j+1}, \ldots, m_k}_{<0} \\
 &=& - [L^{m_1, \ldots, m_k}_{<0} , L^{n+1}]
     + \sum_{j=1}^{k-1} (\delta_{m_1 \ldots m_j} L^{n+1}) \ast L^{m_{j+1}, \ldots, m_k}_{<0}
\eez
so that (\ref{delta_m...L^n<}) also holds for $n+1$. Our second expression for
$\delta_{m_1 \ldots m_k} L^n$ now follows with the help of
\bez
      [ L^{m_1, \ldots, m_k}_{\geq 0} , L^n]
 &=&   [ L^{m_1, \ldots, m_k} - L^{m_1, \ldots, m_k}_{<0} , L^n]
  = - [ L^{m_1, \ldots, m_{k-1}}_{<0} \ast L^{m_k}, L^n ]
     - [ L^{m_1, \ldots, m_k}_{<0} , L^n] \\
 &=& - [ L^{m_1, \ldots, m_{k-1}}_{<0} , L^n ] \ast L^{m_k}
     - [ L^{m_1, \ldots, m_k}_{<0} , L^n] \\
 &=& \Big( \delta_{m_1 \ldots m_{k-1}} L^n
     - \sum_{j=1}^{k-2} (\delta_{m_1 \ldots m_j} L^n) \ast L^{m_{j+1}, \ldots, m_{k-1}}_{<0} \Big)
     \ast L^{m_k}
     - [ L^{m_1, \ldots, m_k}_{<0} , L^n] \\
 &=& \sum_{j=1}^{k-1} (\delta_{m_1 \ldots m_j} L^n) \ast L^{m_{j+1}, \ldots, m_k}
     - [ L^{m_1, \ldots, m_k}_{<0} , L^n] \; .
\eez
\hfill $\blacksquare$
\vskip.2cm

In particular, we have
\be
    \delta_m(\pa)
  = \delta_m (L_{\geq 0})
  = (\delta_m L)_{\geq 0}
  = ( - [ (L^m)_{<0} , L ] )_{\geq 0}
  = 0 \; .
\ee
By induction, using (\ref{delta_m...L^n<}), it follows that
\be
       \delta_{m_1 \ldots m_k}(\pa) = 0  \qquad \quad   k=1,2, \ldots \; .
       \label{delta(pa)=0}
\ee
Using the fact that $\pa^{-1}$ is the inverse of $\pa$, this in turn
implies
\be
   \delta_{m_1 \ldots m_k}(\pa^{-1}) = 0  \qquad \quad   k=1,2, \ldots \; .
       \label{delta(pa^-1)=0}
\ee
The main result of this subsection is stated next.

\begin{proposition}
\label{proposition:delta-rl}
\be
       \delta_{m_1 \ldots m_k} \ell(\alpha)
 &=& \ell(P_{m_1 \ldots m_k} \circ \alpha)
     - \sum_{j=0}^{k-1} \ell(P_{m_1 \ldots m_j} \circ \alpha) \vtr r(P_{m_{j+1} \ldots m_k})
        \label{delta-ell}  \\
      \delta_{m_1 \ldots m_k} r(\alpha)
 &=& r(P_{m_1 \ldots m_k} \circ \alpha)
     + \sum_{j=1}^k \ell(P_{m_1 \ldots m_j}) \vtr r(P_{m_{j+1} \ldots m_k} \circ \alpha)
      \; .      \label{delta-r}
\ee
\hfill $\blacksquare$
\end{proposition}

The remainder of this subsection is devoted to the proof of this proposition.
Let us define
\be
       \delta'_{m_1 \ldots m_k} \ell(\alpha)
 &:=& \ell(P_{m_1 \ldots m_k} \circ \alpha)
     - \sum_{j=0}^{k-1} \ell(P_{m_1 \ldots m_j} \circ \alpha) \vtr r(P_{m_{j+1} \ldots m_k})
        \label{delta'}  \\
      \delta''_{m_1 \ldots m_k} r(\alpha)
 &:=& r(P_{m_1 \ldots m_k} \circ \alpha)
     + \sum_{j=1}^k \ell(P_{m_1 \ldots m_j}) \vtr r(P_{m_{j+1} \ldots m_k} \circ \alpha)
      \; .      \label{delta''}
\ee
We have to show that $\delta'_{m_1 \ldots m_k}$ and $\delta''_{m_1 \ldots m_k}$ coincide
with $\delta_{m_1 \ldots m_k}$ on $\ell(\A(P))$, respectively $r(\A(P))$.

\begin{lemma}
\label{lemma:eq_of_deltas}
\be
     \delta'_{m_1 \ldots m_k} L^n
   = \delta''_{m_1 \ldots m_k} L^n
   = \delta_{m_1 \ldots m_k} L^n  \; .
\ee
\end{lemma}
{\it Proof:} First we note that
\bez
     \delta''_{m_1 \ldots m_k} L^n
   = \delta''_{m_1 \ldots m_k} r(P_n)
   = r(P_{m_1 \ldots m_k} \circ P_n)
     + \sum_{j=1}^k \ell(P_{m_1 \ldots m_j}) \vtr r(P_{m_{j+1} \ldots m_k} \circ P_n)
\eez
which can be further evaluated with the help of (\ref{r(Pm1...mk-c-Pn}),
\bez
      \delta''_{m_1 \ldots m_k} L^n
  &=& - L^{m_1, \ldots, m_k}_{<0}  \ast L^n
      - \sum_{j=1}^{k-1} L^{m_1, \ldots, m_j}_{<0} \ast L^n
          \ast \tilde{L}^{m_{j+1},\ldots,m_k}_{<0}
      + L^n \ast \tilde{L}^{m_1,\ldots,m_k}_{<0} \; .
\eez
Next we use (\ref{tL=L-vtr}) and $(X \vtr Y)_{<0} = - X_{<0} \ast Y_{<0}$ to obtain
\bez
     \delta''_{m_1 \ldots m_k} L^n + [ L^{m_1, \ldots, m_k}_{<0} , L^n ]
 = - \sum_{j=1}^{k-1} [ L^{m_1, \ldots, m_j}_{<0} , L^n ]
          \ast \tilde{L}^{m_{j+1},\ldots,m_k}_{<0}  \; .
\eez
Using this formula, we will prove by induction that $\delta''_{m_1 \ldots m_k} L^n$
equals the right hand side of (\ref{delta_m...L^n<}).
For $k=2$, the last relation reads
\bez
   \delta''_{m_1 m_2} L^n
 = - [ L^{m_1,m_2}_{<0} , L^n ] - [L^{m_1}_{<0} , L^n ] \ast \tilde{L}^{m_2}_{<0}
 = - [ L^{m_1,m_2}_{<0} , L^n ] + (\delta_{m_1} L^n) \ast L^{m_2}_{<0}
\eez
where we used $\delta_m L^n = - [L^m_{<0} , L^n ]$.
Let us now fix $k$ and assume that
$\delta''_{m_1 \ldots m_k} L^n = \delta_{m_1 \ldots m_k} L^n$ holds for
$m_1,\ldots,m_j$ with $2 \leq j \leq k$. Then it also holds for $k+1$ since
\bez
 & & \delta''_{m_1 \ldots m_{k+1}} L^n + [ L^{m_1,\ldots,m_{k+1}}_{<0} , L^n ]
  = - \sum_{j=1}^k [ L^{m_1, \ldots, m_j}_{<0} , L^n ]
      \ast \tilde{L}^{m_{j+1}, \ldots, m_{k+1}}_{<0}  \\
 &=& \sum_{j=1}^k (\delta_{m_1 \ldots m_j} L^n)
      \ast \tilde{L}^{m_{j+1}, \ldots, m_{k+1}}_{<0}
     - \sum_{j=1}^k \sum_{l=1}^{j-1} (\delta_{m_1 \ldots m_l} L^n)
      \ast L^{m_{l+1}, \ldots, m_j}_{<0} \ast \tilde{L}^{m_{j+1}, \ldots, m_{k+1}}_{<0} \\
 &=& \sum_{j=1}^k (\delta_{m_1 \ldots m_j} L^n)
      \ast \tilde{L}^{m_{j+1}, \ldots, m_{k+1}}_{<0}
     - \sum_{l=1}^{k-1} (\delta_{m_1 \ldots m_l} L^n)
      \ast \sum_{j=l+1}^k L^{m_{l+1}, \ldots, m_j}_{<0}
      \ast \tilde{L}^{m_{j+1}, \ldots, m_{k+1}}_{<0}  \\
 &=& \sum_{j=1}^{k} (\delta_{m_1 \ldots m_j} L^n) \ast L^{m_{j+1}, \ldots, m_{k+1}}_{<0}
\eez
using again the `negative' part of (\ref{tL=L-vtr}) in the form
\bez
    \sum_{j=l+1}^k L^{m_1,\ldots,m_j}_{<0} \ast \tilde{L}^{m_{j+1}, \ldots, m_{k+1}}_{<0}
  = \tilde{L}^{m_{l+1}, \ldots, m_{k+1}}_{<0} - L^{m_{l+1}, \ldots, m_{k+1}}_{<0} \; .
\eez
The equality $\delta'_{m_1\ldots m_k} = \delta_{m_1\ldots m_k}$ is obtained in the same way.
\hfill $\blacksquare$

\begin{lemma}
\label{lemma:delta(L^n-ast-rgeq0)}
The following are identities for all $n \in \mathbb{N}$,
\be
   \delta'_{m_1 \ldots m_k} ( \ell(\alpha)_{<0} \ast L^n )
 &=& \sum_{j=0}^k \Big( \delta'_{m_1 \ldots m_j} \ell(\alpha) \Big)_{<0}
     \ast \delta_{m_{j+1} \ldots m_k} L^n  \label{delta'(ell<0-ast-L^n)}  \\
   \delta''_{m_1 \ldots m_k} ( L^n \ast r(\alpha)_{\geq 0} )
 &=& \sum_{j=0}^k (\delta_{m_1 \ldots m_j} L^n)
     \ast \Big( \delta''_{m_{j+1} \ldots m_k} r(\alpha) \Big)_{\geq 0}  \; .
        \label{delta''(L^n-ast-rgeq0)}
\ee
\end{lemma}
{\it Proof:} Using (\ref{r(Pm1...mk-c-(Pn<a)}), we obtain
\bez
      \delta''_{m_1 \ldots m_k} r(P_n \prec \alpha)
   = - L^n \ast r(P_{m_1 \ldots m_k} \circ \alpha)_{\geq 0}
     - \sum_{j=1}^{k} L^{m_1, \ldots, m_j}_{\geq 0} \ast L^n
         \ast r( P_{m_{j+1} \ldots m_k} \circ \alpha)_{\geq 0}
\eez
and thus
\bez
     \delta''_{m_1 \ldots m_k} r(P_n \prec \alpha)
   = - L^n \ast \delta''_{m_1 \ldots m_k} r(\alpha)_{\geq 0}
     - \sum_{j=1}^{k} [ L^{m_1, \ldots, m_j}_{\geq 0} , L^n ]
         \ast r( P_{m_{j+1} \ldots m_k} \circ \alpha)_{\geq 0} \; .
\eez
Now we eliminate the commutator via (\ref{delta_m...L^n>}) to get
\bez
 & &   \delta''_{m_1 \ldots m_k} r(P_n \prec \alpha)  \\
 &=& - L^n \ast \delta''_{m_1 \ldots m_k} r(\alpha)_{\geq 0}
     - \sum_{j=1}^{k} (\delta_{m_1 \ldots m_j} L^n)
           \ast r( P_{m_{j+1} \ldots m_k} \circ \alpha)_{\geq 0} \\
 & & + \sum_{j=2}^{k} \sum_{l=1}^{j-1} (\delta_{m_1 \ldots m_l} L^n)
      \ast L^{m_{l+1}, \ldots, m_j}_{\geq 0} \ast r( P_{m_{j+1} \ldots m_k}
      \circ \alpha)_{\geq 0} \\
 &=& - L^n \ast \delta''_{m_1 \ldots m_k} r(\alpha)_{\geq 0}
     - \sum_{j=1}^{k} (\delta_{m_1 \ldots m_j} L^n)
           \ast r( P_{m_{j+1} \ldots m_k} \circ \alpha)_{\geq 0} \\
 & & + \sum_{l=1}^{k-1} (\delta_{m_1 \ldots m_l} L^n)
         \ast \sum_{j=l+1}^k \ell(P_{m_{l+1} \ldots m_j})_{\geq 0}
           \ast r( P_{m_{j+1} \ldots m_k} \circ \alpha)_{\geq 0} \\
 &=& - L^n \ast \delta''_{m_1 \ldots m_k} r(\alpha)_{\geq 0}
     - (\delta_{m_1 \ldots m_k} L^n) \ast r(\alpha)_{\geq 0} \\
 & & - \sum_{l=1}^{k-1} (\delta_{m_1 \ldots m_l} L^n) \ast \Big(
       r( P_{m_{l+1} \ldots m_k} \circ \alpha)
        + \sum_{j=l+1}^k \ell(P_{m_{l+1} \ldots m_j})
           \vtr r( P_{m_{j+1} \ldots m_k} \circ \alpha) \Big)_{\geq 0} \\
 &=& - L^n \ast \delta''_{m_1 \ldots m_k} r(\alpha)_{\geq 0}
     - (\delta_{m_1 \ldots m_k} L^n) \ast r(\alpha)_{\geq 0}
     - \sum_{l=1}^{k-1} (\delta_{m_1 \ldots m_l} L^n)
        \ast ( \delta''_{m_{l+1} \ldots m_k} r(\alpha) )_{\geq 0} \\
 &=& - \sum_{l=0}^k (\delta_{m_1 \ldots m_l} L^n)
        \ast ( \delta''_{m_{l+1} \ldots m_k} r(\alpha) )_{\geq 0} \; .
\eez
The proof of (\ref{delta''(L^n-ast-rgeq0)}) is completed by inserting
$r(P_n \prec \alpha) = - L^n \ast r(\alpha)_{\geq 0}$. The other identity
can be proved in a similar way.
\hfill $\blacksquare$
\vskip.2cm

For the moment, let us simply write $\delta'$ instead of $\delta'_{m_1 \ldots m_k}$.
By iterative use of (\ref{delta'(ell<0-ast-L^n)}), respectively (\ref{delta''(L^n-ast-rgeq0)}),
we find
\bez
     \delta' \ell(P_{n_1 \ldots n_j})
 &=& (-1)^{j-1} \sum (\delta'_{(1)} L^{n_1}) \overleftarrow{R} \ast
     \ldots \ast (\delta'_{(j-1)} L^{n_{j-1}})\overleftarrow{R} \ast (\delta'_{(j)} L^{n_j}) \\
     \delta'' r(P_{n_1 \ldots n_j})
 &=& (-1)^{j-1} \sum (\delta_{(1)} L^{n_1}) \ast \overrightarrow{R} (\delta''_{(2)} L^{n_2})
     \ast \ldots \ast \overrightarrow{R} (\delta''_{(j)} L^{n_j})
\eez
using the projection operators defined in proposition~\ref{proposition:rightR-to-leftR}
and a Sweedler notation.
According to lemma~\ref{lemma:eq_of_deltas} we may drop the primes on the right hand
sides of these equations. Using the generalized derivation property of
$\delta_{m_1 \ldots m_k}$, we obtain
\bez
    \delta' \ell(P_{n_1 \ldots n_j}) = \delta \ell(P_{n_1 \ldots n_j})
    \qquad\quad
    \delta'' r(P_{n_1 \ldots n_j}) = \delta r(P_{n_1 \ldots n_j}) \; .
\eez
Since the elements $P_{n_1 \ldots n_j}$ span $\A(P)$, this proves
proposition~\ref{proposition:delta-rl}.

\subsection{Commutativity of the generalized derivations}

\begin{lemma}
\be
   \ell(P_{m_1 \ldots m_k} \circ \alpha)
 = \delta_{m_1 \ldots m_k} \ell(\alpha)
   + \sum_{j=0}^{k-1} \delta_{m_1 \ldots m_j} \ell(\alpha) \vtr \ell(P_{m_{j+1} \ldots m_k}) \; .
   \label{ell(Pm1...mk-c-a)}
\ee
\end{lemma}
{\it Proof:} by induction. For $k=1$ this follows directly from (\ref{delta-ell})
using $r(P_m) = \ell(P_m)$.
Let us fix $k$ and assume that (\ref{ell(Pm1...mk-c-a)}) holds for $1 \leq k' \leq k$.
Starting with (\ref{delta-ell}), we obtain
\bez
 & & \ell(P_{m_1\ldots m_{k+1}} \circ \alpha) - \delta_{m_1\ldots m_{k+1}} \ell(\alpha)
  = \sum_{j=0}^k  \ell(P_{m_1\ldots m_j} \circ \alpha) \vtr r(P_{m_{j+1} \ldots m_{k+1}}) \\
 &=& \sum_{j=0}^k \delta_{m_1\ldots m_j} \ell(\alpha) \vtr r(P_{m_{j+1} \ldots m_{k+1}})
    + \sum_{j=0}^k \sum_{l=0}^{j-1} \delta_{m_1\ldots m_l} \ell(\alpha)
        \vtr \ell(P_{m_{l+1} \ldots m_j}) \vtr r(P_{m_{j+1} \ldots m_{k+1}})  \\
 &=& \sum_{j=0}^k \delta_{m_1\ldots m_j} \ell(\alpha) \vtr r(P_{m_{j+1} \ldots m_{k+1}})
    + \sum_{l=0}^{k-1} \delta_{m_1\ldots m_l} \ell(\alpha)
      \vtr \sum_{j=l+1}^{k} \ell(P_{m_{l+1} \ldots m_j}) \vtr r(P_{m_{j+1} \ldots m_{k+1}})  \\
 &=& \sum_{j=0}^k \delta_{m_1\ldots m_j} \ell(\alpha) \vtr r(P_{m_{j+1} \ldots m_{k+1}})
    + \sum_{l=0}^{k-1} \delta_{m_1\ldots m_l} \ell(\alpha)
      \vtr ( \ell(P_{m_{l+1} \ldots m_{k+1}}) - r(P_{m_{l+1} \ldots m_{k+1}}) )  \\
 &=& \delta_{m_1\ldots m_k} \ell(\alpha) \vtr r(P_{m_{k+1}})
     + \sum_{j=0}^{k-1} \delta_{m_1\ldots m_j} \ell(\alpha)
      \vtr \ell(P_{m_{j+1} \ldots m_{k+1}}) \\
 &=& \sum_{j=0}^k \delta_{m_1\ldots m_j} \ell(\alpha)
      \vtr \ell(P_{m_{j+1} \ldots m_{k+1}})  \; .
\eez
Hence (\ref{ell(Pm1...mk-c-a)}) also holds for $k+1$.
\hfill $\blacksquare$
\vskip.2cm

\begin{proposition}
\be
   \delta_{m_1 \ldots m_k} \big( \ell(\alpha) \vtr r(\beta) \big)
 = \sum_{j=0}^k \ell(P_{m_1 \ldots m_j} \circ \alpha) \vtr r(P_{m_{j+1} \ldots m_k} \circ \beta)
   \; .     \label{delta(vtr)}
\ee
\end{proposition}
{\it Proof:} With the help of (\ref{ell(Pm1...mk-c-a)}), we find
\bez
 & & \sum_{j=0}^k \ell(P_{m_1 \ldots m_j} \circ \alpha) \vtr r(P_{m_{j+1} \ldots m_k} \circ \beta) \\
 &=& \sum_{j=0}^k \delta_{m_1\ldots m_j} \ell(\alpha) \vtr r(P_{m_{j+1} \ldots m_k} \circ \beta)
     + \sum_{j=1}^k \sum_{l=0}^{j-1} \delta_{m_1\ldots m_l} \ell(\alpha)
        \vtr \ell(P_{m_{l+1}\ldots m_j}) \vtr r(P_{m_{j+1} \ldots m_k} \circ \beta) \\
 &=& \sum_{j=0}^k \delta_{m_1\ldots m_j} \ell(\alpha) \vtr r(P_{m_{j+1} \ldots m_k} \circ \beta)
     + \sum_{l=0}^{k-1} \delta_{m_1\ldots m_l} \ell(\alpha)
    \vtr \sum_{j=l+1}^k \ell(P_{m_{l+1}\ldots m_j}) \vtr r(P_{m_{j+1} \ldots m_k} \circ \beta) \\
 &=& \sum_{l=0}^{k-1} \delta_{m_1\ldots m_l} \ell(\alpha)
    \vtr \Big( r(P_{m_{j+1} \ldots m_k} \circ \beta)
     + \sum_{j=l+1}^k \ell(P_{m_{l+1}\ldots m_j}) \vtr r(P_{m_{j+1} \ldots m_k} \circ \beta) \Big) \\
 & & + \delta_{m_1\ldots m_k} \ell(\alpha) \vtr r(\beta) \\
 &=& \sum_{l=0}^{k-1} \delta_{m_1\ldots m_l} \ell(\alpha)
        \vtr \delta_{m_{j+1} \ldots m_k} r(\beta)
     + \delta_{m_1\ldots m_k} \ell(\alpha) \vtr r(\beta)  \\
 &=& \sum_{l=0}^k \delta_{m_1\ldots m_l} \ell(\alpha)
        \vtr \delta_{m_{j+1} \ldots m_k} r(\beta) \; .
\eez
This equals $\delta_{m_1 \ldots m_k} \big( \ell(\alpha) \vtr r(\beta) \big)$ by
use of the generalized derivation rule (\ref{delta-genderiv}).
\hfill $\blacksquare$

\begin{theorem}
\label{theorem:deltas-commute}
\be
    [ \delta_{m_1 \ldots m_k} , \delta_{n_1 \ldots n_l} ] = 0  \; .
\ee
\end{theorem}
{\it Proof:} Using (\ref{delta-r}), we obtain
\bez
     \delta_{m_1\ldots m_k} \delta_{n_1\ldots n_l} r(\alpha)
 &=& r(P_{m_1\ldots m_k} \circ P_{n_1\ldots n_l} \circ \alpha)
     + \sum_{j=1}^k \ell(P_{m_1\ldots m_j})
        \vtr r(P_{m_{j+1}\ldots m_k} \circ P_{n_1\ldots n_l} \circ \alpha) \\
 & & + \sum_{j=1}^l  \delta_{m_1\ldots m_k} \Big( \ell(P_{n_1\ldots n_j})
        \vtr r(P_{n_{j+1}\ldots n_l} \circ \alpha) \Big)
\eez
where, according to (\ref{delta(vtr)}),
\bez
 \lefteqn{ \delta_{m_1 \ldots m_k} \Big( \ell(P_{n_1 \ldots n_j})
        \vtr r(P_{n_{j+1} \ldots n_l} \circ \alpha) \Big) } \qquad & &  \nonumber \\
  &=& \sum_{p=0}^k \ell(P_{m_1 \ldots m_p} \circ P_{n_1 \ldots n_j})
      \vtr r(P_{m_{p+1} \ldots m_k} \circ P_{n_{j+1} \ldots n_l} \circ \alpha)  \\
  &=& \sum_{p=1}^k \ell(P_{m_1 \ldots m_p} \circ P_{n_1 \ldots n_j})
      \vtr r(P_{m_{p+1} \ldots m_k} \circ P_{n_{j+1} \ldots n_l} \circ \alpha) \\
  & & + \ell(P_{n_1 \ldots n_j}) \vtr r(P_{m_1 \ldots m_k} \circ P_{n_{j+1} \ldots n_l}
          \circ \alpha) \; .
\eez
The commutativity of the $\circ$ product now implies that
$[\delta_{m_1 \ldots m_k} , \delta_{n_1 \ldots n_l}] = 0$ on $r(\A(P))$. A similar
argument shows that this also holds on $\ell(\A(P))$. The generalized derivation
property (\ref{delta-genderiv}) extends this commutation relation to the algebra
generated by $\ell(\A(P)) \cup r(\A(P))$ and $\pa^{-1}$, taking (\ref{delta(pa^-1)=0})
into account (and using the product $\ast$ and the projection
$(\;)_{\geq 0}$). But this reaches the whole of $\mathcal{R}$.
\hfill $\blacksquare$

\subsection{Taking the residue}
In this subsection we explore the properties of the map $\Phi$ defined in
(\ref{Phi}). According to (\ref{tL=L-vtr}) and (\ref{res-vtr}) we also have
$\Phi(\alpha) = \res(r(\alpha))$. An immediate consequence of (\ref{ell(P_n)}) is
\be
    \Phi(P_n) = \res(L^n)    \label{Phi(Pn)}
\ee
and from the definition (\ref{L^m...m-def}) we get
\be
    \Phi(P_{m_1 \ldots m_k}) = \res(L^{m_1, \ldots, m_k})  \; .
\ee

\begin{proposition}
\be
    \Phi(\alpha \prec \beta) &=& -\res(\ell(\alpha) \ast r(\beta)_{\geq 0}) \label{Phi(a<b)} \\
    \Phi(\alpha \succ \beta) &=& \res(\ell(\alpha) \ast r(\beta)_{<0}) \; . \label{Phi(a>b)}
\ee
\end{proposition}
{\it Proof:} For $\beta \in \A^1(P)$, it is sufficient to consider
\bez
   \res( \ell(\alpha \prec P_n) )
 = - \res( \ell(\alpha)_{<0} \ast r(P_n) )
 = - \res( \ell(\alpha) \ast r(P_n)_{\geq 0} )
\eez
by use of (\ref{ell(P_n)}) and (\ref{ell(a<P_n)}).
Let us assume that (\ref{Phi(a<b)}) holds for $\beta \in \A(P)$ of degree $\leq n$,
and for all $\alpha \in \A(P)$. Then (\ref{Phi(a<b)}) also holds for $\beta \in \A(P)$
of degree $n+1$ since
\bez
     \res( \ell(\alpha \prec (P_m \prec \beta) ) )
 &=& \res( \ell( (\alpha \prec P_m) \prec \beta) )
  = - \res( \ell(\alpha \prec P_m) \ast r(\beta)_{\geq 0} ) \\
 &=& \res( \ell(\alpha)_{<0} \ast L^m \ast r(\beta)_{\geq 0} )
  = - \res( \ell(\alpha)_{<0} \ast r(P_m \prec \beta) ) \\
 &=& - \res( \ell(\alpha) \ast r(P_m \prec \beta)_{\geq 0} )  \; .
\eez
The proof of the second relation proceeds in the same way.
\hfill $\blacksquare$

\begin{theorem}
\label{theorem:Phi-htimes-ast-homom}
$\Phi$ has the following homomorphism property:
\be
   \Phi(\alpha \htimes \beta) = \Phi(\alpha) \ast \Phi(\beta) \; .
\ee
\end{theorem}
{\it Proof:}
\bez
    \Phi(\alpha \htimes \beta) &=& -\Phi(\alpha \prec P \succ \beta)
  = \res \big( \ell(\alpha)_{<0} \ast L \ast r(\beta)_{<0} \big) \\
 &=& \res \big( \ell(\alpha)_{<0} \ast \pa \, r(\beta)_{<0} \big)
  = \res( \ell(\alpha) ) \ast \res( r(\beta) )  \; .
\eez
\hfill $\blacksquare$

\begin{lemma}
\be
    \delta_{m_1 \ldots m_k} \, \res X
  = \res \, \delta_{m_1 \ldots m_k} X  \qquad \quad
    \forall \, X \in \mathcal{R}   \; .
\ee
\end{lemma}
{\it Proof:} Using the identity $\res X =  ( X_{<0} \, \pa )_{\geq 0}$
and writing simply $\delta$ for $\delta_{m_1 \ldots m_k}$, we have
\bez
     \delta \, \res X
   = \delta ( X_{<0} \, \pa )_{\geq 0}
   = ( \delta ( X_{<0} \, \pa) )_{\geq 0}
   = ( (\delta X_{<0}) \, \pa )_{\geq 0}
   = ( (\delta X)_{<0} \, \pa )_{\geq 0}
   = \res \, \delta X
\eez
where we used (\ref{delta-geqX}), (\ref{delta-genderiv}), and
(\ref{delta(pa)=0}).
\hfill $\blacksquare$

\begin{proposition}
\label{proposition:deltaPhi=Phi(Pcirc)}
\be
    \delta_{m_1 \ldots m_k} \Phi(\alpha) = \Phi(P_{m_1 \ldots m_k} \circ \alpha) \; .
    \label{deltaPhi=PhiP}
\ee
\end{proposition}
{\it Proof:} Taking the residue of (\ref{delta-r}) and using (\ref{res-vtr}), leads to
\bez
    \delta_{m_1 \ldots m_k} \Phi(\alpha)
 = \res ( \delta_{m_1 \ldots m_k} r(\alpha) )
 = \res \, r( P_{m_1 \ldots m_k} \circ \alpha ) = \Phi( P_{m_1 \ldots m_k} \circ \alpha )
\eez
\hfill $\blacksquare$

\section{Back to the (x)ncKP hierarchy}
\label{section:xncKP}
\setcounter{equation}{0}
The formalism developed in the preceding section will now be applied to recover
properties of the ncKP and xncKP hierarchies from certain sets of algebraic identities
in $\A(P)$.

\subsection{The ncKP hierarchy}
\label{subsection:ncKP}
Since according to theorem~\ref{theorem:deltas-commute} the $\delta_n$, $n \in \mathbb{N}$,
are commuting derivations, we may set $\delta_n = \pa_{t_n}$ on $\mathcal{R}$.
The equations
\be
    L_{t_n} = \delta_n L \qquad \quad n=1,2, \ldots
\ee
are then compatible. These are the defining relations (\ref{KP-Lax}) of the ncKP hierarchy.
An immediate consequence is
\be
    \Phi(P_n) = \pa_{t_n} \phi   \label{ncKP-Phi(Pn)}
\ee
which is (\ref{phi_tn-res}). Furthermore, proposition~\ref{proposition:deltaPhi=Phi(Pcirc)}
leads to $\Phi(P_n \circ \alpha) = \pa_{t_n} \Phi(\alpha)$. In the following, the fundamental
homomorphism property $\Phi(\alpha \htimes \beta) = \Phi(\alpha) \ast \Phi(\beta)$
(theorem~\ref{theorem:Phi-htimes-ast-homom}) will also play an important role.
Applying $\Phi$, for example, to the identity (\ref{KP-Pid}), results in the ncKP
equation (\ref{ncKPpot}).
\vskip.1cm

Let us recall the definitions
\be
     (L^n)_{<0}
   = - \sum_{m=1}^\infty \sigma^{(n)}_m \ast L^{-m}
   = \sum_{m=1}^\infty L^{-m} \ast \eta^{(n)}_m \; .
\ee
of coefficients $\sigma^{(n)}_m$ and $\eta^{(n)}_m$ from \cite{DMH04ncKP}, where
iteration formulae for the (x)ncKP hierarchy equations were derived in terms of them.
The $\sigma$-coefficients frequently appeared in treatments of the `commutative'
KP hierarchy (see \cite{MSS90}, for example).

\begin{theorem}
\be
    \Phi(U_n) = u_n \qquad \quad
    \Phi(C^{(m)}_n) = \sigma^{(m)}_n \qquad \quad
    \Phi(H^{(m)}_n) = \eta^{(m)}_n
\ee
with $U_n, \, C^{(m)}_n, \, H^{(m)}_n$ defined in section~\ref{section:special-rels}.
\end{theorem}
{\it Proof:} Using (\ref{delta-r}), (\ref{vtr}), and $\delta_1 = [\pa , \cdot ]$,
we obtain
\bez
    r(P \circ \alpha)_{\geq 0} = (\delta_1 r(\alpha) - L \vtr r(\alpha))_{\geq 0}
  = ( \pa \, r(\alpha) - r(\alpha) \, \pa
    - \pa \, r(\alpha)_{\geq 0} )_{\geq 0} \; .
\eez
Since $[ \pa , X_{<0} ]_{\geq 0} = 0$ for all $X \in \mathcal{R}$, this implies
$r(P \circ \alpha)_{\geq 0} = - r(\alpha)_{\geq 0} \, \pa$
which can be applied iteratively to the expression
\bez
   r(U_n) = (-1)^n r(P \prec P^{\circ (n-2)})
          = - (-1)^n L \ast r(P \circ P^{\circ(n-3)})_{\geq 0}
\eez
to yield $\Phi(U_n) = \res( r(U_n) ) = \res( L \, \pa^{n-2} ) = u_n$.

With the help of (\ref{Phi(a<b)}) and (\ref{r(P<>a)}), the second relation of the
theorem is obtained as follows,
\bez
     \Phi(C^{(m)}_n)
 &=& (-1)^n \, \res( \ell(P_m \prec P^{\prec n-1} )
  = (-1)^{n+1} \, \res( \ell(P_m)_{<0} \ast r(P^{\prec n-1}) ) \\
 &=& (-1)^n \, \sum_{k=1}^\infty  \sigma^{(m)}_k \ast \res( L^{-k} \ast r(P^{\prec n-1}) ) \\
 &=& (-1)^{n+1} \, \sum_{k=1}^\infty \sigma^{(m)}_k \ast \res( L^{-k+1}
       \ast r(P^{\prec n-2})_{\geq 0} ) \\
 &=& (-1)^{n+1} \, \sum_{k=1}^\infty \sigma^{(m)}_k \ast \res( (L^{-k+1})_{<0}
       \ast r(P^{\prec n-2}) )  \\
 &=& (-1)^{n+1} \, \sum_{k=2}^\infty \sigma^{(m)}_k \ast \res( L^{-k+1} \ast r(P^{\prec n-2}) )
  = \ldots \\
 &=& \sum_{k=n-1}^\infty \sigma^{(m)}_k \ast \res( L^{n-2-k} \ast r(P) )
  = \sum_{k=n}^\infty \sigma^{(m)}_k \ast \res( L^{n-1-k} )
  = \sigma^{(m)}_n
\eez
since $\res(L^{-l}) = 1$ if $l=1$ and $\res(L^{-l}) = 0$ if $l>1$.
The last relation of the theorem is verified in a similar way
(see also the proof of theorem~\ref{theorem:Phi(C,H)}).
\hfill $\blacksquare$
\vskip.2cm

By application of the above results, making use of theorem~\ref{theorem:Phi-htimes-ast-homom}
and proposition~\ref{proposition:deltaPhi=Phi(Pcirc)}, $\Phi$ maps the identity (\ref{PcircP-id})
to
\be
    \pa_{t_m} \pa_{t_n} \phi = \sigma^{(n)}_{m+1} + \eta^{(n)}_{m+1}
     + \sum_{r=1}^{m-1} (\sigma^{(n)}_{m-r} \ast \pa_{t_r}\phi
     + \pa_{t_r}\phi \ast \eta^{(n)}_{m-r})
\ee
which is (5.31) in \cite{DMH04ncKP}. Via (the image under $\Phi$ of) algebraic
relations obtained in section~\ref{section:special-rels} this equation determines
iteratively a `complete' set of ncKP hierarchy equations in the sense that any equation
for $\phi$ which arises from the hierarchy can be expressed as a combination of such
equations.\footnote{With the choice $m=1$, after an $x$-integration the last equation
can be solved for $\phi_{t_n}$ if $n>2$ \cite{DMH04ncKP}.}
Hence, the ncKP hierarchy lies in the image of a set of identities in $\A(P)$ under
the map $\Phi$.
According to results in section~\ref{section:special-rels}, we know that the
respective set of identities in $\A(P)$ can be built from $P_m$, $m \in \mathbb{N}$,
solely by use of the products $\circ$ and $\htimes$. We expect that also
the following statement holds.
\vskip.1cm

\noindent
{\bf Conjecture.} {\em All} identities in $\A(P)$, which are built from $P_m$,
$m \in \mathbb{N}$, only with the help of the products $\circ$ and $\htimes$,
are mapped by $\Phi$ to ncKP equations (expressed in terms of the potential $\phi$).
\hfill $\blacksquare$
\vskip.2cm

If there were such an identity which is \emph{not} mapped by $\Phi$ to an ncKP equation,
we know that it would be mapped to an interesting equation since the latter would
be solvable via the ansatz described in the introduction and thus admit multiple
`soliton' solutions. We believe, however, that the ncKP hierarchy exhausts the
corresponding possibilities (under the restrictions stated in the conjecture).

\subsection{Extension of the Moyal-deformed ncKP hierarchy}
\label{subsection:xncKP}
According to (\ref{deltaPhi=PhiP}),
\be
     \vartheta_{mn} := \frac{1}{2} ( \delta_{mn} - \delta_{nm} )
\ee
satisfies
\be
   \Phi(A_{mn} \circ \alpha) = \vartheta_{mn} \Phi(\alpha)
\ee
with $A_{mn}$ defined in (\ref{A_mn}), and (\ref{delta-genderiv}) leads to
\be
   \vartheta_{mn}(X \ast Y)
 = (\vartheta_{mn} X) \ast Y + X \ast \vartheta_{mn} Y
   + \frac{1}{2} ( \delta_m X \ast \delta_n Y - \delta_n X \ast \delta_m Y)
\ee
which allows us to set $\vartheta_{mn} = \pa_{\theta_{mn}}$
on $\mathcal{R}$ (where $\pa_{\theta_{mn}}$ is the partial derivative with respect
to the deformation parameter $\theta_{mn}$ entering the Moyal $\ast$-product (\ref{Moyal})),
provided that also $\delta_n$ is set equal to $\pa_{t_n}$ (which yields the
ncKP hierarchy).
Since, according to theorem~\ref{theorem:deltas-commute}, the $\vartheta_{mn}$,
$m,n \in \mathbb{N}$, commute with each other and also with the $\delta_n$,
$n \in \mathbb{N}$, the equations
\be
    L_{\theta_{mn}} = \vartheta_{mn} L     \label{ext-ncKP}
\ee
are compatible and extend the Moyal-deformed ncKP hierarchy.\footnote{The xncKP flow
given by (\ref{ext-ncKP}) for fixed $m,n$ only commutes with the corresponding flow
of the same equation with $m,n$ replaced by another pair $r,s$ of natural numbers,
if the ncKP equations associated with the evolution parameters $t_m,t_n,t_r,t_s$
are satisfied (see also \cite{DMH04ncKP}). The proof of theorem~\ref{theorem:deltas-commute}
clearly manifests this dependence of `second order' flows on those of
`first order'. }
In this way, one recovers the extension of the Moyal-deformed ncKP hierarchy obtained
in \cite{DMH04hier} and further explored in \cite{DMH04ncKP}.
\vskip.1cm

 From (\ref{delta_m...L^n>}) we obtain
\be
    \delta_{mn} L^r = [(L^{m,n})_{\geq 0} , L^r] - (\delta_m L^r) \ast (L^n)_{\geq 0}
\ee
and therefore
\be
    \vartheta_{mn} L^r
 = [W^{(m,n)},L^r]_\ast + {1 \over 2} ( \delta_n L^r \ast (L^m)_{\geq 0}
     - \delta_m L^r \ast(L^n)_{\geq 0})   \label{varth-L^r}
\ee
with
\be
    W^{(m,n)}
  := {1 \over 2} (L^{m,n} - L^{n,m})_{\geq 0}
  = {1 \over 2} \Big( (L^n)_{<0} \ast L^m - (L^m)_{<0} \ast L^n \Big)_{\geq 0}
\ee
using $L^{m,n} = -(L^m)_{< 0} \ast L^n$ in the last step.
\vskip.1cm

Replacing $\vartheta_{mn}$ by $\pa_{\theta_{mn}}$ in (\ref{varth-L^r}) for $r=1$, taking the
residue and performing an $x$-integration, leads to
\be
    \pa_{\theta_{mn}} \phi
  = \frac{1}{2} \res( L^{m,n} - L^{n,m} )
  = \Phi( A_{mn} ) \; .
\ee
\vskip.1cm

By application of $\Phi$ to identities in $\A(P)$ involving besides $P_m$ also $A_{mn}$,
and otherwise built with the products $\circ$ and $\htimes$ only,
we obtain explicit xncKP equations beyond those of the ncKP hierarchy. In fact, applying
$\Phi$ to (\ref{Amn-id}), we reach all those equations, since we recover (5.30) in
\cite{DMH04ncKP}.
Recalling results of section~\ref{section:special-rels}, this proves that there
is a set of identities in $\A(P)$, which can be expressed solely in terms of $P_m$,
$A_{mn}$, $m,n \in \mathbb{N}$, and the products $\circ$ and $\htimes$, such
that $\Phi$ maps it to a complete set of xncKP equations for the potential $\phi$.
Probably \emph{all} identities built in this way are mapped by $\Phi$ to xncKP equations.
\vskip.1cm

\noindent
{\it Remark.} It is well-known (see \cite{Gutt+Rawn99}, for example) that by means of
an equivalence transformation
\be
        f \ast' g = D^{-1} \Big( (D f) \ast (D g) \Big)
\ee
with an invertible operator $D$
one can eliminate a possible \emph{symmetric} part of the deformation parameters
$\theta_{mn}$ from the $\ast$-product. Let us see how the algebra $\A(P)$ reflects
this fact. For the moment, let us generalize $\theta_{mn}$
to $t_{mn}$ by adding a symmetric part. From the definition of the main product $\circ$,
we have the identity
\be
    P_{mn} + P_{nm} = P_m \circ P_n - P_{m+n}   \label{Pmn-symm-red-id}
\ee
in $\A(P)$. This is mapped by $\Phi$ to the \emph{linear} equation
\be
    \phi_{t_{mn}} + \phi_{t_{nm}} = \phi_{t_m t_n} - \phi_{t_{m+n}}
\ee
which is equivalent to
\be
  \phi_{t_{mn}} = \phi_{\theta_{mn}} + \frac{1}{2} ( \phi_{t_m t_n} - \phi_{t_{m+n}} )
  \label{phi_tmn_thetamn}
\ee
and allows to express the partial derivative with respect to the
symmetric part of $t_{mn}$ in terms of partial derivatives with
respect to the variables $t_n$. We may therefore restrict our considerations
to the antisymmetric combination $A_{mn} = (P_{mn}-P_{nm})/2$ and thus the
antisymmetric part $\theta_{mn}$ of $t_{mn}$.
\hfill $\blacksquare$

\section{Beyond Moyal deformation: XncKP hierarchy}
\label{section:XncKP}
\setcounter{equation}{0}
In this section we replace the Moyal product by an associative $\ast$-product
which may be regarded as including all (at least in the present framework)
possible deformations. This leads us to an extension of the ncKP hierarchy which is
even bigger than the xncKP hierarchy.

\subsection{Maximal deformation $\ast$-product}
Now we allow the coefficients of $L$ to depend on variables
$t_{(r)} = \{ t_{m_1 \ldots m_r} | \, m_1, \ldots, m_r =1,2,\ldots \}$,
$r=1,2,\ldots$.
In the following we write $\ast$ for the $n \to \infty$ limit of the associative product
$\ast_n$ defined in appendix~C (where $x^{\mu_1 \ldots \mu_r}$ has to be replaced by
$t_{m_1 \ldots m_r}$). Then (\ref{ast_n-diff}) reads
\be
    (f \ast g)_{t_{m_1 \ldots m_r}}
  = f_{t_{m_1 \ldots m_r}} \ast g + f \ast g_{t_{m_1 \ldots m_r}}
    + \sum_{k=1}^{r-1} f_{t_{m_1 \ldots m_k}} \ast g_{t_{m_{k+1} \ldots m_r}}
    \label{ext-ast-deriv-rule}
\ee
and the first of these differentiation rules are
\bez
    (f \ast g)_{t_m} &=& f_{t_m} \ast g + f \ast g_{t_m} \\
    (f \ast g)_{t_{mn}} &=& f_{t_{mn}} \ast g + f \ast g_{t_{mn}} + f_{t_m} \ast g_{t_n} \\
    (f \ast g)_{t_{mnr}} &=& f_{t_{mnr}} \ast g + f \ast g_{t_{mnr}} + f_{t_m} \ast g_{t_{nr}}
      + f_{t_{mn}} \ast g_{t_r} \; .
\eez
Applying them repeatedly, we find, e.g.,
\be
    (f \ast g \ast h)_{t_{mnr}}
 &=& f_{t_{mnr}} \ast g \ast h + f \ast g_{t_{mnr}} \ast h
     + f \ast g \ast h_{t_{mnr}} \nonumber \\
 & & + f_{t_{mn}} \ast g_{t_r} \ast h + f_{t_{mn}} \ast g \ast h_{t_r}
     + f_{t_m} \ast g_{t_{nr}} \ast h + f_{t_m} \ast g \ast h_{t_{nr}} \nonumber \\
 & & + f \ast g_{t_{mn}} \ast h_{t_r} + f \ast g_{t_m} \ast h_{t_{nr}}
     + f_{t_m} \ast g_{t_n} \ast h_{t_r} \; .
\ee
 For $\xi_i = \sum_{m=1}^\infty t_m \, p_i^m$ with parameters $p_i$ we obtain,
for example,
\be
    e^{\xi_1} \ast e^{\xi_2} \ast e^{\xi_3}
 = e^{\xi_1 + \xi_2 + \xi_3 + \xi_{12} + \xi_{13} + \xi_{23} + \xi_{123}}
\ee
where
\be
    \xi_{i_1 \ldots i_r} := \sum_{m_1, \ldots, m_r=1}^\infty t_{m_1 \ldots m_r} \,
      p_{i_1}^{m_1} \ldots p_{i_r}^{m_r} \; .
\ee
More generally,
\be
    e^{\xi_1} \ast \ldots \ast e^{\xi_N}
 = \exp \Big( \sum_{r=1}^N \, \sum_{1 \leq i_1< \ldots <i_r \leq N}
    \xi_{i_1 \ldots i_r} \Big)
\ee
which implies
\be
     (e^{\xi_1} \ast \ldots \ast e^{\xi_N})_{t_{m_1 \ldots m_r}}
 &=& \Big(\sum_{1 \leq i_1 < \ldots <i_r \leq N}
   p_{i_1}^{m_1} \ldots p_{i_r}^{m_r} \Big) \, e^{\xi_1} \ast \ldots \ast e^{\xi_N}
   \nonumber \\
 &=& \Sigma_N(P_{m_1} \prec \ldots \prec P_{m_r}) \, e^{\xi_1} \ast \ldots \ast e^{\xi_N}
\ee
using (\ref{Sigma(A<...<A)}) in the last step.

\subsection{The XncKP hierarchy}
Comparing the generalized derivation rule (\ref{delta-genderiv}) with
(\ref{ext-ast-deriv-rule}) and recalling theorem~\ref{theorem:deltas-commute},
it is consistent to set
$\pa_{t_{m_1 \ldots m_r}} = \delta_{m_1 \ldots m_r}$ on $\mathcal{R}$.
Then (\ref{delta_m...L^n<}), respectively (\ref{delta_m...L^n>}), leads to
\be
     \pa_{t_{m_1 \ldots m_r}} L^n
 &=& - [L^{m_1, \ldots, m_r}{}_{<0},L^n]_\ast
     + \sum_{k=1}^{r-1} (\pa_{t_{m_1 \ldots m_k}} L^n)
            \ast L^{m_{k+1}, \ldots, m_r}{}_{<0} \nonumber \\
 &=& [L^{m_1, \ldots, m_r}{}_{\geq 0},L^n]_\ast
     - \sum_{k=1}^{r-1} (\pa_{t_{m_1 \ldots m_k}} L^n)
            \ast L^{m_{k+1}, \ldots, m_r}{}_{\geq 0}    \label{L^n-XncKP-flows}
\ee
where $L^{m_1, \ldots, m_r}$ is given by (\ref{L^m...m}) in terms of $L$.
For $n=1$, these are the generalized Lax equations
\be
      L_{t_{m_1 \ldots m_r}}
  &=& [L^{m_1, \ldots, m_r}{}_{\geq 0},L]
     - \sum_{k=1}^{r-1} L_{t_{m_1 \ldots m_k}}
            \ast L^{m_{k+1}, \ldots, m_r}{}_{\geq 0}  \nonumber \\
  &=& - [L^{m_1, \ldots, m_r}{}_{<0},L]_\ast
      + \sum_{k=1}^{r-1} L_{t_{m_1 \ldots m_k}} \ast L^{m_{k+1}, \ldots, m_r}{}_{<0}
      \label{XncKP}
\ee
which (as a consequence of theorem~\ref{theorem:deltas-commute}) define a hierarchy
of commuting flows which we call \emph{XncKP hierarchy}.
It is easy to see that they are the integrability conditions of the linear system
\be         \label{linsys}
    L \ast \psi = \la \psi  \qquad \quad
    \psi_{t_{m_1 \ldots m_r}} = L^{m_1, \ldots, m_r}{}_{\geq 0} \ast \psi \; .
\ee
\vskip.1cm

Taking the residue of (\ref{XncKP}), after an $x$-integration we find
\be
      \phi_{t_{m_1 \ldots m_r}} = \res(L^{m_1, \ldots, m_r})
    = \Phi(P_{m_1 \ldots m_r}) \; .    \label{XncKP-phi_t...}
\ee
Introducing coefficients $\sigma^{(m_1, \ldots, m_r)}_k$ via
\be
    L^{m_1, \ldots, m_r}{}_{<0}
 = (-1)^r \sum_{k=1}^\infty \sigma^{(m_1, \ldots, m_r)}_k \ast L^{-k}
\ee
(\ref{XncKP-phi_t...}) takes the form
\be
    \phi_{t_{m_1 \ldots m_r}} = (-1)^r \, \sigma_1^{(m_1, \ldots, m_r)}  \; .
                 \label{XncKP-phi_t...=sigma}
\ee
With the help of $L^{m_1, \ldots, m_{r+1}} = - L^{m_1, \ldots, m_r}{}_{<0}
\ast L^{m_{r+1}}$ one obtains the iteration formula
\be
   \sigma^{(m_1, \ldots, m_{r+1})}_k = \sigma^{(m_1, \ldots, m_r)}_{k+m_{r+1}}
   - \sum_{l=1}^{m_{r+1}-1} \sigma^{(m_1, \ldots, m_r)}_l \ast \sigma^{(m_{r+1}-l)}_k
   \label{sigma-recur}
\ee
which corresponds to the identity (\ref{cmm}) in $\A(P)$. The coefficients $\sigma^{(m)}_k$
already appeared in section~\ref{subsection:ncKP} (see also (5.7) and (5.8) in \cite{DMH04ncKP}).
An example from the set of equations (\ref{XncKP-phi_t...=sigma}) is
\be
   \phi_{t_{1,2,1}} &=& - \sigma^{(1,2,1)}_1 = - \sigma^{(1,2)}_2
   = - \sigma^{(1)}_4 + \sigma^{(1)}_1 \ast \sigma^{(1)}_2  \nonumber \\
  &=& \frac{1}{4} \phi_{t_4} - \frac{1}{3} \phi_{t_1 t_3}
     - \frac{1}{8} \phi_{t_2 t_2} + \frac{1}{4} \phi_{t_1 t_1 t_2}
     - \frac{1}{24} \phi_{t_1 t_1 t_1 t_1}
     + \frac{1}{2} \phi_{t_1} \ast ( \phi_{t_2} - \phi_{t_1 t_1} ) \; .  \label{phi_t121}
\ee
\vskip.1cm

In a similar way, defining $\eta$-coefficients via
\be
    r(P_{m_r} \succ \ldots \succ P_{m_1})_{<0}
  = \sum_{k=1}^\infty L^{-k} \ast \eta^{(m_1, \ldots, m_r)}_k
\ee
one obtains the expression
\be
    \eta^{(m_1, \ldots, m_{r+1})}_k = \eta^{(m_1, \ldots, m_r)}_{k+m_{r+1}}
      + \sum_{l=1}^{m_{r+1}-1} \eta^{(m_{r+1}-l)}_k \ast \eta^{(m_1, \ldots,m_r)}_l
\ee
which corresponds to the identity (\ref{H-recursion}) in $\A(P)$. In fact,
(\ref{sigma-recur}) and the last relation follow directly from the corresponding
relations in $\A(P)$ by use of the following result.

\begin{theorem}
\label{theorem:Phi(C,H)}
\be
    \Phi(H^{(m_1, \ldots, m_r)}_k) = \eta^{(m_1, \ldots, m_r)}_k  \qquad \quad
    \Phi(C^{(m_1, \ldots, m_r)}_k) = \sigma^{(m_1, \ldots, m_r)}_k \; .
\ee
\end{theorem}
{\it Proof:} With the help of (\ref{Phi(a>b)}) and (\ref{ell(a<>P)}), we obtain
\bez
     \Phi(H^{(m_1, \ldots, m_r)}_k)
 &=& \Phi( H_{k-1} \succ P_{m_r} \succ \ldots \succ P_{m_1} )
  =  \res ( \ell(H_{k-1}) \ast r(P_{m_r} \succ \ldots \succ P_{m_1})_{<0} )  \\
 &=& \sum_{l=1}^\infty \res( \ell(P^{\succ k-1}) \ast L^{-l} ) \ast \eta^{(m_1,\ldots,m_r)}_l \\
 &=& \sum_{l=1}^\infty \res( \ell(P^{\succ k-2})_{\geq 0} \ast L^{1-l} )
        \ast \eta^{(m_1,\ldots,m_r)}_l \\
 &=& \sum_{l=1}^\infty \res( \ell(P^{\succ k-2}) \ast (L^{1-l})_{<0} )
        \ast \eta^{(m_1,\ldots,m_r)}_l   \\
 &=& \sum_{l=2}^\infty \res( \ell(P^{\succ k-2}) \ast L^{1-l} )
        \ast \eta^{(m_1,\ldots,m_r)}_l
  = \ldots   \\
 &=& \sum_{l=k-1}^\infty \res( \ell(P) \ast L^{k-2-l} )
        \ast \eta^{(m_1,\ldots,m_r)}_l
  = \sum_{l=k}^\infty \res( L^{k-1-l} ) \ast \eta^{(m_1,\ldots,m_r)}_l \; .
\eez
The second relation of the theorem is verified in a similar way.
\hfill $\blacksquare$
\vskip.2cm

Explicit equations of the XncKP hierarchy are more generally obtained by application
of $\Phi$ to identities in $\A(P)$ built from any subset of the elements
$P_{m_1 \ldots m_k}$ and the products $\circ$ and $\htimes$ (using
(\ref{XncKP-phi_t...}), theorem~\ref{theorem:Phi-htimes-ast-homom}
and proposition~\ref{proposition:deltaPhi=Phi(Pcirc)}).
\vskip.1cm

There is a redundancy in the parameters $t_{m_1 \ldots m_k}$.
The remark at the end of section~\ref{subsection:xncKP}, which also applies
to the more general $\ast$-product under consideration, shows that we may
drop the symmetric part of $t_{mn}$.
But now there are further identities in $\A(P)$ which lead to \emph{linear}
equations for $\phi$ and allow to eliminate partial derivatives of $\phi$
with respect to certain combinations of the variables $t_{m_1 \ldots m_k}$ for
fixed $k$. For example, with the help of (\ref{A-c-a-cw-b}), (\ref{A1cA2})
and (\ref{Pmn-symm-red-id}), we obtain
\be
     P_m \circ P_n \circ P_r
 &=& P_{mnr} + P_{mrn} + P_{nrm} + P_{nmr} + P_{rmn} + P_{rnm} \nonumber \\
 & & + P_m \circ P_{n+r} + P_n \circ P_{m+r} + P_r \circ P_{m+n} - 2 \, P_{m+n+r}
     \label{Pmnr-symm-red-id}
\ee
which is mapped by $\Phi$ to
\be
     \phi_{t_{mnr}} + \phi_{t_{mrn}} + \phi_{t_{nrm}} + \phi_{t_{nmr}} + \phi_{t_{rmn}}
      + \phi_{t_{rnm}}
 &=& \phi_{t_m t_n t_r} - \phi_{t_m t_{n+r}} - \phi_{t_n t_{m+r}} - \phi_{t_r t_{m+n}}
     \nonumber \\
 & & + 2 \, \phi_{t_{m+n+r}}
\ee
as a consequence of which the totally symmetric part of $t_{mnr}$ turns out to
be redundant. In particular, the last equation implies
\be
    \phi_{t_{mmm}} = \frac{1}{6} \phi_{t_m t_m t_m} - \frac{1}{2} \phi_{t_m t_{2m}}
                     + \frac{1}{3} \phi_{t_{3m}} \; .  \label{phi_tmmm}
\ee
A similar calculation yields
\be
    P_{mnr} - P_{mrn} + P_{nrm} + P_{nmr} - P_{rmn} - P_{rnm}
  = 2 \, (P_m \circ A_{nr} - A_{n,m+r} + A_{r,m+n})  \label{P_mnr-red-id}
\ee
and anti-symmetrization with respect to $m,n,r$ leads to
\be
    P_{mnr} - P_{mrn} + P_{nrm} - P_{nmr} + P_{rmn} - P_{rnm}
  = 2 \, (P_m \circ A_{nr} + P_n \circ A_{rm} + P_r \circ A_{mn})
\ee
so that, in particular, the totally antisymmetric part of $t_{mnr}$ is redundant.
Of course, (\ref{P_mnr-red-id}) determines further redundancies. These are
given by
\be
    P_{mmr} - P_{rmm} = P_m \circ A_{mr} - A_{m,m+r} + A_{r,2m}  \qquad \quad r \neq m
    \label{P_mmr-red-id}
\ee
and additional relations with $m,n,r$ pairwise different.
\vskip.1cm

Let us look at some concrete examples. Application of $\Phi$ to the identity
\be
      P_{1,1,1} + P_{1,2} + P \htimes P
    = P \prec P \prec P + P \prec P_2 + P \htimes P
    = 0
\ee
leads to the nonlinear XncKP equation
\be
    \phi_{t_{1,1,1}} + \phi_{t_{1,2}} + \phi_{t_1} \ast \phi_{t_1} = 0 \; .
\ee
By use of the linear equation (\ref{phi_tmmm}) this becomes
\be
    {1 \over 3} \phi_{t_3} - {1 \over 2} \phi_{t_1 t_2}
    + {1 \over 6} \phi_{t_1 t_1 t_1} + \phi_{t_{1,2}} + \phi_{t_1} \ast \phi_{t_1} = 0
    \label{XncKP-3}
\ee
which, with the help of the linear equation (\ref{phi_tmn_thetamn}), is turned into
an xncKP equation,
\be
    \phi_{\theta_{1,2}} - {1 \over 6} (\phi_{t_3} - \phi_{t_1 t_1 t_1})
    + \phi_{t_1} \ast \phi_{t_1} = 0 \; .
\ee
Of course, this equation is obtained more directly from the identity
\be
   A_{1,2} - \frac{1}{6} ( P_3 - P^{\circ 3} ) + P \htimes P = 0 \; .
\ee
\vskip.1cm

Furthermore, the identity
\be
    P_{1,2,1} = P \prec P_2 \prec P
              = P \prec P \succ P \prec P - P \prec P \prec P \prec P
              = - P \hat{\times} P^{\prec 2} - P^{\prec 4}
\ee
leads to the nonlinear XncKP equation
\be
   \phi_{t_{1,2,1}} = - \phi_{t_1} \ast \phi_{t_{1,1}} - \phi_{t_{1,1,1,1}}
\ee
where we should substitute the following expressions (obtained from (\ref{Cn_Schur}),
for example),
\be
   \phi_{t_{1,1}} &=& - {1 \over 2} \phi_{t_2} + {1 \over 2} \phi_{t_1 t_1} \\
   \phi_{t_{1,1,1,1}} &=& - {1 \over 4} \phi_{t_4} + {1 \over 3} \phi_{t_1 t_3}
     + {1 \over 8} \phi_{t_2 t_2} - {1 \over 4} \phi_{t_1 t_1 t_2}
     + {1 \over 24} \phi_{t_1 t_1 t_1 t_1}
\ee
which results in (\ref{phi_t121}).
Expressions for $\phi_{t_{1,1,2}}$ and $\phi_{t_{2,1,1}}$ are then obtained with
the help of linear equations given above. We may take the view, however, that the
dependence of $\phi$ on $t_{1,1,2}$, respectively $t_{2,1,1}$, is redundant (after
selection of the variable $t_{1,2,1}$).

\subsection{Reductions}
Let us impose the constraint $(L^N)_{<0} = 0$ for some fixed $N
\in \mathbb{N}$ which is known to reduce the KP hierarchy to the
$N$th Gelfand-Dickey hierarchy (see \cite{Dick03}, for example).
It immediately follows from (\ref{L^n-XncKP-flows}) that all
equations of the \emph{XncKP} hierarchy preserve this constraint.
Another immediate consequence is $(L^{kN})_{<0} = 0$ and thus
$L_{t_{kN}} = 0$ for all $k \in \mathbb{N}$. Moreover,
(\ref{L^m...m}) shows that
\be
    L^{kN,m_2,\ldots,m_r} = 0  \qquad \quad   k \geq 1, \quad  r \geq 2
\ee
which, by use of (\ref{XncKP}), implies
\be
    L_{t_{kN, m_2 \ldots m_r}} = 0  \qquad \quad  k \geq 1, \quad  r \geq 2 \; .
\ee
Furthermore,
\be
      L^{m_1, \ldots, m_r, kN}{}_{<0}
  &=& ( ( L^{m_1, \ldots, m_{r-1}}{}_{<0} \ast L^{m_r} )_{<0} \ast L^{kN} )_{<0}  \nonumber \\
  &=& ( L^{m_1, \ldots, m_{r-1}}{}_{<0} \ast L^{m_r} \ast L^{kN} )_{<0}
      - ( ( L^{m_1, \ldots, m_{r-1}}{}_{<0} \ast L^{m_r} )_{\geq 0} \ast (L^{kN})_{\geq 0} )_{<0}
        \nonumber \\
  &=& - L^{m_1, \ldots, m_r+kN}{}_{<0} \qquad \quad  k \geq 1, \quad  r \geq 1
      \label{Nred-relation}
\ee
by use of (\ref{ell(a<P_n)}). With the help of (\ref{XncKP}), this leads to
\be
    L_{t_{m_1 \ldots m_r, kN}} = - L_{t_{m_1 \ldots m_r+kN}}
    \qquad k \geq 1, \quad  r \geq 1 \; .
\ee
Moreover, using (\ref{L^m...m}) and (\ref{Nred-relation}), we obtain
\be
      L^{m_1, \ldots, m_{l-1}, kN, m_{l+1}, \ldots, m_r}
   = - L^{m_1, \ldots, m_{l-1}+kN, m_{l+1}, \ldots, m_r}
     \qquad  l = 2, \ldots, r-1, \, r \geq 3, \, k \geq 1
\ee
and thus
\be
    L_{t_{m_1 \ldots m_{l-1}, kN, m_{l+1} \ldots m_r}}
  = - L_{t_{m_1 \ldots, m_{l-1}+kN, m_{l+1} \ldots m_r}}
     \qquad  l = 2, \ldots, r-1 , \, r \geq 3, \, k \geq 1  \; .
\ee

As an example, let us consider the KdV reduction $(L^2)_{<0} = 0$. In this case we have
$\phi_{t_2} = 0$ and $\phi_{t_{1,2}} = - \phi_{t_3}$, so that (\ref{XncKP-3}) reduces to
\be
    \phi_{t_3} = {1 \over 4} \phi_{t_1 t_1 t_1} + {3 \over 2} \phi_{t_1} \ast \phi_{t_1}
    \label{pot-ncKP-eq}
\ee
which is the potential ncKdV equation. Furthermore, (\ref{phi_t121}) reduces to
\be
      \phi_{t_{3,1}}
   = - \phi_{t_{1,2,1}}
   =  \frac{1}{3} \phi_{t_1 t_3}
     + \frac{1}{24} \phi_{t_1 t_1 t_1 t_1}
     + \frac{1}{2} \phi_{t_1} \ast \phi_{t_1 t_1}
\ee
and (\ref{P_mmr-red-id}) leads to the linear equation
\be
    \phi_{t_{1,3}} = - \phi_{t_{1,1,2}} = \phi_{\theta_{1,3}} + \frac{1}{2} \phi_{t_1 t_3}
\ee
with the help of which, and use of (\ref{pot-ncKP-eq}), the previous equation yields
the xncKdV equation
\be
    \phi_{\theta_{1,3}} + \frac{1}{4} \, [ \phi_{t_1} , \phi_{t_1 t_1} ] = 0 \; .
\ee

\subsection{Generalized Sato-Wilson equations and Birkhoff factorization}
The ncKP hierarchy can be formulated alternatively in terms of the Sato-Wilson equations
\be
    W_{t_m} = - (L^m)_{<0} \ast W     \label{Sato-Wilson}
\ee
with (the dressing operator)
\be
    W = 1 + \sum_{n=1}^\infty w_n \, \pa^{-n} \; .
\ee
Since $t_1=x$ and $L_{\geq 0} = \pa$, the case $m=1$ leads to
$W_x = - L_{<0} \ast W = \pa W - L \ast W$.
Hence $L \ast W = W \, \pa$ or $L = W \ast \pa \, W^{-1}$, since $W$ is invertible.
The Sato-Wilson equations now take the form
\be
    W_{t_m} = - (W \ast \pa^m \, W^{-1})_{<0} \ast W \; .
\ee
These equations imply the Lax form (\ref{KP-Lax}) of the ncKP equations
(see also \cite{DMH04ncKP}).
\vskip.1cm

An obvious generalization of the above Sato-Wilson equations is given by
\be
    W_{t_{m_1 \ldots m_r}} = - L^{m_1,\ldots,m_r}{}_{<0} \ast W \; .   \label{g-Sato-Wilson}
\ee
They indeed imply the generalized Lax equations (\ref{XncKP}), as can be demonstrated
by application of $\pa_{t_{m_1 \ldots m_r}}$ to $L \ast W = W \pa$.
 From (\ref{g-Sato-Wilson}) we find
\bez
     W_{t_{m_1 \ldots m_r}} \ast W^{-1}
 &=& - L^{m_1,\ldots,m_r}{}_{<0}
  = (L^{m_1,\ldots,m_{r-1}}{}_{<0} \ast L^{m_r})_{<0} \nonumber \\
 &=& (L^{m_1,\ldots,m_{r-1}}{}_{<0} \ast W \ast \pa^{m_r} W^{-1})_{<0} \nonumber \\
 &=& - (W_{t_{m_1 \ldots m_{r-1}}} \ast \pa^{m_r} W^{-1})_{<0}
\eez
and thus the equivalent inductive form
\be  \label{indsw}
    W_{t_{m_1 \ldots m_r}} = - (W_{t_{m_1 \ldots m_{r-1}}} \ast \pa^{m_r} W^{-1})_{<0}
    \ast W \; .
\ee
This can be rewritten as
\be
    W_{t_{m_1 \ldots m_r}} = (W_{t_{m_1 \ldots m_{r-1}}} \ast \pa^{m_r} W^{-1})_{\geq 0}
    \ast W - W_{t_{m_1 \ldots m_{r-1}}} \, \pa^{m_r}
\ee
and thus
\be
    (W \ast e^{\hat{\xi}})_{t_{m_1 \ldots m_r}} = L^{m_1,\ldots,m_r}{}_{\geq 0}
    \ast (W\ast e^{\hat{\xi}})
\ee
where $\hat{\xi} = \sum_{n \geq 1} t_n \, \pa^n$. Following \cite{Taka94} (see also
\cite{DMH04extMoyal,Saka04}), this leads to the Birkhoff factorization
(generalized Riemann-Hilbert problem, see \cite{HSS92,Seme02} for example)
\be
    W(t) \ast e^{\hat{\xi}(t)} = Y(t) \ast W(0)    \label{birk}
\ee
with $Y = Y_{\geq 0}$. This is equivalent to
\be
    e^{\hat{\xi}} \, W(0)^{-1} = W(t)^{-1} \ast Y(t)
\ee
since $W(t) \in G_-$ and $Y(t) \in G_+$, for the group $G = G_- \, G_+$ of $\Psi$DOs.
\vskip.1cm

Conversely, acting with $\pa_{t_{m_1 \ldots m_r}}$ on (\ref{birk}), we get
\be
     (W(t)_{t_{m_1 \ldots m_r}} + W(t)_{t_{m_1 \ldots m_{r-1}}} \pa^{m_r}) \ast e^{\hat{\xi}(t)}
  = Y(t)_{t_{m_1 \ldots m_r}} \ast Y(t)^{-1} \ast W(t) \ast e^{\hat{\xi}(t)}
\ee
and thus
\be
       W(t)_{t_{m_1 \ldots m_r}} \ast W(t)^{-1}
     + W(t)_{t_{m_1 \ldots m_{r-1}}} \ast \pa^{m_r} W(t)^{-1}
  = Y(t)_{t_{m_1\ldots m_r}}\ast Y(t)^{-1} \; .
\ee
Taking the $\mathcal{R}_{<0}$ part, noting that $(Y(t)_{t_{m_1 \ldots m_r}} \ast Y(t)^{-1})_{<0} = 0$
and $W(t)_{t_{m_1 \ldots m_r}} \ast W(t)^{-1} = (W(t)_{t_{m_1 \ldots m_r}} \ast W(t)^{-1})_{<0}$,
one recovers (\ref{indsw}). Hence, the Birkhoff factorization (\ref{birk}) is equivalent
to the XncKP hierarchy equations (\ref{XncKP}).
Via (\ref{birk}) the space of solutions of the XncKP hierarchy is determined
from the same initial data $W(0)$ as in the KP case \cite{Saka04}.

\section{Conclusions}
\label{section:concl}
\setcounter{equation}{0}
Some crucial steps in this work are sketched in the following diagram.
\[ \begin{CD}
  \A(P)   @>\Psi>> \A(P/Q)  \\
  @V{\Phi}VV       @VV{\Sigma_N}V \\
  \mbox{XncKP}     @>\mbox{trace method}>> \begin{minipage}{2cm} \mbox{algebraic sum} \\
                                           \mbox{identities} \end{minipage}
   \end{CD} \]
Our central object is the algebra $\A(P)$ generated by a single element $P$ and supplied
with certain associative products, which in particular give rise to a (mixable) shuffle product
(and a Rota-Baxter algebra structure). The map $\Psi$ embeds it into a corresponding algebra
generated by two independent commuting elements $P,Q$.
Identities in $\A(P)$ are then mapped by $\Psi$ to identities in the latter algebra.
These in turn are sent by $\Sigma_N$ to algebraic sum identities in variables $p_n, q_n$,
$n = 1, \ldots, N$. Since $N \in \mathbb{N}$ is arbitrary, this results in
families of identities. Such identities were actually the starting point of this work.
In the introduction we explained how algebraic identities of this kind emerge from the
equations of the (nc)KP hierarchy via the `trace method' \cite{Okhu+Wada83}.
It remained to find those families of identities in $\A(P)$ which correspond to
KP equations. This is where the map $\Phi$ entered the stage. We found identities
in $\A(P)$ which are mapped by $\Phi$ to KP equations and the whole hierarchy of
KP equations expressed in the potential $\phi$ is recovered in this way (after setting
the derivations $\delta_n$ equal to partial derivatives $\pa_{t_n}$).
\vskip.1cm

Moreover, we found further families of identities and showed that these determine
extensions of the ncKP hierarchy with deformed products. The xncKP hierarchy
\cite{DMH04hier,DMH04ncKP} is rediscovered in this way. But we even discovered a
new (XncKP) hierarchy which extends the xncKP hierarchy after deforming the product in
a more general way.
\vskip.1cm

The XncKP hierarchy contains linear equations and it seems that their existence
is related to equivalence transformations of the $\ast$-product, which can be used
to reduce the amount of deformation parameters (which correspond to evolution `times' of
the generalized hierarchy). This relation has not been sufficiently clarified in
this work.
\vskip.1cm

The fact that $(\mathcal{R} , ( \; )_{\geq 0})$ (and also $(\mathcal{R} , ( \; )_{<0})$)
is a Rota-Baxter algebra (see appendix~A), suggests to generalize the results
of section \ref{section:PsiDOs} towards other Rota-Baxter algebras.\footnote{We
may e.g. replace $\mathcal{R}$ by an algebra of Laurent series in an indeterminate $\lambda$,
as in the AKNS hierarchy example \cite{DMH04hier,DMH04extMoyal}.}
\vskip.1cm

The correspondence between (X)ncKP equations and algebraic identities
presented in this work sets up a bridge between different areas of mathematics.
In view of the appearance of the KP hierarchy in many physical systems and
in various mathematical problems, this should be an interesting new tool
for further explorations. In particular, the KP hierarchy has deep relations
with string theory (see \cite{AGR87,Sait87,Mula88},
for example) and shows up in related models like topological field theories
\cite{Witt91,Dijk92,Lee03} and matrix models \cite{Bono+Xion92,KKN99}.
We should also mention its appearance in Seiberg-Witten theory \cite{Eguc+Yang96} and
a relation with random matrices \cite{ASvM98}. We expect that
deformations and extensions of the KP hierarchy will play a similar role and that
interesting generalizations of these results can be achieved. Indeed, some motivation
to study (Moyal-) deformations of the KP hierarchy originated from the following fact.
In string theory, $D$-branes with a non-vanishing $B$-field are effectively described
in a low energy limit by a Moyal-deformed Yang-Mills theory \cite{Seib+Witt99,Lee00}.
Corresponding noncommutative instantons \cite{Doug+Nekr01} are solutions of a
Moyal-deformed self-dual Yang-Mills equation, from which Moyal-deformed soliton equations
result by reductions, as in the classical case (see \cite{ACH03}, for example).
Such deformed soliton equations provide us with interesting examples of
noncommutative field theories \cite{Doug+Nekr01}.
\vskip.1cm

Within the framework of integrable systems our results suggest an apparently
new method, namely to look for (series of) algebraic identities of a certain
type in order to construct hierarchies of soliton equations. The XncKP hierarchy
presented in this work was in fact discovered in this way.
\vskip.1cm

\noindent
{\it Acknowledgments.} A. D. would like to thank D. Drivaliaris for useful discussions.
F. M.-H. thanks K. Ebrahimi-Fard for helpful comments.

\renewcommand{\theequation} {\Alph{section}.\arabic{equation}}

\section*{Appendix A: Rota-Baxter operators}
\setcounter{section}{1}
\setcounter{equation}{0}
\addcontentsline{toc}{section}{\numberline{}Appendix A: Rota-Baxter operators}
We recall the Rota-Baxter relation of weight\footnote{Via multiplication of the Rota-Baxter
operator by $\mathrm{q}^{-1}$, we can always achieve that a non-vanishing weight constant
becomes equal to $1$. In this sense, the weight constant is `relatively unimportant'
\cite{Rota+Smit72}.}
$\mathrm{q}$ on a ring $\mathbb{A}$:
\be
   R(a) \, R(b) = R( R(a) \, b + a \, R(b) ) - \mathrm{q} \, R(a \, b)
\ee
(see \cite{Baxt60,Rota69I,Atki63,Rota+Smit72,Rota95}).
A (not exhaustive) class of Rota-Baxter operators is obtained by the
following construction \cite{Atki63,Rota+Smit72}.
Given an endomorphism $\Lambda : \mathbb{A} \to \mathbb{A}$ of an algebra
$\mathbb{A}$, i.e., $\Lambda(a b) = \Lambda(a) \Lambda(b)$ for all $a,b \in \mathbb{A}$,
\be
    R := \sum_{r \geq 1} \Lambda^r
\ee
(assuming convergence, or nilpotence for some power of $\Lambda$) defines a Rota-Baxter
operator of weight $-1$. Note also that $\mathrm{id}+R$ is then a Rota-Baxter
operator of weight $1$.
An important example, already presented by G.~Baxter \cite{Baxt60}, is provided by
the \emph{standard Baxter algebra} \cite{Rota69I,Rota95} of a set of generators
$\{ a, b, c, \ldots \}$ which are infinite sequences $a = (a_1,a_2,\ldots)$,
with componentwise multiplication $a b = (a_1 b_1, a_2 b_2, \ldots)$ and the Rota-Baxter
operator given by
\be
    R(a_1,a_2,a_3,\ldots) = (0,a_1,a_1+a_2,a_1+a_2+a_3,\ldots)
\ee
which is (\ref{R(alpha)-ps-calc}).
This is of the above form with the shift operator
$\Lambda(a_1,a_2,\ldots) := (0,a_1,a_2,\ldots)$. The standard Baxter algebra is naturally
isomorphic to the free Baxter algebra on the same set of generators \cite{Rota69I,Rota95}.
The standard (or free) Baxter algebra with a single generator is isomorphic to the
algebra of symmetric functions \cite{Rota95}.
Another example is obtained by choosing $(\Lambda f)(x) := f(q \, x)$ on functions of
a variable $x$, where $q$ is a parameter. $R$ is then the Jackson $q$-integral \cite{Rota95}.
\vskip.1cm

The following theorem \cite{Atki63} provides us with further examples and,
in particular, shows that $(\mathcal{R} , \ast, ( \; )_{\geq 0})$ and
$(\mathcal{R} , \ast, ( \; )_{<0})$ are Rota-Baxter algebras.
\vskip.2cm

\noindent
{\bf Theorem.} Let $(\mathbb{A},+,\cdot)$ be a (with respect to the product $\cdot$ not
necessarily commutative and not necessarily associative) ring.
The following conditions are equivalent: \\
(i) There is a Rota-Baxter operator $R$ of weight $1$ on $\mathbb{A}$ which is a
group homomorphism of addition. \\
(ii) There are two subrings $\mathbb{A}_{\pm}$ of $\mathbb{A}$ and a subring $\mathbb{B}$
of $\mathbb{A}_+ \times \mathbb{A}_-$ (supplied with a ring structure in the obvious way by
componentwise addition and multiplication) such that each element $a \in \mathbb{A}$
has a unique decomposition $a = a_+ + a_-$ with $(a_+,a_-) \in \mathbb{B}$.

An \emph{idempotent} Rota-Baxter operator $R$ of weight $1$ is equivalent to a
\emph{direct sum} decomposition, i.e., $\mathbb{A} = \mathbb{A}_+ \oplus \mathbb{A}_-$.
\vskip.1cm
\noindent
{\it Proof:} Let us assume that (i) holds. Define $\mathbb{A}_+ := R(\mathbb{A})$ and
$\mathbb{A}_- := (\mathrm{id}-R)(\mathbb{A})$. By assumption, $R(a)+R(b) = R(a+b)$.
Furthermore, the Rota-Baxter relation
$R(a) \, R(b) = R( R(a) \, b + a \, R(b) + a \, b)$ implies
$R(a) \, R(b) \in R(\mathbb{A}) = \mathbb{A}_+$, so that $\mathbb{A}_+$ is a subring.
Moreover, since $\mathrm{id} - R$ satisfies the same Rota-Baxter relation,
$\mathbb{A}_-$ is also a subring. This supplies $\mathbb{A}_+ \times \mathbb{A}_-$
with a ring structure. Now
\bez
    (R(a) \, R(b) , (\mathrm{id}-R)(a) \, (\mathrm{id}-R)(b) ) = ( R(c) , (\mathrm{id}-R)(c) )
\eez
with $c := a \, R(b) + R(a) \, b - a \, b$ shows that there is a subring
$\mathbb{B}$ of $\mathbb{A}_+ \times \mathbb{A}_-$ with the properties specified in (ii).

Conversely, if (ii) holds, $R(a) := a_+$ defines a homomorphism $R$ with respect to
the operation $+$ and we have $a_- = (\mathrm{id} - R)(a)$. Now we compare the decomposition
\bez
   a \, R(b) + R(a) \, b - a \, b
 = R( a \, R(b) + R(a) \, b - a \, b ) + (\mathrm{id} - R)( a \, R(b) + R(a) \, b - a \, b )
\eez
with the identity
\bez
     a \, R(b) + R(a) \, b - a \, b
   = R(a) \, R(b) - (\mathrm{id}-R)(a) \, (\mathrm{id}-R)(b)
\eez
where, as a consequence of the subring properties, the first term on the
right hand side lies in $\mathbb{A}_+$ and the second in $\mathbb{A}_-$.
Since the decomposition of an element of $\mathbb{A}$ is unique, this implies
\bez
     R( a \, R(b) + R(a) \, b - a \, b ) = R(a) \, R(b)
\eez
which is the Rota-Baxter relation (of weight $1$).

If $R$ is idempotent, i.e., $R^2 = R$, one easily verifies that
$\mathbb{A}_+ \cap \mathbb{A}_- = \{ 0 \}$. Conversely, given
$\mathbb{A} = \mathbb{A}_+ \oplus \mathbb{A}_-$, the projections onto the
subrings define idempotent Rota-Baxter operators.
\hfill $\blacksquare$
\vskip.2cm

The theorem also holds with `ring' replaced by `$\mathbb{K}$-algebra' if $R$ is
$\mathbb{K}$-linear.
If one of the conditions of the theorem is fulfilled, the \emph{classical $\mathbf{R}$-matrix}
given by
\be
    \mathbf{R}(a) := a_+ - a_-
\ee
(which generalizes the Hilbert transform) satisfies
\be
   \mathbf{R}(a) \, \mathbf{R}(b) = \mathbf{R}( \mathbf{R}(a) \, b + a \, \mathbf{R}(b) ) - a \, b
\ee
called `Poincar{\'e}-Bertrand formula' in \cite{Seme84} and `modified Rota-Baxter relation'
in \cite{EGK04int1,EGK04int2}. Passing over to commutators, this yields the modified
Yang-Baxter equation \cite{Seme84}. The product $\vtr$ used in section~\ref{section:PsiDOs}
can be expressed as follows \cite{Seme84},
\be
    a \vtr b = a_+ \, b_+ - a_- \, b_-
             = \frac{1}{2} \Big( \mathbf{R}(a) \, b + a \, \mathbf{R}(b) \Big) \; .
\ee
In terms of the Rota-Baxter operator given by $R(a) = a_+$, we have the following
expression,
\be
    a \vtr b = R(a) \, b + a \, R(b) - a \, b  \; .
\ee
Such a product, determined by a Rota-Baxter operator of weight $1$, has been called
`double product' in \cite{EGK04Spitzer} (see also \cite{EGK04int1,EGK04int2}).
It is associative as a consequence of the Rota-Baxter relation.
\vskip.1cm

We also refer to \cite{Guo00diff,Guo01umbral,AGLK03,Ebra+Guo04_RB} for explorations
of Rota-Baxter algebras.
In particular, according to \cite{Ebra02} any Rota-Baxter algebra defines a
dendriform trialgebra (see \cite{Loda+Ronc04}, for example).\footnote{Although
the notation used in work on dendriform algebras looks similar to the notation used
in section~\ref{section:basic}, one should note that the operations
defining a dendriform algebra are \emph{not} associative whereas our products are
associative.}

\section*{Appendix B: Some realizations of the algebra $\cal A$}
\setcounter{section}{2}
\setcounter{equation}{0}
\addcontentsline{toc}{section}{\numberline{}Appendix B: Some realizations of the algebra $\cal A$}
In this appendix we briefly describe some realizations of the algebraic structure
introduced in section~\ref{section:basic}, different from our main example of partial
sum calculus in section~\ref{section:ps}.
\vskip.1cm

\noindent
\textbf{Posets.} A poset $\mathcal{P}$ is a set with a binary relation
$i \leq j$ for $i,j \in \mathcal{P}$, such that
\renewcommand{\labelenumi}{\roman{enumi}.}
\begin{enumerate}
\item for all $i$: $i \leq i$,
\item if $i \leq j$ and $j \leq i$, then $i=j$,
\item if $i \leq j$ and $j \leq k$, then $i \leq k$.
\end{enumerate}
Let us write $i<j$ for $i \leq j$ and $i \neq j$. A finite nonempty subset
$\{i_1,\ldots,i_n\}$ of $\mathcal{P}$ will be called a chain, if $i_1 < \ldots < i_n$.
A chain $I$ always has a smallest element $\min(I)$ and a greatest element $\max(I)$ .
The set $\mathcal{C}$ of chains of $\mathcal{P}$ is `graded' with respect to the number of
elements of the chains. Let $\A$ be the free vector space generated by $\mathcal{C}$
over $\mathbb{K}$ with basis vectors $\{ e_I \, | \, I \in \mathcal{C} \}$.
We define the algebraic structure as in the case of partial sums:
\be
    e_I \bullet e_J := \left\{\begin{array}{ll} e_{I\cup J} & \hbox{if $\max(I)=\min(J)$} \\
     0 & \hbox{otherwise}
    \end{array}\right.
\ee
\be
    e_I \prec e_J := \left\{\begin{array}{ll} e_{I \cup J} & \hbox{if $\max(I)<\min(J)$} \\
     0 & \hbox{otherwise}
    \end{array}\right.
\ee
and thus
\be
    e_I \succ e_J = \left\{\begin{array}{ll} e_{I \cup J} & \hbox{if $\max(I) \leq \min(J)$} \\
     0 & \hbox{otherwise}
    \end{array}\right.
\ee
 From these rules we find
\be
  e_I \circ e_J := \left\{\begin{array}{ll} e_{I \cup J} & \hbox{if $I \cup J \in \mathcal{C}$} \\
     0 & \hbox{otherwise}
    \end{array}\right.
\ee
For  a finite poset $\mathcal{P}$, we define a map $\Sigma : \A \to \mathbb{K}$ by
$\Sigma(e_I) = 1$ for all $I \in \mathcal{C}$. Then, for
$A_a = \sum_{i \in \mathcal{P}} a_{a,i} \, e_i$, $a=1, \ldots, r$, we obtain
\be
    \Sigma(A_1 \circ \ldots \circ A_r) = \sum_{i_1, \ldots, i_r \in \mathcal{P}}
    c_{\{i_1, \ldots, i_r\}} a_{1,i_1} \cdots a_{r,i_r}
\ee
with
\bez
    c_{\{i_1, \ldots, i_r\}} := \left\{\begin{array}{ll} 1 & \hbox{if $\{ i_1,\ldots,i_r \}
    \in \mathcal{C}$ } \\   0 & \hbox{otherwise}
    \end{array}\right.
\eez
A special example of a poset is given by a \emph{rooted tree}, which possesses a distinguished
element, the `root', from which there is a unique path to any other element. The ordering
of nodes along a path obviously defines poset relations $<$ and $\leq$.
Then $R(A) := \sum_{n \in \mathcal{P}} ( \sum_{k<n} a_k ) \, e_n$,
where $A = \sum_{n \in \mathcal{P}} a_n \, e_n$, defines a Rota-Baxter
operator of weight $-1$ for the algebra $(\A^1,\bullet)$.
Hence, with any rooted tree a Rota-Baxter algebra, and thus also a
dendriform trialgebra \cite{Ebra02}, is associated.
\vskip.1cm

\noindent
\textbf{The tensor product algebra of an associative algebra.}
Let $(\A^1,\bullet)$ be any associative algebra over $\mathbb{K}$, and
$\A^r := \A^1 \otimes \ldots \otimes \A^1$ ($r$-fold tensor product over $\mathbb{K}$).
Then $\A = \bigoplus_{r \geq 1} \A^r$ with the tensor product $\otimes$ is an associative algebra.
The product $\bullet$ extends to an associative product in $\A$ by setting
\be
    (A_1 \otimes \ldots \otimes A_r) \bullet (B_1 \otimes \ldots \otimes B_s)
  := A_1 \otimes \ldots \otimes (A_r \bullet B_1) \otimes \ldots \otimes B_s
\ee
for all $A_1, \ldots, A_r, B_1, \ldots, B_s \in \A^1$.
Let us now define a new associative product by
\be
    \alpha \succ \beta = \alpha \otimes \beta + \alpha \bullet \beta
    \qquad \forall \alpha, \beta \in \A  \; .
\ee
Identifying $\otimes$ with $\prec$ in our general formalism, the main product
$\circ$ becomes a `mixable shuffle product', as considered in \cite{Guo+Keig00shuffle}.

If $(\A^1,\bullet)$ is unital with unit $E$, we can define an operator $R : \A \to \A$ by
$R(\alpha) := E \otimes \alpha$. This implies $R(\alpha) \bullet R(\beta) = R(\alpha \otimes \beta)$.
The quasi-shuffle property leads to
\be
    R(\alpha) \circ R(\beta) = R(\alpha \circ R(\beta) + R(\alpha) \circ \beta
    + \alpha \circ \beta)
\ee
so that $R$ is a Rota-Baxter operator of weight $-1$. The algebra $(\A,\circ,R)$ is the
free Rota-Baxter algebra on $\A^1$ (of weight $-1$) \cite{Guo+Keig00shuffle}.
The operator $R$ satisfies
\be
    R(\alpha) \bullet R(\beta) + R^2(\alpha \bullet \beta)
  = R(\alpha \bullet R(\beta) + R(\alpha) \bullet \beta)
\ee
with respect to the $\bullet$-product. This is the condition in \cite{Fuch97}
for the map $R$ to be \emph{hereditary} and called \emph{associative Nijenhuis relation}
in \cite{CGM00,Ebra04,EGK04int1,EGK04int2}.
In fact, the following stronger identity holds,
\be
    R(\alpha \bullet \beta) = R(\alpha) \bullet \beta \; .
\ee

\section*{Appendix C: $\ast_n$ products}
\setcounter{section}{3}
\setcounter{equation}{0}
\addcontentsline{toc}{section}{\numberline{}Appendix C: $\ast_n$ products}
On the space of analytic functions (or formal power series) of the collection
$x = (x^{(1)}, x^{(2)}, \ldots, x^{(n)})$
of variables $x^{(1)} = \{x^\mu\}, \, x^{(2)} = \{x^{\mu\nu}\}, \,
x^{(3)} = \{x^{\mu\nu\rho}\}, \ldots, x^{(n)}=\{x^{\mu_1\ldots\mu_n}\}$
(where the indices run over some discrete set)
we introduce a product $\ast_n$ via\footnote{Here and in the following we should replace
$\pa/\pa x^{\mu_1 \ldots \mu_k}$ by $1$ if $k=0$ and $\pa/\pa x^{\mu_{k+1} \ldots \mu_r}$
by $1$ if $k=r$.}
\be
    (f \ast_n g)(x)
 := \left. \exp \left( \sum_{r=1}^n x^{\mu_1 \ldots \mu_r}
      \sum_{k=0}^r \frac{\pa}{\pa x_1^{\mu_1 \ldots \mu_k}} \,
      \frac{\pa}{\pa x_2^{\mu_{k+1} \ldots \mu_r}} \right) \,
      f(x_1) \, g(x_2)
      \right|_{x_1 = x_2 = 0}
\ee
using the summation convention with respect to the indices $\mu_k$. Obviously,
\bez
    x^{\mu_1 \ldots \mu_r} \ast_n x^{\nu_1 \ldots \nu_s}
  = x^{\mu_1 \ldots \mu_r} \, x^{\nu_1 \ldots \nu_s} + x^{\mu_1 \ldots \mu_r \nu_1 \ldots \nu_s}
    \label{ast_n-mult-x}
\eez
where the last term should be set to zero if the number of indices exceeds $n$.
For $n=1$ the product $\ast_n$ coincides with the ordinary one since
\bez
    (f \ast_1 g)(x^{(1)}) = \left. \exp\left( x^\mu \left( \frac{\pa}{\pa x_1^\mu}
       + \frac{\pa}{\pa x_2^\mu} \right) \right)
         f(x_1) \, g(x_2) \right|_{x_1=x_2=0} = f(x^{(1)}) \, g(x^{(1)}) \; .
\eez
For $n=2$ we find
\bez
     (f \ast_2 g)(x^{(1)},x^{(2)})
 &=& \exp\Bigg( x^\mu \left( \frac{\pa}{\pa x_1^\mu}
      + \frac{\pa}{\pa x_2^\mu} \right) \nonumber \\
 & & \left. \left. + x^{\mu\nu} \left( \frac{\pa}{\pa x_1^{\mu\nu}}
      + \frac{\pa}{\pa x_2^{\mu\nu}} + \frac{\pa}{\pa x_1^\mu} \, \frac{\pa}{\pa x_2^\nu}
        \right) \right) f(x^{(1)}_1,x^{(2)}_1) \, g(x^{(1)}_2,x^{(2)}_2) \right|_{x_1=x_2=0}
        \nonumber \\
 &=& \left. \exp\left( x^{\mu\nu} \frac{\pa}{\pa x_1^\mu} \, \frac{\pa}{\pa x_2^\nu} \right)
     f(x^{(1)}+x^{(1)}_1,x^{(2)}) \, g(x^{(1)} + x^{(1)}_2,x^{(2)}) \right|_{x_1 = x_2 =0}
\eez
which is the usual Moyal product (if the symmetric part of $x^{\mu \nu}$ vanishes).
\vskip.2cm

\noindent
{\bf Proposition.}
{\em The $\ast_n$-product is associative. }
\vskip.1cm
\noindent
{\it Proof:} According to the definition of the $\ast_n$-product, we have
\bez
    (f \ast_n (g \ast_n h))(x)
 &=& \exp \left( \sum_{r=1}^n x^{\mu_1 \ldots \mu_r} \sum_{k=0}^r
     {\pa \over \pa x_1^{\mu_1 \ldots \mu_k}}
     {\pa \over \pa x_2^{\mu_{k+1} \ldots \mu_r}} \right) \times   \nonumber \\
 & & \left. \exp \left( \sum_{s=1}^n x_2^{\nu_1 \ldots \nu_s} \sum_{l=0}^s
     {\pa \over \pa x_3^{\nu_1 \ldots \nu_l}}
     {\pa \over \pa x_4^{\nu_{l+1} \ldots \nu_s}} \right)
     f(x_1) g(x_3) h(x_4) \right|_{x_1=x_2=0 \atop x_3=x_4=0} \; .
\eez
This depends on $x_2$ only through the second exponential.
On functions which do \emph{not} dependent on $x_2$, we find
\bez
  & &  \exp \left( \sum_{r=1}^n x^{\mu_1 \ldots \mu_r}
       \sum_{k=0}^r {\pa \over \pa x_1^{\mu_1 \ldots \mu_k}}
       {\pa \over \pa x_2^{\mu_{k+1} \ldots \mu_r}} \right)
       \exp \left( \sum_{s=1}^n x_2^{\nu_1 \ldots \nu_s}
       \sum_{l=0}^s {\pa \over \pa x_3^{\nu_1 \ldots\nu_l}}
       {\pa \over \pa x_4^{\nu_{l+1} \ldots \nu_s}} \right)    \\
  &=&  \exp \left( \sum_{r=1}^n x^{\mu_1 \ldots \mu_r}
       \sum_{k=0}^r {\pa \over \pa x_1^{\mu_1 \ldots \mu_k}}
       \sum_{l=k}^r {\pa \over \pa x_3^{\mu_{k+1} \ldots \mu_l}}
       {\pa \over \pa x_4^{\mu_{l+1} \ldots \mu_r}} \right)
       \exp \left( \sum_{s=1}^n x_2^{\nu_1 \ldots \nu_s}
       \sum_{l=0}^s {\pa \over \pa x_3^{\nu_1 \ldots\nu_l}}
       {\pa \over \pa x_4^{\nu_{l+1} \ldots \nu_s}} \right)    \\
  &=& \exp \left( \sum_{s=1}^n x_2^{\nu_1 \ldots \nu_s}
      \sum_{l=0}^s {\pa \over \pa x_3^{\nu_1 \ldots \nu_l}}
      {\pa \over \pa x_4^{\nu_{l+1} \ldots \nu_s}} \right)
      \exp \left( \sum_{r=1}^n x^{\mu_1 \ldots \mu_r} S^{1,3,4}_{\mu_1 \ldots \mu_r} \right)
\eez
where
\bez
    \lefteqn{S^{1,3,4}_{\mu_1 \ldots \mu_r}
 := {\pa \over \pa x_1^{\mu_1 \ldots \mu_r}}
    + {\pa \over \pa x_3^{\mu_1 \ldots \mu_r}}
    + {\pa \over \pa x_4^{\mu_1 \ldots \mu_r}}
    + \sum_{0<k<l< r} {\pa \over \pa x_1^{\mu_1 \ldots \mu_k}}
       {\pa \over \pa x_3^{\mu_{k+1} \ldots \mu_l}}
       {\pa \over \pa x_4^{\mu_{l+1} \ldots \mu_r}}}
 & & \nonumber \\
 & & + \sum_{0<k<r} {\pa \over \pa x_1^{\mu_1 \ldots \mu_k}}
         {\pa \over \pa x_3^{\mu_{k+1} \ldots \mu_r}}
     + \sum_{0<k<r} {\pa \over \pa x_1^{\mu_1 \ldots \mu_k}}
         {\pa \over \pa x_4^{\mu_{k+1} \ldots \mu_r}}
     + \sum_{0<k<r} {\pa \over \pa x_3^{\mu_1 \ldots \mu_k}}
         {\pa \over \pa x_4^{\mu_{k+1} \ldots \mu_r}}
\eez
is completely symmetric in the labels 1,3,4. As a consequence, we obtain
\bez
   (f \ast_n (g \ast_n h))(x)
 = \exp \left( \sum_{r=1}^n x^{\mu_1 \ldots \mu_r} S^{1,3,4}_{\mu_1 \ldots \mu_r} \right)
    f(x_1) g(x_3) h(x_4) \Bigg|_{x_1=x_3=x_4=0}
\eez
and a similar calculation yields the same expression for $((f \ast_n g) \ast_n h)(x)$.
\hfill $\blacksquare$
\vskip.2cm

Partial differentiation with respect to $x^{\mu_1 \ldots \mu_r}$ acts on a
$\ast_n$-product as follows,
\be
   {\pa \over \pa x^{\mu_1 \ldots \mu_r}}(f \ast_n g)
 = \sum_{k=0}^r {\pa f \over \pa x^{\mu_1 \ldots \mu_k}} \ast_n
     {\pa g \over \pa x^{\mu_{k+1} \ldots \mu_r}}
   \qquad \quad r \leq n \; .     \label{ast_n-diff}
\ee

Suppose we impose an additional condition of the form
\be
        a_{\mu_1 \ldots \mu_r} \, x^{\mu_1 \ldots \mu_r} = 0
\ee
with constants $a_{\mu_1 \ldots \mu_r}$ on the deformation
parameters. The multiplication rule (\ref{ast_n-mult-x})
then leads to the further compatibility conditions
\be
  0 &=& x^\nu \ast_n (a_{\mu_1 \ldots \mu_r} x^{\mu_1 \ldots \mu_r})
     =  a_{\mu_1 \ldots \mu_r} x^\nu \ast_n x^{\mu_1 \ldots \mu_r} \nonumber \\
    &=& x^\nu (a_{\mu_1 \ldots \mu_r} x^{\mu_1 \ldots \mu_r})
        + a_{\mu_1 \ldots \mu_r} \, x^{\nu \mu_1 \ldots \mu_r}
     = a_{\mu_1 \ldots \mu_r} \, x^{\nu \mu_1 \ldots \mu_r} \; .
\ee
More generally, for all $p,q = 0,1,2 \ldots$ we find
\be
   a_{\mu_1 \ldots \mu_r} \, x^{\nu_1 \ldots \nu_p \mu_1 \ldots \mu_r \rho_1 \ldots \rho_q} = 0
   \; .
\ee

\section*{Appendix D: Left $\A(P)$-modules and Baker-Akhiezer functions}
\setcounter{section}{4}
\setcounter{equation}{0}
\addcontentsline{toc}{section}{\numberline{}Appendix D: Left $\A(P)$-modules and
Baker-Akhiezer functions}
Let $M$ be a left $\A$-module, so that $\alpha \prec m$ and $\alpha \bullet m$ are defined
for all $\alpha \in \A$ and $m \in M$ with the following associativity relations,
\be
    (\alpha \prec \beta) \prec m &=& \alpha \prec (\beta \prec m)  \qquad\quad
    (\alpha \bullet \beta) \bullet m = \alpha \bullet (\beta \bullet m)  \\
    (\alpha \prec \beta) \bullet m &=& \alpha \prec (\beta \bullet m) \qquad\quad
    (\alpha \bullet \beta) \prec m = \alpha \bullet (\beta \prec m) \, .
\ee
We further assume that $M$ is graded, i.e., $M = \bigoplus_{r \geq 0} M^r$
with $\A^r \prec M^s \subseteq M^{r+s}$ and $\A^r \bullet M^s \subseteq M^{r+s-1}$,
and that $M$ is completely determined by $M^0$ and $\prec$, so that
$M^r \subseteq \A^r \prec M^0$. This reduces the left actions on $M$ to
the definitions of $A \prec \chi$ and $A \bullet \chi$ for $A \in \A^1$ and $\chi \in M^0$.
Let $\alpha \succ m := \alpha \bullet m + \alpha \prec m$.
Furthermore, for $\chi \in M^0$, we set
\be
    \alpha \circ \chi := \alpha \succ \chi     \label{defcrc}
\ee
(which does \emph{not} hold for general $m \in M$).
The product $\circ$ then extends via the quasi-shuffle properties
\be
      (A \succ \alpha) \circ (B \succ m)
  &=& A \succ [\alpha \circ (B \succ m)] + B \succ [(A \succ \alpha) \circ m]
      - A \bullet B \succ \alpha \circ m  \qquad  \\
      (A \prec \alpha) \circ (B \prec m)
  &=& A \prec [\alpha \circ (B \prec m)] + B \prec [(A \prec \alpha) \circ m]
      + A \bullet B \prec \alpha \circ m  \\
      (A \succ \alpha) \circ (B \prec m)
  &=& A \succ [\alpha \circ (B \prec m)]
      + B \prec [(A \succ \alpha) \circ m]
\ee
which are consistent with (\ref{defcrc}).
By induction, one obtains
\be
    \alpha \circ (\beta \circ m) = (\alpha \circ \beta) \circ m
\ee
for $\alpha,\beta \in \A$ and $m \in M$. In fact, the proof is rather tedious and requires
several generalizations of results obtained for the algebra $\A$.
\vskip.1cm

In the following, we concentrate on a graded left $\A(P)$-module $M$.
Let $M_\mathcal{R}$ be the left module of $\mathcal{R}$ containing the Baker-Akhiezer
function of the ncKP hierarchy. We define a map $\tilde{\ell}: M \to M_\mathcal{R}$ by
\be
    \tilde{\ell}(\alpha \prec \chi) := - \ell(\alpha)_{<0} \ast \tilde{\ell}(\chi)
    \qquad\quad
    \tilde{\ell}(\alpha \succ \chi) := \ell(\alpha)_{\geq 0} \ast \tilde{\ell}(\chi)
    \label{tl-def}
\ee
for all $\alpha \in \A$ and $\chi \in M^0$. This leads to
\be
    \tilde{\ell}(\alpha \bullet \chi) = \ell(\alpha) \ast \tilde{\ell}(\chi) \; .
\ee
Furthermore, we define linear operators $\delta_{m_1 \ldots m_r}$ on $M_\mathcal{R}$
by setting
\be
     \delta_{m_1 \ldots m_r} \tilde{\ell}(m)
  := \tilde{\ell}( P_{m_1 \ldots m_r} \circ m)
\ee
and requiring the generalized derivation rule
\be
   \delta_{m_1 \ldots m_r} ( X \ast \tilde{\ell}(\chi) )
 = \sum_{k=0}^r (\delta_{m_1 \ldots m_k} X)
      \ast \delta_{m_{k+1} \ldots m_r} \tilde{\ell}(\chi)
\ee
for all $X \in \mathcal{R}$.
Using (\ref{defcrc}) and (\ref{tl-def}), we obtain
\be
    \delta_{m_1 \ldots m_r} \tilde{\ell}(\chi)
 &=& \tilde{\ell}( P_{m_1 \ldots m_r} \circ \chi )
  = \tilde{\ell}( P_{m_1 \ldots m_r} \succ \chi )
  = \ell( P_{m_1 \ldots m_r} )_{\geq 0} \ast \tilde{\ell}( \chi ) \nonumber \\
 &=& L^{m_1,\ldots m_r}{}_{\geq 0}
    \ast \tilde{\ell}(\chi) \; .  \label{lmn}
\ee
Let us call $\chi \in M^0$ a \emph{Baker-Akhiezer element} if it satisfies
\be
    P \bullet \chi = \la \, \chi
\ee
with $\la \in \mathbb{K}$.
Acting with $\tilde{\ell}$ on this equation leads to
\be
    L \ast \tilde{\ell}(\chi) = \la \, \tilde{\ell}(\chi) \; .
\ee
Together with (\ref{lmn}), this is equivalent to the linear system (\ref{linsys}).


\begin{thebibliography}{10}

\bibitem{EGR97}
Etingof P, Gelfand I and Retakh V 1997
Factorization of differential operators, quasideterminants, and
nonabelian Toda field equations
{\em Math. Research Lett.} {\bf 4} 413--425

\bibitem{Kupe00}
Kupershmidt B A 2000
KP or mKP {\em Mathematical Surveys and Monographs} vol 78
(Providence, RI: American Math. Soc.)

\bibitem{Dick03}
Dickey L A 2003
{\em Soliton Equations and Hamiltonian Systems}
(Singapore: World Scientific)

\bibitem{Hama03b}
Hamanaka M 2003
Commuting flows and conservation laws for noncommutative Lax hierarchies
{\em Preprint} hep-th/0311206

\bibitem{DMH04hier}
Dimakis A and M\"uller-Hoissen F 2004
Extension of noncommutative soliton hierarchies
{\em J. Phys. A: Math. Gen.} {\bf 37} 4069--4084

\bibitem{DMH04ncKP}
Dimakis A and M\"uller-Hoissen F 2004
Explorations of the extended ncKP hierarchy
{\em J. Phys. A: Math. Gen.} {\bf 37} 10899--10930

\bibitem{DMH04extMoyal}
Dimakis A and M\"uller-Hoissen F 2004
Extension of Moyal-deformed hierarchies of soliton equations
{\em XI International Conference Symmetry Methods in Physics}
ed \v{C} Burdik, O Navr{\'a}til and S Po\v{s}ta (Dubna: JINR),
{\em Preprint} nlin.SI/0408023

\bibitem{Okhu+Wada83}
Okhuma K and Wadati M 1983
The Kadomtsev-Petviashvili equation: the trace method and the
soliton resonances
{\em J. Phys. Soc. Japan} {\bf 52} 749--760

\bibitem{Pani01}
Paniak L D 2001
Exact noncommutative KP and KdV multi-solitons
{\em Preprint} hep-th/0105185

\bibitem{Groe46}
Groenewold H J 1946
On the principles of elementary quantum mechanics
{\em Physica} {\bf 12} 405--460

\bibitem{Moya49}
Moyal J E 1949
Quantum mechanics as a statistical theory
{\em Proc. Cambridge Phil. Soc.} {\bf 45} 99--124

\bibitem{BFFLS78}
Bayen F, Flato M, Fronsdal C, Lichnerowicz A and Sternheimer D 1978
Deformation theory and quantization
{\em Ann. Phys.} {\bf 111} 61--151

\bibitem{Duet+Fred01}
D\"utsch M and Fredenhagen K 2001
Perturbative algebraic field theory, and deformation quantization
{\em Fields Institute Communications} {\bf 30} 151--160

\bibitem{Baxt60}
Baxter G 1960
An analytic problem whose solution follows from a simple algebraic identity
{\em Pacific J. Math.} {\bf 10} 731--742

\bibitem{Rota69I}
Rota G C 1969
Baxter algebras and combinatorial identities. I
{\em Bull. Amer. Math. Soc.} {\bf 75} 325--329

\bibitem{Rota+Smit72}
Rota G C and Smith D A 1972
Fluctuation theory and Baxter algebras
{\em Symposia Mathematica} {\bf IX} 179--201

\bibitem{Rota95}
Rota G C 1995
Baxter operators, an introduction
{\em Gian-Carlo Rota on Combinatorics, Introductory Papers and Commentaries}
ed J P S Kung (Boston: Birkh\"auser) pp 504--512

\bibitem{Conn+Krei99}
Connes A and Kreimer D 1999
Renormalization in quantum field theory and the Riemann-Hilbert problem
{\em JHEP} 9909:024

\bibitem{Krei03}
Kreimer D 2003
New mathematical structures in renormalizable quantum field theories
{\em Annals Phys.} {\bf 303} 179--202

\bibitem{Figu+Grac04}
Figueroa H and Gracia-Bondia J M 2004
The uses of Connes and Kreimer's algebraic formulation of renormalization theory
{\em Int. J. Mod. Phys. A} {\bf 19} 2739--2754

\bibitem{Manc04}
Manchon D 2004
Hopf algebras, from basics to applications to renormalization
{\em Preprint} math.QA/0408405

\bibitem{EGK04int1}
Ebrahimi-Fard K, Guo L and Kreimer D 2004
Integrable renormalization I: the ladder case
{\em J. Math. Phys.} {\bf 45} 3758--3769

\bibitem{EGK04int2}
Ebrahimi-Fard K, Guo L and Kreimer D 2004
Integrable renormalization II: the general case
{\em Preprint} hep-th/0403118

\bibitem{Seme84}
Semenov-Tian-Shansky M A 1984
What is a classical $r$-matrix?
{\em Funct. Anal. Appl.} {\bf 17} 259--272

\bibitem{Seme02}
Semenov-Tian-Shansky M A 2002
Integrable systems and factorization problems
{\em Preprint} nlin.SI/0209057

\bibitem{Fadd+Takh87}
Faddeev L D and Takhtajan L A 1987
{\em Hamiltonian Methods in the Theory of Solitons}
(Berlin: Springer)

\bibitem{Guo+Keig00shuffle}
Guo L and Keigher W 2000
Baxter algebras and shuffle products
{\em Adv. Math.} {\bf 150} 117--149

\bibitem{Guo00diff}
Guo L 2000
Baxter algebras and differential algebras
{\em Differential algebra and related topics}
ed L Guo, W F Keigher, P J Cassidy and W Y Sit
(Newark, NJ: World Scientific) pp 281--305

\bibitem{Swee69}
Sweedler M E 1969
{\em Hopf Algebras}
(New York: Benjamin)

\bibitem{Hoff00}
Hoffman M E 2000
Quasi-shuffle products
{\em J. Algebraic Combin.} {\bf 11} 49--68

\bibitem{Guo00free}
Guo L 2000
Properties of free Baxter algebras
{\em Adv. Math.} {\bf 151} 346--374

\bibitem{Guo+Keig00Bax-compl}
Guo L and Keigher W 2000
On free Baxter algebras: completions and the internal construction
{\em Adv. Math.} {\bf 151} 101--127

\bibitem{Ebra+Guo04}
Ebrahimi-Fard K and Guo L 2004
Quasi-shuffles, mixable shuffles and Hopf algebras
{\em Preprint}

\bibitem{Carl49}
Carlitz L 1949
Some properties of Hurwitz series
{\em Duke Math. J.} {\bf 16} 285--295

\bibitem{Macd79}
Macdonald I G 1979
{\em Symmetric functions and Hall polynomials}
(Oxford: Clarendon Press)

\bibitem{Spit56}
Spitzer F 1956
A combinatorial lemma and its application to probability theory
{\em Trans. Amer. Math. Soc.} {\bf 82} 323--339

\bibitem{Cart72}
Cartier P 1972
On the structure of free Baxter algebras
{\em Adv. Math.} {\bf 9} 253--265

\bibitem{EGK04Spitzer}
Ebrahimi-Fard K, Guo L and Kreimer D 2004
Spitzer's identity and the algebraic Birkhoff decomposition in pQFT
{\em J. Phys. A: Math. Gen.} {\bf 37} 11037--11052

\bibitem{Olve+Shak92}
Olver P J and Shakiban C 1992
Dissipative decomposition of partial differential equations
{\em Rocky Mountain J. Math.} {\bf 22} 1483--1510

\bibitem{Stem85}
Stembridge J R 1985
A characterization of supersymmetric polynomials
{\em J. Algebra} {\bf 95} 439--444

\bibitem{MNR81}
Metropolis N, Nicoletti G and Rota G C 1981
A new class of symmetric functions
{\em Math. Anal. Appl., Part B, Adv. Math. Suppl. Stud.}
{\bf 7B} 563--575

\bibitem{Mole+Reta04}
Molev A and Retakh V 2004
Quasideterminants and Casimir elements for the general Lie superalgebra
{\em Intern. Math. Res. Notes} (13) 611--619

\bibitem{MSS90}
Matsukidaira J, Satsuma J and Strampp W 1990
Conserved quantities and symmetries of KP hierarchy
{\em J. Math. Phys.} {\bf 31} 1426--1434

\bibitem{Gutt+Rawn99}
Gutt S and Rawnsley J 1999
Equivalence of star products on a symplectic manifold; an
introduction to Deligne's \v{C}ech cohomology classes
{\em J. Geom. Phys.} {\bf 29} 347--392

\bibitem{Taka94}
Takasaki K 1994
Nonabelian KP hierarchy with Moyal algebraic coefficients
{\em J. Geom. Phys.} {\bf 14} 332--364

\bibitem{Saka04}
Sakakibara M 2004
Factorization methods for noncommutative KP and Toda hierarchy
{\em J. Phys. A: Math. Gen.} {\bf 37} L599--L604

\bibitem{HSS92}
Haak G, Schmidt M and Schrader R 1992
Group theoretic formulation of the Segal-Wilson approach to
integrable systems with applications
{\em Rev. Math. Phys.} {\bf 4} 451--499

\bibitem{AGR87}
Alvarez-Gaum{\'e} L, Gomez C and Reina C 1987
Loop groups, Grassmanians and string theory
{\em Phys. Lett. B} {\bf 190} 55--62 \\
Alvarez-Gaum{\'e} L, Gomez C and Reina C 1987
New methods in string theory
{\em Superstrings `87}
ed L Alvarez-Gaum{\'e}, M B Green, M T Grisaru, R Iengo and E Sezgin
(Teaneck, NJ: World Scientific Publishing) pp 341--422

\bibitem{Sait87}
Saito S 1987
String amplitudes as solutions to soliton equations
{\em Phys. Rev. D} {\bf 36} 1819--1826 \\
Saito S 1987
String theories and Hirota's bilinear difference equation
{\em Phys. Rev. Lett.} {\bf 59} 1798--1801 \\
Sogo K 1987
A way from string to soliton - introduction of KP coordinate to
  string amplitudes -
{\em J. Phys. Soc. Japan} {\bf 56} 2291--2297 \\
Yamaguchi H and Saito S 2004
A realization of matrix KP hierarchy by coincident D-brane states
{\em Preprint} hep-th/0412056

\bibitem{Mula88}
Mulase M 1988
KP equations, strings, and the Schottky problem
{\em Algebraic Analysis} vol II
ed M Kashiwara and T Kawai
(Boston: Academic Press) pp 473--492

\bibitem{Witt91}
Witten E 1991
Two-dimensional gravity and intersection theory on moduli space
{\em Surveys Diff. Geom.} {\bf 1} 243--310 \\
Witten E 1993
Algebraic geometry associated with matrix models of two-dimensional gravity
{\em Topological methods in modern mathematics}
ed L R Goldberg and A V Phillips
(Houston, TX: Publish or Perish) pp 235--269

\bibitem{Dijk92}
Dijkgraaf R 1992
Intersection theory, integrable hierarchies and topological field theory
{\em Preprint} hep-th/9201003

\bibitem{Lee03}
Lee Y-P 2003
Witten's conjecture, Virasoro conjecture, and invariance of tautological equations
{\em Preprint} math/0311100

\bibitem{Bono+Xion92}
Bonora L and Xiong C S 1992
An alternative approach to KP hierarchy in matrix models
{\em Phys. Lett. B} {\bf 285} 191--198

\bibitem{KKN99}
Kazakov V, Kostov I and Nekrasov N 1999
D-particles, matrix integrals and KP hierarchy
{\em Nucl. Phys. B} {\bf 557} 413--442

\bibitem{Eguc+Yang96}
Eguchi T and Yang S K 1996
A new description of the {$E_6$} singularity
{\em Preprint} hep-th/9612086

\bibitem{ASvM98}
Adler M, Shiota T and van Moerbeke P 1998
Random matrices, Virasoro algebras, and noncommutative KP
{\em Duke Math. J.} {\bf 94} 379--431

\bibitem{Seib+Witt99}
Seiberg N and Witten E 1999
String theory and noncommutative geometry
{\em JHEP} 9909:032

\bibitem{Lee00}
Lee T 2000
Noncommutative Dirac-Born-Infeld action for $D$-brane
{\em Phys.Lett. B} {\bf 478} 313--319

\bibitem{Doug+Nekr01}
Douglas M R and Nekrasov N A 2001
Noncommutative field theory
{\em Rev. Mod. Phys.} {\bf 73} 977--1029

\bibitem{ACH03}
Ablowitz M J, Chakravarty S and Halburd R G 2003
Integrable systems and reductions of the self-dual Yang-Mills equations
{\em J. Math. Phys.} {\bf 44} 3147--3173

\bibitem{Atki63}
Atkinson F V 1963
Some aspects of Baxter's functional equation
{\em J. Math. Anal. Appl.} {\bf 7} 1--30

\bibitem{Guo01umbral}
Guo L 2001
Baxter algebras and the umbral calculus
{\em Adv. Applied Math.} {\bf 27} 405--426

\bibitem{AGLK03}
Andrews G E, Guo L, Keigher W and Ono K 2003
Baxter algebras and Hopf algebras
{\em Trans. Amer. Math. Soc.} {\bf 355} 4639--4656

\bibitem{Ebra+Guo04_RB}
Ebrahimi-Fard K and Guo L 2004
Rota-Baxter algebras, dendriform algebras and Poincar{\'e}-Birkhoff-Witt theorem
{\em Preprint}

\bibitem{Ebra02}
Ebrahimi-Fard K 2002
Loday-type algebras and the Rota-Baxter relation
{\em Lett. Math. Phys.} {\bf 61} 139--147

\bibitem{Loda+Ronc04}
Loday J-L and Ronco M A 2004
Trialgebras and families of polytopes
{\em Contemp. Math.} {\bf 346} 369--398

\bibitem{Fuch97}
Fuchssteiner B 1997
Compatibility in abstract algebraic structures
{\em Algebraic Aspects of Integrable Systems}
ed A S Fokas and I M Gelfand
(Boston: Birkh\"auser) pp 131--141

\bibitem{CGM00}
Cari{\~n}ena J, Grabowski J and Marmo G 2000
Quantum bi-Hamiltonian systems
{\em J. Mod. Phys. A} {\bf 15} 4797--4810

\bibitem{Ebra04}
Ebrahimi-Fard K 2004
On the associative {N}ijenhuis relation
{\em Electronic J. Comb.} {\bf 11} R38

\end{thebibliography}
\end{document}